# Stationary state distribution and efficiency analysis of the Langevin equation *via* real or virtual dynamics

Dezhang Li[1], Xu Han[1], Yichen Chai[1], Cong Wang[2], Zhijun Zhang[1], Zifei Chen[1], Jian Liu[1, a)], Jiushu Shao[2, b)]

1. Beijing National Laboratory for Molecular Sciences, Institute of Theoretical and Computational Chemistry, College of Chemistry and Molecular Engineering, Peking University, Beijing 100871, China
2. College of Chemistry and Center for Advanced Quantum Studies, Key Laboratory of Theoretical and Computational Photochemistry, Ministry of Education, Beijing Normal University, Beijing 100875, China

a) Electronic mail: jianliupku@pku.edu.cn

b) Electronic mail: jiushu@bnu.edu.cn

**Abstract**

Langevin dynamics has become a popular tool to simulate the Boltzmann equilibrium distribution. When the repartition of the Langevin equation involves the exact realization of the Ornstein-Uhlenbeck noise, in addition to the conventional density evolution, there exists another type of discrete evolution that may not correspond to a continuous, real dynamical counterpart. This *virtual* dynamics case is also able to produce the desired stationary distribution. Different types of repartition lead to different numerical schemes, of which the accuracy and efficiency are investigated through studying the harmonic oscillator potential, an analytical solvable model. By analyzing the asymptotic distribution and characteristic correlation time that are derived by either directly solving the discrete equations of motion or using the related phase space propagators, it is shown that the optimal friction coefficient resulting in the minimum characteristic correlation time depends on the time interval chosen in the numerical implementation. When the recommended "middle" scheme is employed, both analytical and numerical results demonstrate that for good numerical performance in efficiency as well as accuracy, one may choose a friction coefficient in a wide range from around the optimal value to the high friction limit.

## I. Introduction

The behavior of complicated systems, either social or natural, is often described by the stochastic differential equation (SDE). The numerical simulation of SDE is nowadays an important theme in such diverse fields from finance, ecology, to physics and chemistry[1-3]. In the present paper we will analytically derive the stationary phase-space density of a Brownian system described by SDEs according to different numerical algorithms, while the characteristic correlation time will also be analytically obtained in the harmonic limit. While the former indicates the accuracy of the algorithm, the latter suggests the efficiency.

Assume the (time-independent) Hamiltonian of the system $H$ to be of standard Cartesian form

$$H = \mathbf{p}^T \mathbf{M}^{-1} \mathbf{p} / 2 + U(\mathbf{x}) \quad . \tag{1}$$

Here $\mathbf{M}$ is the diagonal "mass matrix" with elements $\{m_j\}$, and $\mathbf{p}$ and $\mathbf{x}$ are the momentum and coordinate vectors, respectively. $N$ is the number of particles and $3N$ is the total number of degrees of freedom. ($3N$ becomes one when a one-dimensional one-particle system is studied.) It is also assumed that the system is in the heat bath at temperature $T$. Define $\beta = 1/k_B T$ with $k_B$ as the Boltzmann constant.

Brownian motion of the system may be described by the Langevin equation[4]

$$d\mathbf{x}_t = \mathbf{M}^{-1}\mathbf{p}_t dt , \tag{2}$$

$$d\mathbf{p}_t = -\nabla_{\mathbf{x}_t} U(\mathbf{x}_t) dt - \boldsymbol{\gamma}\mathbf{p}_t dt + \boldsymbol{\sigma}\mathbf{M}^{1/2} d\mathbf{W}_t , \tag{3}$$

where $\mathbf{W}_t$ is a vector of $3N$-dimensional independent Wiener processes, $\boldsymbol{\gamma}$ is often a diagonal friction matrix with positive elements, and $\boldsymbol{\sigma} = \sqrt{2/\beta}\,\boldsymbol{\gamma}^{1/2}$. Here and in the following a function $F$ of time $t$ ($F(t)$) is also denoted as $F_t$ for abbreviation. Note that the relation between the matrix $\boldsymbol{\sigma}$ and the friction matrix $\boldsymbol{\gamma}$ is based on the fluctuation-dissipation theorem, which guarantees that the steady state of the Langevin system satisfies the Boltzmann distribution $e^{-\beta H(\mathbf{x},\mathbf{p})}$. Various numerical algorithms have been proposed to use the Langevin equation as a type of thermostat to obtain the desired Boltzmann distribution. While some algorithms were originally proposed without employing the Lie-Trotter splitting[5-11], some researchers involved the splitting to derive integrators for Langevin dynamics[12-25]. Leimkuhler and Matthews have recently compared a few numerical algorithms for Langevin dynamics in the high friction limit[20-22] for their performances in accuracy.

In addition to accuracy, sampling efficiency is often another important factor to consider when the Langevin equation is employed as a type of thermostat. The main purpose of the paper is to present two theoretical approaches to study both the accuracy and efficiency of the numerical algorithm. One approach is derived by solving the discrete equations of motion (the trajectory-based approach), and the other by using the related phase space propagators (the phase space propagator approach). Most practical algorithms employ second-order (splitting) schemes, because higher order (splitting) ones[11, 26-34] are more complicated and do not necessarily offer more economic algorithms for general molecular systems, because they involve either more force calculations or second-order derivatives (or even higher order derivatives) of the potential energy surface[25]. We then focus the exploration on various second-order algorithms (more accurately, on algorithms that lead to second-order accuracy for the stationary distribution). The paper is organized as follows: In Section II we use the Langevin equation to justify the Fokker-Planck equation (or the forward Kolmogorov equation) that the phase-space density $\rho(\mathbf{x},\mathbf{p},t)$ satisfies, and then show the stationary state of $\rho(\mathbf{x},\mathbf{p},t)$ is the Boltzmann distribution. The procedure will also be used to analyze the stationary state and characteristic correlation time for a numerical algorithm when time is discretized. Section III summarizes three types of repartition and their schemes of numerical algorithms. Section IV then offers two different approaches to obtain the stationary state for each scheme (for both real and virtual dynamics if the latter is available) for the harmonic system, while Section V analyzes the characteristic correlation time and suggests the optimal value of the friction coefficient for the underlying algorithm. Section VI demonstrates some numerical examples beyond the harmonic limit to verify the results derived from the analytical analysis. Our conclusion remarks are presented in Section VII.

## II. Boltzmann distribution as the stationary state of Langevin dynamics

Note that $\rho(\mathbf{x},\mathbf{p},t) \equiv \langle \delta(\mathbf{x}_t - \mathbf{x})\delta(\mathbf{p}_t - \mathbf{p}) \rangle$, where $\langle \ \rangle$ denotes the stochastic averaging over the Wiener process $\mathbf{W}_t$. With the chain rule and Eqs. (2) and (3) the derivative of $\rho$ with respect to time $t$ is

$$\frac{\partial}{\partial t}\rho(\mathbf{x},\mathbf{p},t) = -\nabla_{\mathbf{x}} \cdot \langle \dot{\mathbf{x}}_t \delta(\mathbf{x}_t - \mathbf{x})\delta(\mathbf{p}_t - \mathbf{p}) \rangle - \nabla_{\mathbf{p}} \cdot \langle \dot{\mathbf{p}}_t \delta(\mathbf{x}_t - \mathbf{x})\delta(\mathbf{p}_t - \mathbf{p}) \rangle \tag{4}$$

$$\begin{aligned} &= -\mathbf{M}^{-1}\nabla_{\mathbf{x}} \cdot \langle \mathbf{p}_t \delta(\mathbf{x}_t - \mathbf{x})\delta(\mathbf{p}_t - \mathbf{p}) \rangle + \nabla_{\mathbf{p}} \cdot \left\langle \nabla_{\mathbf{x}_t} U(\mathbf{x}_t)\delta(\mathbf{x}_t - \mathbf{x})\delta(\mathbf{p}_t - \mathbf{p}) \right\rangle \\ &\quad + \boldsymbol{\gamma}\nabla_{\mathbf{p}} \cdot \langle \mathbf{p}_t \delta(\mathbf{x}_t - \mathbf{x})\delta(\mathbf{p}_t - \mathbf{p}) \rangle - \boldsymbol{\sigma}\mathbf{M}^{1/2}\nabla_{\mathbf{p}} \cdot \langle \boldsymbol{\eta}_t \delta(\mathbf{x}_t - \mathbf{x})\delta(\mathbf{p}_t - \mathbf{p}) \rangle \end{aligned} \tag{5}$$

where $\boldsymbol{\eta}_t$ formally stands for $d\mathbf{W}_t/dt$ and is called the white noise vector. The property of the $\delta$-function allows us to obtain

$$\langle \mathbf{x}_t \delta(\mathbf{x}_t - \mathbf{x})\delta(\mathbf{p}_t - \mathbf{p}) \rangle = \mathbf{x}\rho(\mathbf{x},\mathbf{p},t) \quad ,$$

$$\left\langle \nabla_{\mathbf{x}_t} U(\mathbf{x}_t)\delta(\mathbf{x}_t - \mathbf{x})\delta(\mathbf{p}_t - \mathbf{p}) \right\rangle = \nabla_{\mathbf{x}} U(\mathbf{x})\rho(\mathbf{x},\mathbf{p},t) \quad ,$$

and

$$\langle \mathbf{p}_t \delta(\mathbf{x}_t - \mathbf{x})\delta(\mathbf{p}_t - \mathbf{p}) \rangle = \mathbf{p}\rho(\mathbf{x},\mathbf{p},t) \quad .$$

Upon resorting to the Furutsu-Novikov theorem[3] and Eq. (3), there is

$$\begin{aligned}\left\langle \boldsymbol{\eta}_t \delta(\mathbf{x}_t - \mathbf{x})\delta(\mathbf{p}_t - \mathbf{p})\right\rangle &= -\frac{1}{2}\nabla_{\mathbf{p}} \left\langle \frac{\delta \mathbf{p}_t}{\delta \boldsymbol{\eta}_t} \delta(\mathbf{x}_t - \mathbf{x})\delta(\mathbf{p}_t - \mathbf{p})\right\rangle \\ &= -\frac{1}{2}\boldsymbol{\sigma}\mathbf{M}^{1/2}\nabla_{\mathbf{p}}\rho(\mathbf{x},\mathbf{p},t) \ .\end{aligned} \tag{6}$$

Substituting these relations into Eq. (5) using a simplified notation $\rho$ for the density distribution, we obtain

$$\frac{\partial}{\partial t}\rho = -\left(\mathbf{M}^{-1}\mathbf{p}\right)\cdot\nabla_{\mathbf{x}}\rho + \nabla_{\mathbf{x}}U(\mathbf{x})\cdot\nabla_{\mathbf{p}}\rho + \left(\boldsymbol{\gamma}\nabla_{\mathbf{p}}\right)\cdot\left(\mathbf{p}\rho\right) + \left(\frac{1}{2}\boldsymbol{\sigma}^2\mathbf{M}\nabla_{\mathbf{p}}\right)\cdot\nabla_{\mathbf{p}}\rho \ , \tag{7}$$

which is the Fokker-Planck equation (or the forward Kolmogorov equation) for Langevin dynamics. One can simply recast Eq. (7) as $\partial\rho/\partial t = \mathcal{L}\rho$, where the relevant Kolmogorov operator for the right-hand side (RHS) of Eq. (7) is

$$\mathcal{L}\rho = -\left(\mathbf{M}^{-1}\mathbf{p}\right)\cdot\nabla_{\mathbf{x}}\rho + \nabla_{\mathbf{x}}U(\mathbf{x})\cdot\nabla_{\mathbf{p}}\rho + \nabla_{\mathbf{p}}\cdot\left(\boldsymbol{\gamma}\mathbf{p}\rho\right) + \frac{1}{\beta}\nabla_{\mathbf{p}}\cdot\left(\boldsymbol{\gamma}\mathbf{M}\nabla_{\mathbf{p}}\rho\right) \ . \tag{8}$$

It is straightforward to show that the Boltzmann distribution $e^{-\beta H(\mathbf{x},\mathbf{p})}$ is a stationary state for

$$\partial\rho/\partial t = 0 \ . \tag{9}$$

Given an initial distribution, the transient behavior of the density distribution can in principle be computed by using Eq. (7). But we are only interested in the steady state for the Brownian system described by the difference equation resulted from Eqs. (2) and (3) as time is discretized, because the result may be useful for optimizing numerical algorithms. In the following we will give a brief introduction to the frequently used numerical algorithms for solving Eqs. (2) and (3) and study the corresponding stationary states for the harmonic system. A uniform time interval (or step size) $\Delta t$ will be adopted.

## III. Numerical algorithms

Several numerical simulation techniques have been developed for solving the Langevin system. The efficiency strongly depends on the underlying algorithm. A useful strategy to design numerical algorithms is based on the repartition of Eqs. (2) and (3).

**1. First type of repartition**

The first type of repartition of Eqs. (2) and (3) reads

$$\begin{bmatrix} d\mathbf{x}_t \\ d\mathbf{p}_t \end{bmatrix} = \underbrace{\begin{bmatrix} \mathbf{M}^{-1}\mathbf{p}_t \\ 0 \end{bmatrix} dt}_{\mathrm{x}} + \underbrace{\begin{bmatrix} 0 \\ -\nabla_{\mathbf{x}_t} U(\mathbf{x}_t) \end{bmatrix} dt}_{\mathrm{p}} + \underbrace{\begin{bmatrix} 0 \\ -\boldsymbol{\gamma}\mathbf{p}_t dt + \boldsymbol{\sigma}\mathbf{M}^{1/2} d\mathbf{W}_t \end{bmatrix}}_{\mathrm{T}} , \tag{10}$$

which allows one to take full advantage of the "solubility" of the three parts by splitting the evolution in one step into different sub-steps. Suppose the system starts with $\left(\mathbf{x}(t), \mathbf{p}(t)\right)$ at time $t$. When there is only the first term in the RHS of Eq. (10), then exact dynamics leads to the update relation

$$\begin{bmatrix} \mathbf{x}(t+\Delta t) \\ \mathbf{p}(t+\Delta t) \end{bmatrix} = \begin{bmatrix} \mathbf{x}(t) + \mathbf{M}^{-1}\mathbf{p}(t)\Delta t \\ \mathbf{p}(t) \end{bmatrix} . \tag{11}$$

Similarly, the other two solutions corresponding to the 2nd and 3rd terms respectively read

$$\begin{bmatrix} \mathbf{x}(t+\Delta t) \\ \mathbf{p}(t+\Delta t) \end{bmatrix} = \begin{bmatrix} \mathbf{x}(t) \\ \mathbf{p}(t) - \nabla U(\mathbf{x})\big|_{\mathbf{x}=\mathbf{x}(t)} \Delta t \end{bmatrix} , \tag{12}$$

$$\begin{bmatrix} \mathbf{x}(t+\Delta t) \\ \mathbf{p}(t+\Delta t) \end{bmatrix} = \begin{bmatrix} \mathbf{x}(t) \\ e^{-\boldsymbol{\gamma}\Delta t}\mathbf{p}(t) + \boldsymbol{\Omega}(t, \Delta t) \end{bmatrix} , \tag{13}$$

where in Eq. (13)

$$\boldsymbol{\Omega}(t, \Delta t) = \boldsymbol{\sigma}\mathbf{M}^{1/2} \int_t^{t+\Delta t} ds e^{-\boldsymbol{\gamma}(t+\Delta t - s)} \boldsymbol{\eta}_s . \tag{14}$$

It is straightforward to show that $\boldsymbol{\Omega}(t, \Delta t)$ for a fixed time $t$ is a Gaussian random vector with zero mean and diagonal deviation matrix[35]

$$\left\langle \boldsymbol{\Omega}(t, \Delta t) \boldsymbol{\Omega}^T(t, \Delta t) \right\rangle = \frac{1}{2} \boldsymbol{\sigma}^2 \mathbf{M} \boldsymbol{\gamma}^{-1} \left( \mathbf{1} - e^{-2\boldsymbol{\gamma}\Delta t} \right) \tag{15}$$

$$= \frac{1}{\beta} \mathbf{M} \left( \mathbf{1} - e^{-2\boldsymbol{\gamma}\Delta t} \right) , \tag{16}$$

where and in the following $\mathbf{1}$ denotes the unit matrix with a suitable dimension obvious in the context. Note that the third part is the Ornstein-Uhlenbeck (OU) process.

Consider a standard-Gaussian-random-number vector $\boldsymbol{\mu}(t,\Delta t)$ for a fixed time $t$ with zero mean

$$\left\langle \boldsymbol{\mu}(t,\Delta t)\right\rangle = 0 \tag{17}$$

and diagonal deviation matrix

$$\left\langle \boldsymbol{\mu}(t,\Delta t)\boldsymbol{\mu}^T(t,\Delta t)\right\rangle = \mathbf{1} \quad . \tag{18}$$

The Gaussian random vector can be recast as

$$\boldsymbol{\Omega}(t,\Delta t) = \sqrt{\frac{1}{\beta}}\mathbf{M}^{1/2}\left(\mathbf{1}-e^{-2\boldsymbol{\gamma}\Delta t}\right)^{1/2}\boldsymbol{\mu}(t,\Delta t) \quad . \tag{19}$$

In numerical implementation the element $\Omega^{(j)}$ of $\boldsymbol{\Omega}$ can be generated feasibly by using a standard-Gaussian-random-number generator with coefficient $\beta^{-1/2}M_j^{1/2}\left(1-e^{-2\gamma_j\Delta t}\right)^{1/2}$, where $M_j$ and $\gamma_j$ are the $j$-th diagonal elements of $\mathbf{M}$ and $\boldsymbol{\gamma}$, respectively.

We use $e^{\mathcal{L}\Delta t}$ to represent the phase space propagator for Eqs. (2)-(3) or for Eq. (7) when the infinitesimal time interval $dt$ becomes finite as $\Delta t$. Analogously, the phase space propagators for Eqs. (11)-(13) are denoted as $e^{\mathcal{L}_{\mathbf{x}}\Delta t}$, $e^{\mathcal{L}_{\mathbf{p}}\Delta t}$, and $e^{\mathcal{L}_T\Delta t}$, respectively. Here $\mathcal{L}_{\mathbf{x}}$, $\mathcal{L}_{\mathbf{p}}$, and $\mathcal{L}_T$ are the relevant Kolmogorov operators,

$$\mathcal{L}_{\mathbf{x}}\rho = -\left(\mathbf{M}^{-1}\mathbf{p}\right)\cdot\nabla_{\mathbf{x}}\rho \quad , \tag{20}$$

$$\mathcal{L}_{\mathbf{p}}\rho = \nabla_{\mathbf{x}}U(\mathbf{x})\cdot\nabla_{\mathbf{p}}\rho \quad , \tag{21}$$

$$\mathcal{L}_T\rho = \nabla_{\mathbf{p}}\cdot\left(\boldsymbol{\gamma}\mathbf{p}\rho\right)+\frac{1}{\beta}\nabla_{\mathbf{p}}\cdot\left(\boldsymbol{\gamma}\mathbf{M}\nabla_{\mathbf{p}}\rho\right) \quad , \tag{22}$$

where $\rho$ is a density distribution in the phase space.

Different splitting orders for Eq. (10) lead to different algorithms. We have four schemes that are reduced to the conventional velocity-Verlet algorithm ($e^{\mathcal{L}_{\mathbf{p}}\Delta t/2}e^{\mathcal{L}_{\mathbf{x}}\Delta t}e^{\mathcal{L}_{\mathbf{p}}\Delta t/2}$) for constant energy MD when the OU process is not included. Note that the conventional velocity-Verlet algorithm is symplectic.

**1) Middle scheme**

$$e^{\mathcal{L}\Delta t} \approx e^{\mathcal{L}^{\text{Middle}}\Delta t} = e^{\mathcal{L}_{\mathbf{p}}\Delta t/2} e^{\mathcal{L}_{\mathbf{x}}\Delta t/2} e^{\mathcal{L}_{T}\Delta t} e^{\mathcal{L}_{\mathbf{x}}\Delta t/2} e^{\mathcal{L}_{\mathbf{p}}\Delta t/2} \quad . \tag{23}$$

The thermostat process is arranged in the middle. As will be discussed in Section IV, in addition to real dynamics for the Langevin equation as conventionally studied, virtual dynamics may also be proposed in the scheme to obtain the Boltzmann distribution. Note that the BAOAB algorithm proposed by Leimkuhler and Matthews[20] is the real dynamics case of the "middle" scheme.

**2) End scheme**

$$e^{\mathcal{L}\Delta t} \approx e^{\mathcal{L}^{\text{End}}\Delta t} = e^{\mathcal{L}_{T}\Delta t} e^{\mathcal{L}_{\mathbf{p}}\Delta t/2} e^{\mathcal{L}_{\mathbf{x}}\Delta t} e^{\mathcal{L}_{\mathbf{p}}\Delta t/2} \quad . \tag{24}$$

The thermostat process is applied after the velocity-Verlet process. The real time dynamics case was given in Ref. [20].

**3) Beginning scheme**

$$e^{\mathcal{L}\Delta t} \approx e^{\mathcal{L}^{\text{Begin}}\Delta t} = e^{\mathcal{L}_{\mathbf{p}}\Delta t/2} e^{\mathcal{L}_{\mathbf{x}}\Delta t} e^{\mathcal{L}_{\mathbf{p}}\Delta t/2} e^{\mathcal{L}_{T}\Delta t} \quad . \tag{25}$$

The thermostat process is applied before the velocity-Verlet process. The real time dynamics case was presented in Ref. [18].

**4) Side scheme**

$$e^{\mathcal{L}\Delta t} \approx e^{\mathcal{L}^{\text{Side}}\Delta t} = e^{\mathcal{L}_{T}\Delta t/2} e^{\mathcal{L}_{\mathbf{p}}\Delta t/2} e^{\mathcal{L}_{\mathbf{x}}\Delta t} e^{\mathcal{L}_{\mathbf{p}}\Delta t/2} e^{\mathcal{L}_{T}\Delta t/2} \quad . \tag{26}$$

The thermostat process for half an interval $\Delta t/2$ is at each of the two sides (i.e., before and after the velocity-Verlet process). The real time dynamics case was suggested in Ref. [16].

Similarly, one can also obtain the schemes

**5) PV-middle scheme**

$$e^{\mathcal{L}\Delta t} \approx e^{\mathcal{L}^{\text{PV-middle}}\Delta t} = e^{\mathcal{L}_{\mathbf{x}}\Delta t/2} e^{\mathcal{L}_{\mathbf{p}}\Delta t/2} e^{\mathcal{L}_{T}\Delta t} e^{\mathcal{L}_{\mathbf{p}}\Delta t/2} e^{\mathcal{L}_{\mathbf{x}}\Delta t/2} \quad . \tag{27}$$

**6) PV-end scheme**

$$e^{\mathcal{L}\Delta t} \approx e^{\mathcal{L}^{\text{PV-end}}\Delta t} = e^{\mathcal{L}_{T}\Delta t} e^{\mathcal{L}_{\mathbf{x}}\Delta t/2} e^{\mathcal{L}_{\mathbf{p}}\Delta t} e^{\mathcal{L}_{\mathbf{x}}\Delta t/2} \quad . \tag{28}$$

**7) PV-beginning scheme**

$$e^{\mathcal{L}\Delta t} \approx e^{\mathcal{L}^{\text{PV-begin}}\Delta t} = e^{\mathcal{L}_{\mathbf{x}}\Delta t/2} e^{\mathcal{L}_{\mathbf{p}}\Delta t} e^{\mathcal{L}_{\mathbf{x}}\Delta t/2} e^{\mathcal{L}_{T}\Delta t} \quad . \tag{29}$$

**8) PV-side scheme**

$$e^{\mathcal{L}\Delta t} \approx e^{\mathcal{L}^{\text{PV-side}}\Delta t} = e^{\mathcal{L}_{T}\Delta t/2} e^{\mathcal{L}_{\mathbf{x}}\Delta t/2} e^{\mathcal{L}_{\mathbf{p}}\Delta t} e^{\mathcal{L}_{\mathbf{x}}\Delta t/2} e^{\mathcal{L}_{T}\Delta t/2} \quad . \tag{30}$$

The real dynamics cases of the four schemes were earlier studied by Leimkuhler *et al.* [20, 22]. These schemes are reduced to the position-Verlet (PV) algorithm ( $e^{\mathcal{L}_{\mathbf{x}}\Delta t/2} e^{\mathcal{L}_{\mathbf{p}}\Delta t} e^{\mathcal{L}_{\mathbf{x}}\Delta t/2}$ ) for constant energy MD as the OU process is not considered. (The PV algorithm is also symplectic.)

As will be discussed in Section V, the stationary state distribution produced by any one of the eight schemes for a harmonic system is *independent* of the Langevin friction coefficient $\gamma$. More schemes may be proposed with the repartition of Eq. (10), but their stationary state distributions for a harmonic system will often depend on the friction coefficient. [See supplementary material for more discussion.]

**2. Second type of repartition**

An alternative type of repartition of Eqs. (2) and (3) is

$$\begin{bmatrix} d\mathbf{x}_t \\ d\mathbf{p}_t \end{bmatrix} = \underbrace{\begin{bmatrix} 0 \\ -\nabla_{\mathbf{x}_t} U(\mathbf{x}_t)dt - \boldsymbol{\gamma}\mathbf{p}_t dt + \boldsymbol{\sigma}\mathbf{M}^{1/2} d\mathbf{W}_t \end{bmatrix}}_{\text{pT}} + \underbrace{\begin{bmatrix} \mathbf{M}^{-1}\mathbf{p}_t \\ 0 \end{bmatrix} dt}_{\text{x}} \quad . \tag{31}$$

When there is only the first term in Eq. (31), exact dynamics leads to the update relation

$$\begin{bmatrix} \mathbf{x}(t+\Delta t) \\ \mathbf{p}(t+\Delta t) \end{bmatrix} = \begin{bmatrix} \mathbf{x}(t) \\ e^{-\boldsymbol{\gamma}\Delta t}\mathbf{p}(t) - \boldsymbol{\gamma}^{-1}\left(\mathbf{1} - e^{-\boldsymbol{\gamma}\Delta t}\right)\nabla U(\mathbf{x})\big|_{\mathbf{x}=\mathbf{x}(t)} + \boldsymbol{\Omega}(t,\Delta t) \end{bmatrix} \quad , \tag{32}$$

where $\boldsymbol{\Omega}(t,\Delta t)$ is defined in the same way as in Eq. (14) or in Eq. (16). The phase space propagator for the first term in Eq. (31) is $e^{\mathcal{L}_{\mathbf{p}-T}\Delta t}$ for a finite time interval $\Delta t$, where the relevant Kolmogorov operator $\mathcal{L}_{\mathbf{p}-T}$ is

$$\mathcal{L}_{\mathbf{p}-T}\rho = \nabla_{\mathbf{x}}U(\mathbf{x})\cdot\nabla_{\mathbf{p}}\rho + \nabla_{\mathbf{p}}\cdot\left(\boldsymbol{\gamma}\mathbf{p}\rho\right) + \frac{1}{\beta}\nabla_{\mathbf{p}}\cdot\left(\boldsymbol{\gamma}\mathbf{M}\nabla_{\mathbf{p}}\rho\right) \quad . \tag{33}$$

Two schemes for Eq. (31) are then

**9) Middle-pT scheme**

$$e^{\mathcal{L}\Delta t} \approx e^{\mathcal{L}^{\text{Middle-pT}}\Delta t} = e^{\mathcal{L}_{\mathbf{x}}\Delta t/2} e^{\mathcal{L}_{\mathbf{p}-T}\Delta t} e^{\mathcal{L}_{\mathbf{x}}\Delta t/2} \quad , \tag{34}$$

**10) Side-pT scheme**

$$e^{\mathcal{L}\Delta t} \approx e^{\mathcal{L}^{\text{Side-pT}}\Delta t} = e^{\mathcal{L}_{\mathbf{p}-T}\Delta t/2} e^{\mathcal{L}_{\mathbf{x}}\Delta t} e^{\mathcal{L}_{\mathbf{p}-T}\Delta t/2} \quad . \tag{35}$$

Both schemes were proposed by Melchionna in 2007[17].

**3. Third type of repartition**

Analogously, the third type of repartition of Eqs. (2) and (3) is

$$\begin{bmatrix} d\mathbf{x}_t \\ d\mathbf{p}_t \end{bmatrix} = \underbrace{\begin{bmatrix} \mathbf{M}^{-1}\mathbf{p}_t dt \\ -\boldsymbol{\gamma}\mathbf{p}_t dt + \boldsymbol{\sigma}\mathbf{M}^{1/2} d\mathbf{W}_t \end{bmatrix}}_{\text{xT}} + \underbrace{\begin{bmatrix} 0 \\ -\nabla_{\mathbf{x}_t} U(\mathbf{x}_t) \end{bmatrix} dt}_{\text{p}} \quad . \tag{36}$$

The solution to the first term of Eq. (36) reads

$$\begin{bmatrix} \mathbf{x}(t+\Delta t) \\ \mathbf{p}(t+\Delta t) \end{bmatrix} = e^{-\mathbf{K}\Delta t} \begin{bmatrix} \mathbf{x}(t) \\ \mathbf{p}(t) \end{bmatrix} + \int_t^{t+\Delta t} ds\ e^{-\mathbf{K}(t+\Delta t - s)} \mathbf{K}_1 \bar{\boldsymbol{\eta}}_s \quad , \tag{37}$$

where

$$\mathbf{K} = \begin{pmatrix} 0 & -\mathbf{M}^{-1} \\ 0 & \boldsymbol{\gamma} \end{pmatrix},\ \mathbf{K}_1 = \begin{pmatrix} 0 & 0 \\ 0 & \boldsymbol{\sigma}\mathbf{M}^{1/2} \end{pmatrix},\ \bar{\boldsymbol{\eta}}_s = \begin{pmatrix} 0 \\ \boldsymbol{\eta}_s \end{pmatrix} \quad . \tag{38}$$

It is easy to verify that

$$e^{-\mathbf{K}\Delta t} = \begin{pmatrix} \mathbf{1} & \boldsymbol{\gamma}^{-1}\left(\mathbf{1} - e^{-\boldsymbol{\gamma}\Delta t}\right)\mathbf{M}^{-1} \\ 0 & e^{-\boldsymbol{\gamma}\Delta t} \end{pmatrix} \quad . \tag{39}$$

Eq. (37) can be expressed as

$$\begin{bmatrix} \mathbf{x}(t+\Delta t) \\ \mathbf{p}(t+\Delta t) \end{bmatrix} = e^{-\mathbf{K}\Delta t} \begin{bmatrix} \mathbf{x}(t) \\ \mathbf{p}(t) \end{bmatrix} + \bar{\mathbf{\Omega}}(t,\Delta t) \quad , \tag{40}$$

where

$$\bar{\mathbf{\Omega}}(t,\Delta t) = \int_t^{t+\Delta t} ds\ e^{-\mathbf{K}(t+\Delta t-s)} \mathbf{K}_1 \bar{\boldsymbol{\eta}}_s \quad . \tag{41}$$

It is straightforward to show that $\bar{\mathbf{\Omega}}(t,\Delta t)$ is a Gaussian random vector with zero mean and deviation matrix

$$\begin{aligned} \left\langle \bar{\mathbf{\Omega}}(t,\Delta t)\bar{\mathbf{\Omega}}^T(t,\Delta t) \right\rangle &= \int_t^{t+\Delta t} ds\ e^{-\mathbf{K}(t+\Delta t-s)} \begin{pmatrix} 0 & 0 \\ 0 & \frac{2}{\beta}\boldsymbol{\gamma}\mathbf{M} \end{pmatrix} e^{-\mathbf{K}^T(t+\Delta t-s)} \\ &= \frac{1}{\beta} \begin{pmatrix} \boldsymbol{\gamma}^{-2}\left(2\boldsymbol{\gamma}\Delta t - \mathbf{3} + 4e^{-\boldsymbol{\gamma}\Delta t} - e^{-2\boldsymbol{\gamma}\Delta t}\right)\mathbf{M}^{-1} & \boldsymbol{\gamma}^{-1}\left(\mathbf{1} - e^{-\boldsymbol{\gamma}\Delta t}\right)^2 \\ \boldsymbol{\gamma}^{-1}\left(\mathbf{1} - e^{-\boldsymbol{\gamma}\Delta t}\right)^2 & \mathbf{M}\left(\mathbf{1} - e^{-2\boldsymbol{\gamma}\Delta t}\right) \end{pmatrix} \end{aligned} \tag{42}$$

In numerical implementation $\bar{\mathbf{\Omega}}$ is generated by using a standard-Gaussian- random-number generator and a matrix $\mathbf{C}$, i.e.,

$$\bar{\mathbf{\Omega}}(t,\Delta t) = \mathbf{C}\,\boldsymbol{\mu}\left(t,\Delta t\right) \tag{43}$$

where $\boldsymbol{\mu}(t,\Delta t)$ is the standard Gaussian random vector defined by Eqs. (17)-(18) and

$$\mathbf{C} = \mathbf{T}\mathbf{\Lambda}^{1/2} \tag{44}$$

Here

$$\begin{aligned} \mathbf{T} &= \begin{pmatrix} \frac{1}{2}\mathbf{G}_1^{-1}\mathbf{P}_3^{-1}\left(\mathbf{P}_1 - \mathbf{P}_2 - \mathbf{E}\right) & \frac{1}{2}\mathbf{G}_2^{-1}\mathbf{P}_3^{-1}\left(\mathbf{P}_1 - \mathbf{P}_2 + \mathbf{E}\right) \\ \mathbf{G}_1^{-1} & \mathbf{G}_2^{-1} \end{pmatrix} \\ \mathbf{\Lambda} &= \frac{1}{2\beta} \begin{pmatrix} \mathbf{P}_1 + \mathbf{P}_2 - \mathbf{E} & 0 \\ 0 & \mathbf{P}_1 + \mathbf{P}_2 + \mathbf{E} \end{pmatrix} \end{aligned} \tag{45}$$

with the diagonal matrix

$$
\begin{aligned}
&\mathbf{P}_1 = \gamma^{-2}\left(2\gamma\Delta t - \mathbf{3} + 4e^{-\gamma\Delta t} - e^{-2\gamma\Delta t}\right)\mathbf{M}^{-1} \\
&\mathbf{P}_2 = \mathbf{M}\left(\mathbf{1} - e^{-2\gamma\Delta t}\right) \\
&\mathbf{P}_3 = \gamma^{-1}\left(\mathbf{1} - e^{-\gamma\Delta t}\right)^2 \\
&\mathbf{E} = \left[\left(\mathbf{P}_1 - \mathbf{P}_2\right)^2 + 4\mathbf{P}_3{}^2\right]^{1/2} \\
&\mathbf{G}_1 = \left\{\left[\frac{1}{2}\mathbf{P}_3^{-1}\left(\mathbf{P}_1 - \mathbf{P}_2 - \mathbf{E}\right)\right]^2 + \mathbf{1}\right\}^{1/2} \\
&\mathbf{G}_2 = \left\{\left[\frac{1}{2}\mathbf{P}_3^{-1}\left(\mathbf{P}_1 - \mathbf{P}_2 + \mathbf{E}\right)\right]^2 + \mathbf{1}\right\}^{1/2}
\end{aligned}
\tag{46}
$$

The phase space propagator for the first term in Eq. (36) is $e^{\mathcal{L}_{\mathbf{x}-T}\Delta t}$ for a finite time interval $\Delta t$, where the relevant Kolmogorov operator $\mathcal{L}_{\mathbf{x}-T}$ is

$$
\mathcal{L}_{\mathbf{x}-T}\rho = -\left(\mathbf{M}^{-1}\mathbf{p}\right)\cdot\nabla_{\mathbf{x}}\rho + \nabla_{\mathbf{p}}\cdot\left(\boldsymbol{\gamma}\mathbf{p}\rho\right) + \frac{1}{\beta}\nabla_{\mathbf{p}}\cdot\left(\boldsymbol{\gamma}\mathbf{M}\nabla_{\mathbf{p}}\rho\right) \ . \tag{47}
$$

Two schemes for Eq. (36) can be proposed as

**11) Middle-xT scheme**

$$
e^{\mathcal{L}\Delta t} \approx e^{\mathcal{L}^{\text{Middle-xT}}\Delta t} = e^{\mathcal{L}_{\mathbf{p}}\Delta t/2} e^{\mathcal{L}_{\mathbf{x}-T}\Delta t} e^{\mathcal{L}_{\mathbf{p}}\Delta t/2} \ , \tag{48}
$$

**12) Side-xT scheme**

$$
e^{\mathcal{L}\Delta t} \approx e^{\mathcal{L}^{\text{Side-xT}}\Delta t} = e^{\mathcal{L}_{\mathbf{x}-T}\Delta t/2} e^{\mathcal{L}_{\mathbf{p}}\Delta t} e^{\mathcal{L}_{\mathbf{x}-T}\Delta t/2} \ . \tag{49}
$$

The splitting in either scheme was earlier mentioned by Drozdov in 1993[12], although no algorithm was discussed or studied in the literature.

When the thermostat is not included, it is trivial to show that the "middle-xT" and "side-pT" schemes are reduced to the velocity-Verlet algorithm for constant energy MD, while the "side-xT" and "middle-pT" schemes approach the position Verlet algorithm instead.

## IV. Stationary distribution for the harmonic system

We should stress that it is very difficult if not impossible to carry out the error analysis for general systems. Here we only consider the linear one for which the potential function is

$$U(\mathbf{x}) = \frac{1}{2}\left(\mathbf{x}-\mathbf{x}_{eq}\right)^T \mathbf{A}\left(\mathbf{x}-\mathbf{x}_{eq}\right) \quad , \tag{50}$$

where $\mathbf{x}_{eq}$ is a constant vector and $\mathbf{A}$ a constant Hessian matrix. Below we show two approaches to obtain the stationary state distribution.

**1. Phase space propagator approach**

We first present an approach that uses phase space propagators to do the analysis. It is described in detail in Appendix A on how to obtain the stationary state distribution for a one-dimensional harmonic system. We follow Appendix C of Ref. [24] to study the multi-dimensional case for the first eight schemes that use the first type of repartition for the harmonic system [Eq. (50)]. Now Eq. (21) becomes

$$\mathcal{L}_{\mathbf{p}}\rho = (\mathbf{x}-\mathbf{x}_{eq})^T \mathbf{A}\frac{\partial \rho}{\partial \mathbf{p}} \quad . \tag{51}$$

Using the Taylor expansion $e^{\mathcal{L}_{\mathbf{p}}\Delta t} = \sum_{n=0}^{\infty}\frac{1}{n!}\left(\Delta t(\mathbf{x}-\mathbf{x}_{eq})^T \mathbf{A}\nabla_{\mathbf{p}}\right)^n$, it is straightforward to verify

$$e^{\mathcal{L}_{\mathbf{p}}\Delta t} g(\mathbf{p}) = g\left(\mathbf{p}+\mathbf{A}\left(\mathbf{x}-\mathbf{x}_{eq}\right)\Delta t\right) \quad . \tag{52}$$

That is, $e^{\mathcal{L}_{\mathbf{p}}\Delta t}$ is a momentum shift operator. Similarly, Eq. (20) leads to a position shift operator $e^{\mathcal{L}_{\mathbf{x}}\Delta t}$ that has

$$e^{\mathcal{L}_{\mathbf{x}}\Delta t} f(\mathbf{x}) = f(\mathbf{x}-\mathbf{M}^{-1}\mathbf{p}\Delta t) \quad . \tag{53}$$

The OU process [Eq. (13) or Eq. (22)] keeps the Maxwell momentum distribution unchanged, i.e.,

$$e^{\mathcal{L}_T\Delta t} \exp\left\{-\beta\left[\frac{1}{2}\mathbf{p}^T\mathbf{M}^{-1}\mathbf{p}\right]\right\} = \exp\left\{-\beta\left[\frac{1}{2}\mathbf{p}^T\mathbf{M}^{-1}\mathbf{p}\right]\right\} \quad . \tag{54}$$

Consider the density distribution

$$\rho^{\text{Middle}}(\mathbf{x},\mathbf{p}) = \frac{1}{Z_N}\exp\left\{-\beta\left[\frac{1}{2}\mathbf{p}^T\left(\mathbf{M}-\mathbf{A}\frac{\Delta t^2}{4}\right)^{-1}\mathbf{p}+\frac{1}{2}\left(\mathbf{x}-\mathbf{x}_{eq}\right)^T\mathbf{A}\left(\mathbf{x}-\mathbf{x}_{eq}\right)\right]\right\} \quad , \qquad (55)$$

where $Z_N$ is the normalization constant. It is easy to show that Eq. (55) is a stationary density distribution for the "middle" scheme, which satisfies

$$e^{\mathcal{L}^{\text{Middle}}\Delta t}\rho^{\text{Middle}} = \rho^{\text{Middle}} \quad . \qquad (56)$$

Analogously, one may verify that the "side"/"end"/"beginning" schemes share the same stationary density distribution for the harmonic system,

$$\rho^{\text{Side/End/Begin}}(\mathbf{x},\mathbf{p}) = \frac{1}{Z'_N}\exp\left\{-\beta\left[\frac{1}{2}\mathbf{p}^T\mathbf{M}^{-1}\mathbf{p}+\frac{1}{2}\left(\mathbf{x}-\mathbf{x}_{eq}\right)^T(\mathbf{1}-\mathbf{A}\mathbf{M}^{-1}\frac{\Delta t^2}{4})\mathbf{A}\left(\mathbf{x}-\mathbf{x}_{eq}\right)\right]\right\} \quad , \qquad (57)$$

where $Z'_N$ is the normalization constant.

Adopting the same procedure, we obtain the stationary density distribution for the "PV-middle" scheme

$$\rho^{\text{PV-middle}}(\mathbf{x},\mathbf{p}) = \frac{1}{\bar{Z}_N}\exp\left\{-\beta\left[\frac{1}{2}\mathbf{p}^T\left(\mathbf{1}-\mathbf{M}^{-1}\mathbf{A}\frac{\Delta t^2}{4}\right)\mathbf{M}^{-1}\mathbf{p}+\frac{1}{2}\left(\mathbf{x}-\mathbf{x}_{eq}\right)^T\mathbf{A}\left(\mathbf{x}-\mathbf{x}_{eq}\right)\right]\right\} \quad , \qquad (58)$$

where $\bar{Z}_N$ is the normalization constant. The "PV-side"/"PV-end"/"PV-beginning" schemes share the same stationary density distribution for the harmonic system,

$$\rho^{\text{PV-side/PV-end/PV-begin}}(\mathbf{x},\mathbf{p}) = \frac{1}{\bar{Z}'_N}\exp\left\{-\beta\left[\frac{1}{2}\mathbf{p}^T\mathbf{M}^{-1}\mathbf{p}+\frac{1}{2}\left(\mathbf{x}-\mathbf{x}_{eq}\right)^T(\mathbf{1}-\mathbf{A}\mathbf{M}^{-1}\frac{\Delta t^2}{4})^{-1}\mathbf{A}\left(\mathbf{x}-\mathbf{x}_{eq}\right)\right]\right\}, \qquad (59)$$

with the normalization constant $\bar{Z}'_N$. In the harmonic limit, the "middle"/"PV-middle" schemes lead to the exact configurational distribution but not the exact momentum distribution, while the "side"/"end"/"beginning"/"PV-side"/"PV-end"/"PV-beginning" schemes produce the exact momentum distribution but not the exact configurational distribution. Interestingly, the stationary density distribution for the harmonic system obtained from any one of these schemes does not depend on the diagonal friction matrix $\boldsymbol{\gamma}$.

It is important to note that all conclusions above hold as long as Eq. (54) is satisfied. Replace $e^{-\gamma\Delta t}$ by $-e^{-\gamma\Delta t}$ and $e^{-n\gamma\Delta t}$ by $\left(-e^{-\gamma\Delta t}\right)^n$ for any integer $n$ in Eqs. (13) and (16) [i.e., the solution to the OU process]. Eq. (13) then becomes

$$\begin{bmatrix} \mathbf{x}(t+\Delta t) \\ \mathbf{p}(t+\Delta t) \end{bmatrix} = \begin{bmatrix} \mathbf{x}(t) \\ -e^{-\gamma\Delta t}\mathbf{p}(t) + \boldsymbol{\Omega}(t,\Delta t) \end{bmatrix} , \tag{60}$$

where $\boldsymbol{\Omega}(t,\Delta t)$, a Gaussian random vector with zero mean, for which the diagonal deviation matrix is defined by Eq. (16). Interestingly, Eq. (60) does not change the Maxwell momentum distribution, which still satisfies Eq. (54). It then presents another thermostat method based on the Langevin equation, albeit that Eq. (60) is not a physical solution to

$$\begin{bmatrix} d\mathbf{x}_t \\ d\mathbf{p}_t \end{bmatrix} = \begin{bmatrix} 0 \\ -\boldsymbol{\gamma}\mathbf{p}_t dt + \boldsymbol{\sigma}\mathbf{M}^{1/2} d\mathbf{W}_t \end{bmatrix} . \tag{61}$$

Each of the first eight schemes [Eqs. (23)-(30)] in Section III then has two versions. One is the real dynamics case as conventionally used, the other is the *virtual* dynamics case by replacing $e^{\mathcal{L}_T\Delta t}$ by $e^{\mathcal{L}_T^{vir}\Delta t}$ [the phase space propagator for Eq. (60)] in the scheme. The virtual dynamics cases for the eight schemes [Eqs. (23)-(30)] in Section III are denoted as "middle (vir)", "side (vir)", "end (vir)", "beginning (vir)", "PV-middle (vir)", "PV-side (vir)", "PV-end (vir)", and "PV-beginning (vir)", respectively. Any one of the virtual dynamics cases then shares the same stationary density distribution with its corresponding real dynamics case for a finite time interval $\Delta t$ for the harmonic system. More importantly, it is easy to verify that any one of the virtual dynamics cases for the eight schemes [Eqs. (23)-(30)] in Section III leads to the correct Boltzmann distribution $e^{-\beta H(\mathbf{x},\mathbf{p})}$ for any general systems [of the form Eq. (1)] when the time interval $\Delta t$ approaches zero.

Interestingly, in Appendix B we derive the relation between the "middle"/"middle (vir)" scheme and Grønbech-Jensen and Farago's algorithm[10]. The unified theoretical framework suggested in the paper naturally includes our proposed methods, Grønbech-Jensen and Farago's algorithm[10], and Leimkuhler and Matthews's integrators[20, 21].

**2. Trajectory-based approach**

We further employ a trajectory-based approach by directly solving the discrete equations of motion to do the analysis.

We denote $\left(\mathbf{x}_n, \mathbf{p}_n\right)$ for the phase-space point and $\rho_n$ for the density distribution at the *n*-th step. By definition, the density distribution at the *n*-th time step is $\rho_n \equiv \left\langle \delta(\mathbf{x}_n - \mathbf{x})\delta(\mathbf{p}_n - \mathbf{p}) \right\rangle$. In principle, if $\mathbf{x}_n$ and $\mathbf{p}_n$ are known as a functional of the white noise $\boldsymbol{\eta}_t$ or the independent increment process $\boldsymbol{\Omega}(t, \Delta t)$, $\rho_n$ could be evaluated.

We first consider the "middle" scheme. The update of the position and momentum based on this algorithm reads

$$\mathbf{x}_{n+1} = \mathbf{x}_n + \frac{\Delta t}{2}\mathbf{M}^{-1}\left(\mathbf{1} + e^{-\gamma\Delta t}\right)\left[\mathbf{p}_n - \frac{\Delta t}{2}\nabla_{\mathbf{x}_n} U(\mathbf{x}_n)\right] + \frac{\Delta t}{2}\mathbf{M}^{-1}\boldsymbol{\Omega}_n \tag{62}$$

$$\mathbf{p}_{n+1} = e^{-\gamma\Delta t}\left[\mathbf{p}_n - \frac{\Delta t}{2}\nabla_{\mathbf{x}_n} U(\mathbf{x}_n)\right] - \frac{\Delta t}{2}\nabla_{\mathbf{x}_{n+1}} U(\mathbf{x}_{n+1}) + \boldsymbol{\Omega}_n \quad , \tag{63}$$

where $\boldsymbol{\Omega}_n \equiv \boldsymbol{\Omega}(n\Delta t, \Delta t)$ defined by Eqs. (14) and (15) [or Eq. (19)] are taken to be an independent increment process in the numerical implementation, namely $\left\langle \boldsymbol{\Omega}_n \right\rangle = 0$ and

$$\left\langle \mathbf{\Omega}_n \mathbf{\Omega}_{n'}^T \right\rangle = \delta_{nn'} \frac{1}{\beta} \mathbf{M} \left( \mathbf{1} - e^{-2\gamma\Delta t} \right) \quad . \tag{64}$$

Eqs. (62)-(63) for the harmonic system [Eq. (50)] become

$$\mathbf{x}_{n+1} = \mathbf{x}_n + \frac{\Delta t}{2} \mathbf{M}^{-1} \left( \mathbf{1} + e^{-\gamma\Delta t} \right) \left[ \mathbf{p}_n - \frac{\Delta t}{2} \mathbf{A} \left( \mathbf{x}_n - \mathbf{x}_{eq} \right) \right] + \frac{\Delta t}{2} \mathbf{M}^{-1} \mathbf{\Omega}_n \tag{65}$$

$$\mathbf{p}_{n+1} = e^{-\gamma\Delta t} \left[ \mathbf{p}_n - \frac{\Delta t}{2} \mathbf{A} \left( \mathbf{x}_n - \mathbf{x}_{eq} \right) \right] - \frac{\Delta t}{2} \mathbf{A} \left( \mathbf{x}_{n+1} - \mathbf{x}_{eq} \right) + \mathbf{\Omega}_n \quad . \tag{66}$$

For convenience we now deal with the discrete-time evolution in phase space. Let $\mathbf{R}$ denote the phase-space point, namely

$$\mathbf{R}_n \equiv \begin{pmatrix} \mathbf{x}_n \\ \mathbf{p}_n \end{pmatrix} \quad . \tag{67}$$

As the system is driven by the Gaussian process, the resultant dynamics described by the update relations Eqs. (65) and (66) is a linear combination of stochastic Gaussian processes, which is itself a Gaussian, too. Physically, the system finally evolves to the stationary state possessing a fixed averaged phased-space point $\bar{\mathbf{R}} \equiv \lim_{n\to\infty} \bar{\mathbf{R}}_n$ (here $\bar{\mathbf{R}}_n \equiv \left\langle \mathbf{R}_n \right\rangle$ ) and fluctuation correlation matrix $\mathbf{W} \equiv \lim_{n\to\infty} \left\langle \left( \mathbf{R}_n - \bar{\mathbf{R}}_n \right) \left( \mathbf{R}_n - \bar{\mathbf{R}}_n \right)^T \right\rangle$. Because the stationary state assumes a Gaussian distribution, its property is fully determined by the two quantities $\bar{\mathbf{R}}$ and $\mathbf{W}$ that are required to be determined. To calculate $\bar{\mathbf{R}}$, one may take the random average of Eqs. (65) and (66) and let $n$ go to infinity to obtain

$$\bar{\mathbf{x}} = \bar{\mathbf{x}} + \frac{\Delta t}{2} \mathbf{M}^{-1} \left( \mathbf{1} + e^{-\gamma\Delta t} \right) \left[ \bar{\mathbf{p}} - \frac{\Delta t}{2} \mathbf{A} \left( \bar{\mathbf{x}} - \mathbf{x}_{eq} \right) \right] \tag{68}$$

$$\bar{\mathbf{p}} = e^{-\gamma\Delta t} \left[ \bar{\mathbf{p}} - \frac{\Delta t}{2} \mathbf{A} \left( \bar{\mathbf{x}} - \mathbf{x}_{eq} \right) \right] - \frac{\Delta t}{2} \mathbf{A} \left( \bar{\mathbf{x}} - \mathbf{x}_{eq} \right) \quad . \tag{69}$$

From these equations one readily obtains

$$\bar{\mathbf{x}} = \mathbf{x}_{eq} \tag{70}$$

$$\bar{\mathbf{p}} = 0 \quad . \tag{71}$$

It should be stressed that here we have assumed that the stationary state exists so that $\bar{\mathbf{x}} = \lim_{n\to\infty} \bar{\mathbf{x}}_n = \lim_{n\to\infty} \bar{\mathbf{x}}_{n+1}$ and $\bar{\mathbf{p}} = \lim_{n\to\infty} \bar{\mathbf{p}}_n = \lim_{n\to\infty} \bar{\mathbf{p}}_{n+1}$. Although the same procedure, that is, treating the motion in the configuration and momentum space separately according to Eqs. (65) and (66) may be used to establish a set of equations for the equilibrium fluctuation correlation matrices $\mathbf{W}_{\mathbf{xx}} \equiv \lim_{n\to\infty} \left\langle \left(\mathbf{x}_n - \bar{\mathbf{x}}_n\right)\left(\mathbf{x}_n - \bar{\mathbf{x}}_n\right)^T \right\rangle$, $\mathbf{W}_{\mathbf{xp}} \equiv \lim_{n\to\infty} \left\langle \left(\mathbf{x}_n - \bar{\mathbf{x}}_n\right)\left(\mathbf{p}_n - \bar{\mathbf{p}}_n\right)^T \right\rangle = \mathbf{W}_{\mathbf{px}}^T$ , and $\mathbf{W}_{\mathbf{pp}} \equiv \lim_{n\to\infty} \left\langle \left(\mathbf{p}_n - \bar{\mathbf{p}}_n\right)\left(\mathbf{p}_n - \bar{\mathbf{p}}_n\right)^T \right\rangle$ that can be used to construct $\mathbf{W}$, no powerful techniques are available to solve them except for one-dimensional system. For general systems, therefore, we follow a "physical" way of the phase-space dynamics to determine $\mathbf{W}$ after solving $\mathbf{R}_n$. It turns out that the discrete trajectory $\mathbf{R}_n$ may shed more insights on the steady state as well as the stability of the dynamics, which may be hidden otherwise. To solve the phase-space dynamics we recast iteration relations Eqs. (65) and (66) as[10]

$$\mathbf{R}_{n+1} = \tilde{\mathbf{M}}\mathbf{R}_n + \mathbf{F}_0 + \tilde{\mathbf{\Omega}}_n \quad , \tag{72}$$

where

$$\tilde{\mathbf{M}} = \begin{pmatrix} \tilde{\mathbf{M}}_{\mathbf{xx}} & \tilde{\mathbf{M}}_{\mathbf{xp}} \\ \tilde{\mathbf{M}}_{\mathbf{px}} & \tilde{\mathbf{M}}_{\mathbf{pp}} \end{pmatrix} \quad , \tag{73}$$

$$\mathbf{F}_0 = \begin{pmatrix} \left(\frac{\Delta t}{2}\right)^2 \mathbf{M}^{-1}\left(\mathbf{1} + e^{-\gamma\Delta t}\right)\mathbf{A}\mathbf{x}_{eq} \\ \frac{\Delta t}{2}\left[\mathbf{1} - \left(\frac{\Delta t}{2}\right)^2 \mathbf{A}\mathbf{M}^{-1}\right]\left(\mathbf{1} + e^{-\gamma\Delta t}\right)\mathbf{A}\mathbf{x}_{eq} \end{pmatrix} \quad , \tag{74}$$

$$\tilde{\mathbf{\Omega}}_n = \begin{pmatrix} \frac{\Delta t}{2}\mathbf{M}^{-1}\mathbf{\Omega}_n \\ \left[\mathbf{1} - \left(\frac{\Delta t}{2}\right)^2 \mathbf{A}\mathbf{M}^{-1}\right]\mathbf{\Omega}_n \end{pmatrix} \quad , \tag{75}$$

with

$$\tilde{\mathbf{M}}_{\mathbf{xx}} = \mathbf{1} - \left(\frac{\Delta t}{2}\right)^2 \mathbf{M}^{-1}\left(\mathbf{1} + e^{-\gamma\Delta t}\right)\mathbf{A} \tag{76}$$

$$\tilde{\mathbf{M}}_{\mathbf{xp}} = \frac{\Delta t}{2}\mathbf{M}^{-1}\left(\mathbf{1} + e^{-\gamma\Delta t}\right) \tag{77}$$

$$\tilde{\mathbf{M}}_{\mathbf{px}} = -\frac{\Delta t}{2}\left[\mathbf{1} - \left(\frac{\Delta t}{2}\right)^2 \mathbf{A}\mathbf{M}^{-1}\right]\left(\mathbf{1} + e^{-\gamma\Delta t}\right)\mathbf{A} \tag{78}$$

$$\tilde{\mathbf{M}}_{\mathbf{pp}} = e^{-\gamma\Delta t} - \left(\frac{\Delta t}{2}\right)^2 \mathbf{A}\mathbf{M}^{-1}\left(\mathbf{1} + e^{-\gamma\Delta t}\right) \ . \tag{79}$$

Given the initial condition $\mathbf{R}_0$, one can use the iteration relation Eq. (72) to find the solution of $\mathbf{R}_n$,

$$\mathbf{R}_n = \tilde{\mathbf{M}}^n \mathbf{R}_0 + \sum_{j=0}^{n-1} \tilde{\mathbf{M}}^j \left(\mathbf{F}_0 + \tilde{\mathbf{\Omega}}_{n-1-j}\right) \ . \tag{80}$$

Upon taking random average and letting $n \to \infty$, there yields the equilibrium phase-space point $\bar{\mathbf{R}}$ as

$$\bar{\mathbf{R}} = \lim_{n\to\infty} \tilde{\mathbf{M}}^n \mathbf{R}_0 + \sum_{j=0}^{\infty} \tilde{\mathbf{M}}^j \mathbf{F}_0 \ . \tag{81}$$

Here the first term should vanish, which is physically required for an equilibrium state if it exists. The second term, the series can be summed up to yield

$$\bar{\mathbf{R}} = \left(\mathbf{1} - \tilde{\mathbf{M}}\right)^{-1} \mathbf{F}_0 \ . \tag{82}$$

We now resort to the algebra techniques for block matrices to find the expression of the inverse of the matrix $\mathbf{1} - \tilde{\mathbf{M}}$. Consider a block matrix

$$\mathbf{N} = \begin{pmatrix} \mathbf{N}_{11} & \mathbf{N}_{12} \\ \mathbf{N}_{21} & \mathbf{N}_{22} \end{pmatrix} ,$$

where $\mathbf{N}_{11}$, $\mathbf{N}_{12}$, $\mathbf{N}_{21}$, and $\mathbf{N}_{22}$ are $n \times n$, $n \times m$, $m \times n$, and $m \times m$ block matrices. If $\mathbf{N}_{11}$ is invertible, then there is a matrix inversion lemma

$$\mathbf{N}^{-1}=\begin{bmatrix}\mathbf{N}_{11}{}^{-1}+\mathbf{N}_{11}{}^{-1}\mathbf{N}_{12}\left(\mathbf{N}/\mathbf{N}_{11}\right)^{-1}\mathbf{N}_{21}\mathbf{N}_{11}{}^{-1} & -\mathbf{N}_{11}{}^{-1}\mathbf{N}_{12}\left(\mathbf{N}/\mathbf{N}_{11}\right)^{-1}\\ -\left(\mathbf{N}/\mathbf{N}_{11}\right)^{-1}\mathbf{N}_{21}\mathbf{N}_{11}{}^{-1} & \left(\mathbf{N}/\mathbf{N}_{11}\right)^{-1}\end{bmatrix}\quad, \tag{83}$$

where $\mathbf{N}/\mathbf{N}_{11}\equiv\mathbf{N}_{22}-\mathbf{N}_{21}\mathbf{N}_{11}{}^{-1}\mathbf{N}_{12}$ is called the Schur complement of the block $\mathbf{N}_{11}$. In our case the four blocks $\mathbf{N}_{11}=\mathbf{1}-\tilde{\mathbf{M}}_{\mathbf{xx}}$, $\mathbf{N}_{12}=-\tilde{\mathbf{M}}_{\mathbf{xp}}$, $\mathbf{N}_{21}=-\tilde{\mathbf{M}}_{\mathbf{px}}$, and $\mathbf{N}_{22}=\mathbf{1}-\tilde{\mathbf{M}}_{\mathbf{pp}}$ are all $n\times n$. From Eq. (76) we readily obtain the inverse of the block $\mathbf{N}_{11}$,

$$\mathbf{N}_{11}{}^{-1}=\left(\mathbf{1}-\tilde{\mathbf{M}}_{\mathbf{xx}}\right)^{-1}=\left(\frac{\Delta t}{2}\right)^{-2}\mathbf{A}^{-1}\mathbf{M}\left(\mathbf{1}+e^{-\gamma\Delta t}\right)^{-1}\quad .$$

With this as well as Eqs. (77), (78), and (79) we obtain the Schur complement of $\mathbf{N}_{11}$ by a direct matrix multiplication,

$$\begin{aligned}\mathbf{N}/\mathbf{N}_{11}&=\mathbf{1}-\tilde{\mathbf{M}}_{\mathbf{pp}}-\tilde{\mathbf{M}}_{\mathbf{px}}\left(\mathbf{1}-\tilde{\mathbf{M}}_{\mathbf{xx}}\right)^{-1}\tilde{\mathbf{M}}_{\mathbf{xp}}\\&=2\mathbf{1}\ .\end{aligned}\tag{84}$$

Upon substituting into Eq. (83), there is

$$\left(\mathbf{1}-\tilde{\mathbf{M}}\right)^{-1}=\begin{bmatrix}\left(\frac{\Delta t}{2}\right)^{-2}\mathbf{A}^{-1}\mathbf{M}\left\{\left(\mathbf{1}+e^{-\gamma\Delta t}\right)^{-1}-\frac{1}{2}\left[\mathbf{1}-\left(\frac{\Delta t}{2}\right)^{2}\mathbf{M}^{-1}\mathbf{A}\right]\right\} & \frac{1}{2}\left(\frac{\Delta t}{2}\right)^{-1}\mathbf{A}^{-1}\\ -\frac{1}{2}\left(\frac{\Delta t}{2}\right)^{-1}\left[\mathbf{M}-\left(\frac{\Delta t}{2}\right)^{2}\mathbf{A}\right] & \frac{1}{2}\mathbf{1}\end{bmatrix}\ . \tag{85}$$

Inserting this result and Eq. (73) into Eq. (82), we obtain

$$\bar{\mathbf{R}}=\begin{pmatrix}\mathbf{x}_{eq}\\0\end{pmatrix}\quad, \tag{86}$$

which is the same as that given by Eqs. (70) and (71) as it should be. From Eq. (72) it is straightforward to obtain the iteration relation for the fluctuation correlation of $\mathbf{R}_n$, namely,

$$\left\langle\left(\mathbf{R}_{n+1}-\bar{\mathbf{R}}_{n+1}\right)\left(\mathbf{R}_{n+1}-\bar{\mathbf{R}}_{n+1}\right)^{T}\right\rangle=\tilde{\mathbf{M}}\left\langle\left(\mathbf{R}_{n}-\bar{\mathbf{R}}_{n}\right)\left(\mathbf{R}_{n}-\bar{\mathbf{R}}_{n}\right)^{T}\right\rangle\tilde{\mathbf{M}}^{T}+\left\langle\tilde{\mathbf{\Omega}}_n\tilde{\mathbf{\Omega}}_n^{T}\right\rangle\quad . \tag{87}$$

Taking the asymptotic limit $n\to\infty$, we obtain the equation for the stationary fluctuation correlation $\mathbf{W}$,

$$\mathbf{W}=\tilde{\mathbf{M}}\mathbf{W}\tilde{\mathbf{M}}^{T}+\mathbf{\Theta}\quad, \tag{88}$$

which is the discrete Lyapunov equation. Here $\mathbf{\Theta}$ reads

$$\boldsymbol{\Theta} = \begin{bmatrix} \left(\frac{\Delta t}{2}\right)^2 \mathbf{M}^{-1}\mathbf{Q}\mathbf{M}^{-1} & \frac{\Delta t}{2}\mathbf{M}^{-1}\mathbf{Q}\left[\mathbf{1}-\left(\frac{\Delta t}{2}\right)^2 \mathbf{M}^{-1}\mathbf{A}\right] \\ \frac{\Delta t}{2}\left[\mathbf{1}-\left(\frac{\Delta t}{2}\right)^2 \mathbf{A}\mathbf{M}^{-1}\right]\mathbf{Q}\mathbf{M}^{-1} & \left[\mathbf{1}-\left(\frac{\Delta t}{2}\right)^2 \mathbf{A}\mathbf{M}^{-1}\right]\mathbf{Q}\left[\mathbf{1}-\left(\frac{\Delta t}{2}\right)^2 \mathbf{M}^{-1}\mathbf{A}\right] \end{bmatrix} ,$$

where $\mathbf{Q} = \frac{1}{\beta}\mathbf{M}\left(\mathbf{1}-e^{-2\gamma\Delta t}\right)$. We emphasize that for arbitrary $\tilde{\mathbf{M}}$ and $\boldsymbol{\Theta}$ the discrete Lyapunov equation defies an analytical solution, although one can work out an infinite series for $\mathbf{W}$

$$\mathbf{W} = \sum_{j=0}^{\infty} \tilde{\mathbf{M}}^j \boldsymbol{\Theta}\left(\tilde{\mathbf{M}}^T\right)^j \quad . \tag{89}$$

For a better understanding of the procedure, we consider the special case of the harmonic system, the free Brownian particle.

**2A. Free Brownian motion**

A free Brownian system means $\mathbf{A} = 0$ and $\mathbf{x}_{eq} = 0$. In this case the average position given by Eq. (70) is ill-defined. However, the discrete trajectory $\mathbf{R}_n$ expressed as Eq. (80) ($\mathbf{F}_0 = 0$) can be readily calculated because the powers of $\tilde{\mathbf{M}}$ assume simple expressions

$$\tilde{\mathbf{M}}^n = \begin{bmatrix} \mathbf{1} & \frac{\Delta t}{2}\mathbf{M}^{-1}\mathbf{S}_n \\ 0 & e^{-n\gamma\Delta t} \end{bmatrix} ,$$

where $\mathbf{S}_n = \left(\mathbf{1}+e^{-\gamma\Delta t}\right)\left(\mathbf{1}-e^{-n\gamma\Delta t}\right)\left(\mathbf{1}-e^{-\gamma\Delta t}\right)^{-1}$. As a consequence, the position and momentum read respectively,

$$\mathbf{x}_n = \mathbf{x}_0 + \frac{\Delta t}{2}\mathbf{M}^{-1}\mathbf{S}_n\mathbf{p}_0 + \frac{\Delta t}{2}\mathbf{M}^{-1}\sum_{j=0}^{n-1}\left(\mathbf{1}+\mathbf{S}_j\right)\boldsymbol{\Omega}_{n-1-j} \tag{90}$$

$$\mathbf{p}_n = e^{-n\gamma\Delta t}\mathbf{p}_0 + \sum_{j=0}^{n-1} e^{-j\gamma\Delta t}\boldsymbol{\Omega}_{n-1-j} \quad . \tag{91}$$

Taking random average on both sides, one readily obtains

$$\bar{\mathbf{x}}_n = \mathbf{x}_0 + \frac{\Delta t}{2}\mathbf{M}^{-1}\mathbf{S}_n\mathbf{p}_0 \tag{92}$$

$$\bar{\mathbf{p}}_n = e^{-n\gamma\Delta t}\mathbf{p}_0 \quad . \tag{93}$$

As $n \to \infty$, they become the equilibrium values,

$$\bar{\mathbf{x}} = \mathbf{x}_0 + \frac{\Delta t}{2}\mathbf{M}^{-1}\left(\mathbf{1}+e^{-\gamma\Delta t}\right)\left(\mathbf{1}-e^{-\gamma\Delta t}\right)^{-1}\mathbf{p}_0 \tag{94}$$

$$\bar{\mathbf{p}} = 0 \quad . \tag{95}$$

We now derive the stationary density distribution. Define the fluctuation correlation matrix, $\mathbf{W}_{\mathbf{R}_n\mathbf{R}_n} \equiv \left\langle \left(\mathbf{R}_n - \bar{\mathbf{R}}_n\right)\left(\mathbf{R}_n - \bar{\mathbf{R}}_n\right)^T \right\rangle$ for $\mathbf{R}_n = \left(\mathbf{x}_n, \mathbf{p}_n\right)^T$. Using Eqs. (90) and (91) and brute force calculations, we are able to derive the four required fluctuation correlation matrices at every time step,

$$\mathbf{W}_{\mathbf{x}_n\mathbf{x}_n} = \left(\frac{\Delta t}{2}\right)^2 \frac{1}{\beta}\mathbf{M}^{-1}\left(\mathbf{1}+e^{-\gamma\Delta t}\right)^2\left[4n\left(\mathbf{1}-e^{-2\gamma\Delta t}\right)^{-1} + \left(\mathbf{1}-e^{-2n\gamma\Delta t} - 4\left(\mathbf{1}-e^{-n\gamma\Delta t}\right)\right)\left(\mathbf{1}-e^{-\gamma\Delta t}\right)^{-2}\right] \; , \tag{96}$$

$$\mathbf{W}_{\mathbf{x}_n\mathbf{p}_n} = \mathbf{W}_{\mathbf{p}_n\mathbf{x}_n} = \frac{\Delta t}{2}\cdot\frac{1}{\beta}\left(\mathbf{1}+e^{-\gamma\Delta t}\right)\left(\mathbf{1}-2e^{-n\gamma\Delta t}+e^{-2n\gamma\Delta t}\right)\left(\mathbf{1}-e^{-\gamma\Delta t}\right)^{-1} \; , \tag{97}$$

$$\mathbf{W}_{\mathbf{p}_n\mathbf{p}_n} = \frac{1}{\beta}\mathbf{M}\left(\mathbf{1}-e^{-2n\gamma\Delta t}\right) \; , \tag{98}$$

which are all diagonal. We should point out that as $n$ gets large, $\mathbf{W}_{\mathbf{x}_n\mathbf{x}_n}$ becomes linearly proportional to $n$ and eventually divergent for $n \to \infty$. This is typical for a diffusion process. For the stationary density distribution we need to evaluate the inverse of the block matrix consisting of these fluctuation correlation matrices, namely

$$\mathbf{W}^{-1} = \lim_{n\to\infty}\begin{pmatrix} \mathbf{W}_{\mathbf{x}_n\mathbf{x}_n} & \mathbf{W}_{\mathbf{x}_n\mathbf{p}_n} \\ \mathbf{W}_{\mathbf{p}_n\mathbf{x}_n} & \mathbf{W}_{\mathbf{p}_n\mathbf{p}_n} \end{pmatrix}^{-1} \quad .$$

Using the matrix inversion lemma Eq. (83) and recognizing $\mathbf{A}^{-1} = 0$ in this case, we obtain

$$\mathbf{W}^{-1} = \begin{pmatrix} 0 & 0 \\ 0 & \beta\mathbf{M}^{-1} \end{pmatrix} \; , \tag{99}$$

which exactly corresponds to the Boltzmann distribution in the free particle limit.

**2B. One-dimensional case**

We now consider the one-dimensional case where Eq. (50) becomes

$$U(x) = \frac{1}{2}\left(x - x_{\mathrm{eq}}\right)^T A\left(x - x_{\mathrm{eq}}\right) \quad . \tag{100}$$

Note that now the block matrix $\tilde{\mathbf{M}}$ [Eq. (73)] becomes an ordinary $2 \times 2$ matrix with eigenvalues

$$\varepsilon_{1,2} = \frac{T}{2} \pm \sqrt{\frac{T^2}{4} - D} \tag{101}$$

where

$$\begin{aligned} T &= \tilde{M}_{xx} + \tilde{M}_{pp} = \left(1 + e^{-\gamma\Delta t}\right)\left[1 - 2\left(\frac{\Delta t}{2}\right)^2 AM^{-1}\right] , \\ D &= \tilde{M}_{xx}\tilde{M}_{pp} - \tilde{M}_{xp}\tilde{M}_{px} = e^{-\gamma\Delta t} \end{aligned} \tag{102}$$

and eigenvectors

$$\boldsymbol{v}_{1,2} = \frac{1}{\sqrt{\tilde{M}_{xp}^2 + \left(\varepsilon_{1,2} - \tilde{M}_{xx}\right)^2}} \begin{pmatrix} \tilde{M}_{xp} \\ \varepsilon_{1,2} - \tilde{M}_{xx} \end{pmatrix} \quad . \tag{103}$$

Therefore, $\tilde{\mathbf{M}}$ can be diagonalized by $\mathbf{V} = (\boldsymbol{v}_1, \boldsymbol{v}_2)$, namely

$$\mathbf{V}^{-1}\tilde{\mathbf{M}}\mathbf{V} = \begin{pmatrix} \varepsilon_1 & 0 \\ 0 & \varepsilon_2 \end{pmatrix} \equiv \mathbf{\Lambda} \quad , \tag{104}$$

or $\tilde{\mathbf{M}} = \mathbf{V}\mathbf{\Lambda}\mathbf{V}^{-1}$. With this result the powers of $\tilde{\mathbf{M}}$ can be written explicitly,

$$\begin{aligned} \tilde{\mathbf{M}}^n &= \mathbf{V}\mathbf{\Lambda}^n\mathbf{V}^{-1} \\ &= \begin{pmatrix} v_{11}u_{11}\varepsilon_1^n + v_{12}u_{21}\varepsilon_2^n & v_{11}u_{12}\varepsilon_1^n + v_{12}u_{22}\varepsilon_2^n \\ v_{21}u_{11}\varepsilon_1^n + v_{22}u_{21}\varepsilon_2^n & v_{21}u_{12}\varepsilon_1^n + v_{22}u_{22}\varepsilon_2^n \end{pmatrix} , \end{aligned} \tag{105}$$

where $v_{i,j}$ and $u_{i,j} (i, j = 1, 2)$ are the elements of $\mathbf{V}$ and $\mathbf{V}^{-1}$ respectively. Substituting Eq. (105) into Eq. (80), we may obtain the deviations of the position and momentum from their equilibrium values,

$$\Delta x_n = x_n - \overline{x}_n = \sum_{j=0}^{n-1}\left(c_{x1}\varepsilon_1^j + c_{x2}\varepsilon_2^j\right)\Omega_{n-1-j} \tag{106}$$

$$\Delta p_n = p_n - \bar{p}_n = \sum_{j=0}^{n-1} \left( c_{p1}\varepsilon_1^j + c_{p2}\varepsilon_2^j \right) \Omega_{n-1-j} \quad , \tag{107}$$

where these coefficients are

$$c_{x1} = \frac{\Delta t}{2M} v_{11}u_{11} + \left[ 1 - \left( \frac{\Delta t}{2} \right)^2 \frac{A}{M} \right] v_{11}u_{12}$$

$$c_{x2} = \frac{\Delta t}{2M} v_{12}u_{21} + \left[ 1 - \left( \frac{\Delta t}{2} \right)^2 \frac{A}{M} \right] v_{12}u_{22}$$

$$c_{p1} = \frac{\Delta t}{2M} v_{21}u_{11} + \left[ 1 - \left( \frac{\Delta t}{2} \right)^2 \frac{A}{M} \right] v_{21}u_{12}$$

$$c_{p2} = \frac{\Delta t}{2M} v_{22}u_{21} + \left[ 1 - \left( \frac{\Delta t}{2} \right)^2 \frac{A}{M} \right] v_{22}u_{22} \quad .$$

With these expressions we are able to calculate the required fluctuation correlations

$$W_{x_n x_n} = Q \left[ c_{x1}^2 \frac{1-\varepsilon_1^{2n}}{1-\varepsilon_1^2} + 2c_{x1}c_{x2} \frac{1-\varepsilon_1^n \varepsilon_2^n}{1-\varepsilon_1\varepsilon_2} + c_{x2}^2 \frac{1-\varepsilon_2^{2n}}{1-\varepsilon_2^2} \right] \tag{108}$$

$$\begin{aligned} W_{x_n p_n} &= W_{p_n x_n} \\ &= Q \left[ c_{x1}c_{p1} \frac{1-\varepsilon_1^{2n}}{1-\varepsilon_1^2} + \left( c_{x1}c_{p2} + c_{x2}c_{p1} \right) \frac{1-\varepsilon_1^n \varepsilon_2^n}{1-\varepsilon_1\varepsilon_2} + c_{x2}c_{p2} \frac{1-\varepsilon_2^{2n}}{1-\varepsilon_2^2} \right] \end{aligned} \tag{109}$$

$$W_{p_n p_n} = Q \left[ c_{p1}^2 \frac{1-\varepsilon_1^{2n}}{1-\varepsilon_1^2} + 2c_{p1}c_{p2} \frac{1-\varepsilon_1^n \varepsilon_2^n}{1-\varepsilon_1\varepsilon_2} + c_{p2}^2 \frac{1-\varepsilon_2^{2n}}{1-\varepsilon_2^2} \right] \quad , \tag{110}$$

where $Q = \frac{1}{\beta} M \left( 1 - e^{-2\gamma\Delta t} \right)$. It should be stressed that we are only interested in the stable iteration of the phase-space points, which requires $\left| \varepsilon_{1,2} \right| < 1$. Otherwise, the fluctuation correlation matrix elements Eqs. (108), (109), and (110) will be divergent. To complete the involved algebra, it turns out that parameterizing the matrix $\tilde{\mathbf{M}}$ is helpful. To this end, its eigenvalues $\varepsilon_{1,2}$ is recast as[10]

$$\varepsilon_{1,2} = \sqrt{D} e^{\pm\lambda} \tag{111}$$

and

$$\tilde{\mathbf{M}} = \begin{pmatrix} \sqrt{D}\cosh\lambda + \frac{1-D}{2} & b \\ \frac{1}{b}\left[ D\sinh^2\lambda - \frac{(1-D)^2}{4} \right] & \sqrt{D}\cosh\lambda - \frac{1-D}{2} \end{pmatrix}, \tag{112}$$

where $b = \tilde{M}_{xp} = \frac{\Delta t}{2} M^{-1}\left(1+e^{-\gamma\Delta t}\right)$ is not changed. Then the diagonalization matrix $\mathbf{V}$ and the inverse $\mathbf{V}^{-1}$ read

$$\mathbf{V} = \begin{bmatrix} \frac{1}{n_1} & \frac{1}{n_2} \\ \frac{1}{bn_1}\left( \sqrt{D}\sinh\lambda - \frac{1-D}{2} \right) & -\frac{1}{bn_2}\left( \sqrt{D}\sinh\lambda + \frac{1-D}{2} \right) \end{bmatrix} \tag{113}$$

and

$$\mathbf{V}^{-1} = \frac{1}{\det\mathbf{V}} \begin{bmatrix} -\frac{1}{bn_2}\left( \sqrt{D}\sinh\lambda + \frac{1-D}{2} \right) & -\frac{1}{n_2} \\ -\frac{1}{bn_1}\left( \sqrt{D}\sinh\lambda - \frac{1-D}{2} \right) & \frac{1}{n_1} \end{bmatrix}, \tag{114}$$

where the "normalization" constants $n_{1,2}$ are

$$n_{1,2} = \sqrt{1 + \frac{1}{b^2}\left| \pm\sqrt{D}\sinh\lambda - \frac{1-D}{2} \right|^2} \quad .$$

with these expressions we readily obtain

$$c_{x1} = \frac{b}{2\sqrt{D}\sinh\lambda}\frac{1+\sqrt{D}e^{\lambda}}{1+D}$$

$$c_{x2} = -\frac{b}{2\sqrt{D}\sinh\lambda}\frac{1+\sqrt{D}e^{-\lambda}}{1+D}$$

$$c_{p1} = \frac{\sqrt{D}\sinh\lambda - \frac{1-D}{2}}{2\sqrt{D}\sinh\lambda}\frac{1+\sqrt{D}e^{\lambda}}{1+D}$$

$$c_{p2} = \frac{\sqrt{D}\sinh\lambda + \frac{1-D}{2}}{2\sqrt{D}\sinh\lambda}\frac{1+\sqrt{D}e^{-\lambda}}{1+D} \quad .$$

Substituting into Eqs. (108), (109), and (110), we obtain

$$W_{xx} \equiv \lim_{n\to\infty} W_{x_n x_n} = \frac{b^2 Q}{4D\sinh^2\lambda(1+D)^2}\left[\frac{1+\sqrt{D}e^{\lambda}}{1-\sqrt{D}e^{\lambda}} + \frac{1+\sqrt{D}e^{-\lambda}}{1-\sqrt{D}e^{-\lambda}} - \frac{2\left(1+\sqrt{D}e^{\lambda}\right)\left(1+\sqrt{D}e^{-\lambda}\right)}{1-D}\right]$$

$$= \frac{2b^2 Q}{(1+D)^2(1-D)\left(1-\sqrt{D}e^{\lambda}\right)\left(1-\sqrt{D}e^{-\lambda}\right)} \qquad (115)$$

$$= \frac{1}{\beta}A^{-1}$$

$$W_{xp} \equiv \lim_{n\to\infty} W_{x_n p_n} = W_{px}$$

$$= \frac{bQ}{4D\sinh^2\lambda(1+D)^2}\left[\frac{1+\sqrt{D}e^{\lambda}}{1-\sqrt{D}e^{\lambda}}\left(\sqrt{D}\sinh\lambda - \frac{1-D}{2}\right)\right. \qquad (116)$$

$$\left. - \frac{1+\sqrt{D}e^{-\lambda}}{1-\sqrt{D}e^{-\lambda}}\left(\sqrt{D}\sinh\lambda + \frac{1-D}{2}\right) + \left(1+\sqrt{D}e^{\lambda}\right)\left(1+\sqrt{D}e^{-\lambda}\right)\right]$$

$$= 0$$

$$W_{pp} \equiv \lim_{n\to\infty} W_{p_n p_n} = \frac{Q}{4D\sinh^2\lambda(1+D)^2}\left\{\frac{1+\sqrt{D}e^{\lambda}}{1-\sqrt{D}e^{\lambda}}\left(\sqrt{D}\sinh\lambda - \frac{1-D}{2}\right)^2\right.$$

$$+ \frac{1+\sqrt{D}e^{-\lambda}}{1-\sqrt{D}e^{-\lambda}}\left(\sqrt{D}\sinh\lambda + \frac{1-D}{2}\right)^2$$

$$\left. + \frac{2\left(1+\sqrt{D}e^{\lambda}\right)\left(1+\sqrt{D}e^{-\lambda}\right)}{1-D}\left[D\sinh^2\lambda - \frac{(1-D)^2}{4}\right]\right\}$$

$$= -\frac{2Q\left[D\sinh^2\lambda - \frac{(1-D)^2}{4}\right]}{(1+D)^2(1-D)\left(1-\sqrt{D}e^{\lambda}\right)\left(1-\sqrt{D}e^{-\lambda}\right)}$$

$$= \frac{1}{\beta}M\left[1 - \left(\frac{\Delta t}{2}\right)^2 AM^{-1}\right]. \qquad (117)$$

## 2C. Multi-dimensional case

As one can see from the above, the derivation of the stationary density distribution from the known trajectory is tedious even for the one-dimensional harmonic system. The difficulty we are facing is that the techniques used for the one-dimensional case cannot be directly applied to the multi-dimensional case. To make things simpler, we assume that the "middle" scheme gives an exact Boltzmann distribution for the position, which has been revealed from the analysis of the one-dimensional system, the numerical simulation, and the former studies[24, 25]. Besides, there is no fluctuation correlation between the position and momentum in the equilibrium. Therefore, we may assume the following fluctuation correlation matrices

$$\mathbf{W_{xx}} \equiv \lim_{n\to\infty}\left\langle \left(\mathbf{x}_n - \overline{\mathbf{x}}_n\right)\left(\mathbf{x}_n - \overline{\mathbf{x}}_n\right)^T \right\rangle = \frac{1}{\beta}\mathbf{A}^{-1} \tag{118}$$

$$\mathbf{W_{xp}} = \mathbf{W_{px}}^T \equiv \lim_{n\to\infty}\left\langle \left(\mathbf{x}_n - \overline{\mathbf{x}}_n\right)\left(\mathbf{p}_n - \overline{\mathbf{p}}_n\right)^T \right\rangle = 0 \tag{119}$$

and will determine the fluctuation correlation matrix for the momentum $\mathbf{W_{pp}} \equiv \lim_{n\to\infty}\left\langle \left(\mathbf{p}_n - \overline{\mathbf{p}}_n\right)\left(\mathbf{p}_n - \overline{\mathbf{p}}_n\right)^T \right\rangle$ by resorting to the discrete Lyapunov equation, Eq. (88) in which $\mathbf{F}_0 = 0$ now. By some straightforward algebra one readily proves the consistency of the above Ansatz of the equilibrium density and obtains

$$\mathbf{W_{pp}} = \frac{1}{\beta}\left[\mathbf{M} - \left(\frac{\Delta t}{2}\right)^2 \mathbf{A}\right] \quad . \tag{120}$$

Because the resultant dynamics defined by Eqs. (65)-(66) is Gaussian for the harmonic system, the fluctuation correlation matrices Eqs. (118)-(120) and the averaged phase space point defined by Eqs. (70)-(71) lead to the stationary density distribution Eq. (55).

Consider the virtual dynamics case of the “middle” scheme (i.e., “middle (vir)”). Replace the diagonal matrix $e^{-\gamma\Delta t}$ by $-e^{-\gamma\Delta t}$ and $e^{-n\gamma\Delta t}$ by $\left(-e^{-\gamma\Delta t}\right)^n$ for any integer $n$ in the equations presented in the trajectory-based approach. One may follow the same procedure to demonstrate that “middle (vir)” produces the same stationary density distribution as Eq. (55). Similarly, it is straightforward to show that Eqs. (57), (58), and (59) are the stationary density distribution in the harmonic limit for both the real and virtual dynamics cases of the “side”/“end”/“beginning” schemes, that for both cases of the “PV-middle” scheme, and that for both cases of “PV-side”/“PV-end”/“PV-beginning” schemes, respectively. This is consistent with the conclusions from the phase space propagator approach.

**2D. The last four schemes**

The analysis for such as the “middle-xT”/“side-xT”/“middle-pT”/“side-pT” schemes is more tedious, albeit that the algebra is similar. We may employ the trajectory-based approach to re-derive the stationary density distribution for the one-dimensional case [Eq. (100)] for each of the last four schemes. The results agree with the stationary density distributions obtained by the phase space propagator approach as presented in Appendix A-1. (See Table 1.)

The stationary distributions listed for the last four schemes in Table 1 are not well-defined when the friction coefficient approaches infinity. This suggests that the results obtained from the last four schemes are more sensitive to the friction coefficient (than the first eight schemes do).

Finally we note that the trajectory-based approach is capable of analyzing Langevin dynamics algorithms that the phase space propagator approach is inconvenient to deal with. (See Section S5 of supplementary material.)

## V. Characteristic correlation time

The characteristic correlation time is often used to estimate the sampling efficiency. For example, the potential energy autocorrelation function is defined as

$$C_{pot}(t) = \frac{\left\langle \left[U(x(t)) - \langle U(x)\rangle\right]\left[U(x(0)) - \langle U(x)\rangle\right]\right\rangle}{\left\langle \left[U(x) - \langle U(x)\rangle\right]^2 \right\rangle} \quad . \tag{121}$$

The bracket $\langle\ \rangle$ of Eq. (121) denotes the phase space average of the Boltzmann distribution. The characteristic correlation time for the potential correlation function is then given by

$$\tau_{pot} = \int_0^\infty C_{pot}(t)dt \quad . \tag{122}$$

The smaller $\tau_{pot}$ is, the more efficiently the Langevin equation [Eqs. (2)-(3)] explores the potential energy surface and samples the configurational space.

Analogously, the characteristic correlation time of the Hamiltonian autocorrelation function

$$C_{Ham}(t)=\frac{\left\langle\left[H(x(t),p(t))-\left\langle H(x,p)\right\rangle\right]\left[H(x(0),p(0))-\left\langle H(x,p)\right\rangle\right]\right\rangle}{\left\langle\left[H(x,p)-\left\langle H(x,p)\right\rangle\right]^2\right\rangle} \tag{123}$$

is

$$\tau_{Ham}=\int_0^\infty C_{Ham}(t)dt \quad . \tag{124}$$

The smaller $\tau_{Ham}$ is, the more efficiently the Langevin equation [Eqs. (2)-(3)] explores the phase space.

For simplicity we consider the one-dimensional harmonic system [Eq. (100)], where $A=M\omega^2$ is the force constant.

**1. Infinitesimal time interval**

It is easy to follow Refs. [1, 4, 19, 24] to derive $\tau_{pot}$ or $\tau_{Ham}$ for the harmonic system [Eq. (100)] when the time interval is infinitesimal. Here we rather employ the phase space propagator approach presented in Appendix A to re-derive $\tau_{pot}$ or $\tau_{Ham}$, because the derivation procedure may be generalized to the case when the time interval becomes *finite*.

The propagation of the density distribution in the phase space [Eq. (7)] for the one-dimensional harmonic system [Eq. (100)] becomes

$$\frac{\partial\rho}{\partial t}=\mathcal{L}\rho=-\frac{p}{M}\frac{\partial\rho}{\partial x}+M\omega^2\left(x-x_{eq}\right)\frac{\partial\rho}{\partial p}+\frac{\partial}{\partial p}\left(\gamma p\cdot\rho\right)+\frac{\gamma M}{\beta}\frac{\partial^2\rho}{\partial p^2} \quad . \tag{125}$$

Assume that $\rho_t\equiv\rho\left(x,p;t\middle|x_0,p_0;0\right)$ is the solution to Eq. (125). Although the explicit expression of $\rho_t$ is difficult to obtain, we directly analyze the displacement squared autocorrelation function that may be expressed as

$$\left\langle \left[x(0)-x_{eq}\right]^2 \left[x(t)-x_{eq}\right]^2 \right\rangle = \int \rho_0\left(x_0, p_0\right) \rho_t\left(x_0 - x_{eq}\right)^2 \left(x - x_{eq}\right)^2 dx_0 dp_0 dx dp \quad . \tag{126}$$

Because the analysis is for the correlation function of the stationary state, the initial distribution is the Boltzmann distribution

$$\rho_0\left(x_0, p_0\right) = \frac{\beta\omega}{2\pi} \exp\left\{-\beta\left[\frac{p_0^2}{2M} + \frac{1}{2} M\omega^2 \left(x_0 - x_{eq}\right)^2\right]\right\} \quad . \tag{127}$$

Consider the time derivative of Eq. (126), i.e.,

$$\frac{\partial}{\partial t}\left\langle \left[x(0)-x_{eq}\right]^2 \left[x(t)-x_{eq}\right]^2 \right\rangle = \int \rho_0\left(x_0, p_0\right) \frac{\partial \rho_t}{\partial t}\left(x_0 - x_{eq}\right)^2 \left(x - x_{eq}\right)^2 dx_0 dp_0 dx dp \quad . \tag{128}$$

Substituting Eq. (125) into Eq. (128) and applying integration by parts produces

$$\begin{aligned} &\frac{\partial}{\partial t}\left\langle \left[x(0)-x_{eq}\right]^2 \left[x(t)-x_{eq}\right]^2 \right\rangle \\ &= \int \rho_0\left(x_0, p_0\right)\left(2\left(x_0 - x_{eq}\right)^2 \left(x - x_{eq}\right) \frac{p}{M} \rho_t\right) dx_0 dp_0 dx dp \quad . \\ &= \frac{2}{M}\left\langle \left[x(0)-x_{eq}\right]^2 \left[x(t)-x_{eq}\right] p(t) \right\rangle \end{aligned} \tag{129}$$

Similarly, we further obtain

$$\begin{aligned} &\frac{\partial}{\partial t}\left\langle \left[x(0)-x_{eq}\right]^2 \left[x(t)-x_{eq}\right] p(t) \right\rangle \\ &= -M\omega^2 \left\langle \left[x(0)-x_{eq}\right]^2 \left[x(t)-x_{eq}\right]^2 \right\rangle - \gamma \left\langle \left[x(0)-x_{eq}\right]^2 \left[x(t)-x_{eq}\right] p(t) \right\rangle \quad , \\ &\quad + \frac{1}{M}\left\langle \left[x(0)-x_{eq}\right]^2 p^2(t) \right\rangle \end{aligned} \tag{130}$$

and

$$\begin{aligned} &\frac{\partial}{\partial t}\left\langle \left[x(0)-x_{eq}\right]^2 p^2(t) \right\rangle \\ &= -2M\omega^2 \left\langle \left[x(0)-x_{eq}\right]^2 \left[x(t)-x_{eq}\right] p(t) \right\rangle - 2\gamma \left\langle \left[x(0)-x_{eq}\right]^2 p^2(t) \right\rangle + \frac{2\gamma}{\beta^2\omega^2} \end{aligned} \quad . \tag{131}$$

Define

$$\begin{aligned} \boldsymbol{\chi}(t) &= \left(\chi^{(1)}(t), \chi^{(2)}(t), \chi^{(3)}(t)\right)^T \\ &= \left(\left\langle \left[x(0)-x_{eq}\right]^2 \left[x(t)-x_{eq}\right]^2 \right\rangle, \left\langle \left[x(0)-x_{eq}\right]^2 \left[x(t)-x_{eq}\right] p(t) \right\rangle, \left\langle \left[x(0)-x_{eq}\right]^2 p^2(t) \right\rangle\right)^T . \end{aligned}$$

Solving the differential function

$$\dot{\boldsymbol{\chi}} = \tilde{\mathbf{G}}\boldsymbol{\chi} + \tilde{\mathbf{g}} \tag{132}$$

where

$$\tilde{\mathbf{g}} = \left(0, 0, \frac{2\gamma}{\beta^2\omega^2}\right)^T \quad , \tag{133}$$

$$\tilde{\mathbf{G}} = \begin{pmatrix} 0 & \frac{2}{M} & 0 \\ -M\omega^2 & -\gamma & \frac{1}{M} \\ 0 & -2M\omega^2 & -2\gamma \end{pmatrix} \tag{134}$$

with the initial condition

$$\boldsymbol{\chi}(0) = \left(\frac{3}{\beta^2 M^2 \omega^4}, 0, \frac{1}{\beta^2\omega^2}\right)^T \quad , \tag{135}$$

we then obtain

$$\boldsymbol{\chi}(t) = e^{\tilde{\mathbf{G}}t}\left[\tilde{\mathbf{G}}^{-1}\tilde{\mathbf{g}} + \boldsymbol{\chi}(0)\right] - \tilde{\mathbf{G}}^{-1}\tilde{\mathbf{g}} \quad . \tag{136}$$

When $t$ goes to infinity, $\rho_t$ approaches the Boltzmann distribution, i.e.,

$$\rho_t \equiv \rho\left(x, p; t \middle| x_0, p_0; 0\right) \to \frac{\beta\omega}{2\pi}\exp\left\{-\beta\left[\frac{p^2}{2M} + \frac{1}{2}M\omega^2\left(x - x_{eq}\right)^2\right]\right\} \quad . \tag{137}$$

It is then straightforward to verify

$$\boldsymbol{\chi}(\infty) = \left(\left\langle\left(x - x_{eq}\right)^2\right\rangle^2, \ \left\langle\left(x - x_{eq}\right)^2\right\rangle\left\langle\left(x - x_{eq}\right)p\right\rangle, \ \left\langle\left(x - x_{eq}\right)^2\right\rangle\left\langle p^2\right\rangle\right)^T \quad . \tag{138}$$

The eigenvalues of $\tilde{\mathbf{G}}$ are $-\gamma$ , $-\gamma - 2\lambda$ and $-\gamma + 2\lambda$ , where $\lambda = \frac{1}{2}\sqrt{\gamma^2 - 4\omega^2}$ . Because the real parts of the eigenvalues of $\tilde{\mathbf{G}}$ are all negative, from Eq. (136) we obtain

$$\boldsymbol{\chi}(\infty) = -\tilde{\mathbf{G}}^{-1}\tilde{\mathbf{g}} = \left(\frac{1}{\beta^2 M^2 \omega^4}, 0, \frac{1}{\beta^2\omega^2}\right)^T \quad , \tag{139}$$

where

$$\tilde{\mathbf{G}}^{-1} = \begin{pmatrix} -\frac{\gamma^2+\omega^2}{2\gamma\omega^2} & -\frac{1}{M\omega^2} & -\frac{1}{2\gamma M^2\omega^2} \\ \frac{M}{2} & 0 & 0 \\ -\frac{M^2\omega^2}{2\gamma} & 0 & -\frac{1}{2\gamma} \end{pmatrix} . \quad (140)$$

Note that the characteristic correlation time of the potential [Eq. (122)] is

$$\begin{aligned} \tau_{pot} &= \int_0^\infty \frac{\left\langle \left[x(0)-x_{eq}\right]^2 \left[x(t)-x_{eq}\right]^2 \right\rangle - \left\langle \left(x-x_{eq}\right)^2 \right\rangle^2}{\left\langle \left(x-x_{eq}\right)^4 \right\rangle - \left\langle \left(x-x_{eq}\right)^2 \right\rangle^2} dt \\ &= \frac{\int_0^\infty \left[\chi^{(1)}(t)-\chi^{(1)}(\infty)\right] dt}{\chi^{(1)}(0)-\chi^{(1)}(\infty)} \end{aligned} . \quad (141)$$

The integral in the numerator is then

$$\int_0^\infty \left[\boldsymbol{\chi}(t)-\boldsymbol{\chi}(\infty)\right] dt = \int_0^\infty e^{\tilde{\mathbf{G}}t} \left[\tilde{\mathbf{G}}^{-1}\tilde{\mathbf{g}}+\boldsymbol{\chi}(0)\right] dt = -\tilde{\mathbf{G}}^{-1}\left[\tilde{\mathbf{G}}^{-1}\tilde{\mathbf{g}}+\boldsymbol{\chi}(0)\right] . \quad (142)$$

Substituting Eqs. (135), (139), and (140) into Eq. (142) leads to

$$\int_0^\infty \left[\boldsymbol{\chi}(t)-\boldsymbol{\chi}(\infty)\right] dt = \begin{pmatrix} \frac{1}{2}\left(\frac{1}{\gamma}+\frac{\gamma}{\omega^2}\right)\frac{2}{\beta^2 M^2\omega^4} \\ -\frac{1}{\beta^2 M\omega^4} \\ \frac{1}{\beta^2\omega^2\gamma} \end{pmatrix} . \quad (143)$$

Employing Eqs. (136), (139), and (142), we may show that Eq. (141) yields

$$\tau_{pot} = \frac{1}{2}\left(\frac{1}{\gamma}+\frac{\gamma}{\omega^2}\right) \quad (144)$$

for an infinitesimal time interval. The optimal value of the friction coefficient

$$\gamma_{pot}^{\text{opt}} = \omega \quad (145)$$

then leads to the minimum characteristic correlation time[24] for Eq. (144)

$$\tau_{pot}^{\min} = \frac{1}{\omega} . \quad (146)$$

Analogously, it is straightforward to prove that $\tau_{Ham}$ for the harmonic system [Eq. (100)]

is

$$\tau_{Ham} = \frac{1}{\gamma} + \frac{\gamma}{4\omega^2} \tag{147}$$

for an infinitesimal time interval. The optimal value of the friction coefficient

$$\gamma_{Ham}^{\text{opt}} = 2\omega \tag{148}$$

produces the minimum characteristic correlation time for Eq. (124)

$$\tau_{Ham}^{\min} = \frac{1}{\omega} \quad . \tag{149}$$

**2. Finite time interval**

When the time interval $\Delta t$ is finite, the optimal value of the friction coefficient depends on the underlying algorithm. The potential energy autocorrelation function [Eq. (121)] is expressed as

$$C_{pot}(n\Delta t) = \frac{\langle U(n\Delta t)U(0)\rangle - \langle U\rangle^2}{\langle U^2\rangle - \langle U\rangle^2} \quad . \tag{150}$$

The bracket $\langle\ \rangle$ of Eq. (150) denotes the phase space average of the stationary density distribution for the underlying algorithm. Its characteristic correlation time is then

$$\tau_{pot} = \Delta t \sum_{n=0}^{\infty} C_{pot}(n\Delta t) \quad . \tag{151}$$

Analogously, the Hamiltonian autocorrelation function and its characteristic correlation time can also be obtained for the finite time interval $\Delta t$.

Below we use the "middle" scheme [Eq. (23)] as the example.

**2A. Phase space propagator approach**

We may extend the derivation in Section IV-1 to study the characteristic correlation time for the finite time interval.

**1) Real dynamics case in the middle scheme**

The relevant Kolmogorov operators in Eqs. (20)-(22) for the one-dimensional harmonic system [Eq. (100)] become

$$\mathcal{L}_x\rho = -\frac{p}{M}\frac{\partial\rho}{\partial x} \tag{152}$$

$$\mathcal{L}_p\rho = M\omega^2\left(x-x_{eq}\right)\frac{\partial\rho}{\partial p} \tag{153}$$

$$\mathcal{L}_T\rho = \frac{\partial}{\partial p}\left(\gamma p\cdot\rho\right)+\frac{\gamma M}{\beta}\frac{\partial^2\rho}{\partial p^2} \quad . \tag{154}$$

We define the conditional densities

$$\begin{aligned}
&\rho_{n,0}\left(x,p\right)\equiv\rho\left(x,p;n\Delta t\left|x_0,p_0;0\right.\right)=\left(e^{\mathcal{L}^{\text{Middle}}\Delta t}\right)^n\delta\left(x-x_0\right)\delta\left(p-p_0\right)\\
&\rho_{n,1}\left(x,p\right)\equiv e^{\mathcal{L}_p\Delta t/2}\rho_{n,0}\left(x,p\right)\\
&\rho_{n,2}\left(x,p\right)\equiv e^{\mathcal{L}_x\Delta t/2}\rho_{n,1}\left(x,p\right)\\
&\rho_{n,3}\left(x,p\right)\equiv e^{\mathcal{L}_T\Delta t}\rho_{n,2}\left(x,p\right)\\
&\rho_{n,4}\left(x,p\right)\equiv e^{\mathcal{L}_x\Delta t/2}\rho_{n,3}\left(x,p\right)
\end{aligned} \quad , \tag{155}$$

which lead to

$$\rho_{n+1,0}\left(x,p\right)=e^{\mathcal{L}_p\Delta t/2}\rho_{n,4}\left(x,p\right) \quad . \tag{156}$$

Although the explicit expression of $\rho_{n,i}\left(x,p\right)$ $\left(i=\overline{0,4}\right)$ is difficult to obtain, we directly analyze the displacement squared autocorrelation function

$$\begin{aligned}
&\left\langle\left(x_0-x_{eq}\right)^2\left(x_n-x_{eq}\right)^2\right\rangle_i\\
&=\int\rho_0\left(x_0,p_0\right)\rho_{n,i}\left(x,p\right)\left(x_0-x_{eq}\right)^2\left(x-x_{eq}\right)^2dx_0dp_0dxdp \quad \left(i=\overline{0,4}\right)
\end{aligned} \quad . \tag{157}$$

When the “middle” scheme is employed, the initial distribution $\rho_0\left(x_0,p_0\right)$ is the stationary distribution [Eq. (55)], i.e.,

$$\rho_0^{\text{Middle}}\left(x_0,p_0\right)=\frac{1}{Z_N}\exp\left\{-\beta\left[\frac{p_0^2}{2M}\left(1-\frac{\omega^2\Delta t^2}{4}\right)^{-1}+\frac{1}{2}M\omega^2\left(x_0-x_{eq}\right)^2\right]\right\} \tag{158}$$

for the one-dimensional harmonic system [Eq. (100)].

Following the strategy in Part 1 of this section, it is easy to show

$$\boldsymbol{\chi}_{n,1} = \tilde{\mathbf{G}}_1 \boldsymbol{\chi}_{n,0} \quad , \tag{159}$$

$$\boldsymbol{\chi}_{n,2} = \tilde{\mathbf{G}}_2 \boldsymbol{\chi}_{n,1} \quad , \tag{160}$$

$$\boldsymbol{\chi}_{n,3} = \tilde{\mathbf{G}}_3 \boldsymbol{\chi}_{n,2} + \tilde{\tilde{\mathbf{g}}} \quad , \tag{161}$$

$$\boldsymbol{\chi}_{n,4} = \tilde{\mathbf{G}}_2 \boldsymbol{\chi}_{n,3} \quad , \tag{162}$$

$$\boldsymbol{\chi}_{n+1,0} = \tilde{\mathbf{G}}_1 \boldsymbol{\chi}_{n,4} \quad , \tag{163}$$

where

$$\boldsymbol{\chi}_{n,i} = \left( \left\langle \left(x_0 - x_{eq}\right)^2 \left(x_n - x_{eq}\right)^2 \right\rangle_i , \left\langle \left(x_0 - x_{eq}\right)^2 \left(x_n - x_{eq}\right) p_n \right\rangle_i , \left\langle \left(x_0 - x_{eq}\right)^2 p_n^2 \right\rangle_i \right)^T \quad \left(i = \overline{0,4}\right) \quad , \tag{164}$$

$$\tilde{\mathbf{G}}_1 = \begin{pmatrix} 1 & 0 & 0 \\ -M\omega^2 \dfrac{\Delta t}{2} & 1 & 0 \\ M^2\omega^4 \dfrac{\Delta t^2}{4} & -M\omega^2 \Delta t & 1 \end{pmatrix} \quad , \tag{165}$$

$$\tilde{\mathbf{G}}_2 = \begin{pmatrix} 1 & \dfrac{\Delta t}{M} & \dfrac{\Delta t^2}{4M^2} \\ 0 & 1 & \dfrac{\Delta t}{2M} \\ 0 & 0 & 1 \end{pmatrix} \quad , \tag{166}$$

$$\tilde{\mathbf{G}}_3 = \begin{pmatrix} 1 & 0 & 0 \\ 0 & e^{-\gamma \Delta t} & 0 \\ 0 & 0 & e^{-2\gamma \Delta t} \end{pmatrix} \quad , \tag{167}$$

and

$$\tilde{\tilde{\mathbf{g}}} = \left( 0, 0, \frac{1 - e^{-2\gamma \Delta t}}{\beta^2 \omega^2} \right)^T \quad . \tag{168}$$

Substituting Eqs. (159)-(162) into Eq. (163), we obtain

$$\boldsymbol{\chi}_{n+1,0} = \tilde{\mathbf{G}}_1 \tilde{\mathbf{G}}_2 \tilde{\mathbf{G}}_3 \tilde{\mathbf{G}}_2 \tilde{\mathbf{G}}_1 \boldsymbol{\chi}_{n,0} + \tilde{\mathbf{G}}_1 \tilde{\mathbf{G}}_2 \tilde{\tilde{\mathbf{g}}} \quad . \tag{169}$$

A more compact form of Eq. (169) is

$$\boldsymbol{\chi}_{n+1,0} = \bar{\bar{\mathbf{G}}}\boldsymbol{\chi}_{n,0} + \bar{\bar{\mathbf{g}}} \tag{170}$$

with

$$\bar{\bar{\mathbf{G}}} = \tilde{\mathbf{G}}_1\tilde{\mathbf{G}}_2\tilde{\mathbf{G}}_3\tilde{\mathbf{G}}_2\tilde{\mathbf{G}}_1 \tag{171}$$

and

$$\bar{\bar{\mathbf{g}}} = \tilde{\mathbf{G}}_1\tilde{\mathbf{G}}_2\tilde{\tilde{\mathbf{g}}} \quad . \tag{172}$$

When $n$ goes to infinity, $\rho_{n,0}\left(x,p\right)$ approaches the stationary distribution, i.e.,

$$\rho_{n,0}\left(x,p\right) \to \frac{1}{Z_N}\exp\left\{-\beta\left[\frac{p^2}{2M}\left(1-\frac{\omega^2\Delta t^2}{4}\right)^{-1} + \frac{1}{2}M\omega^2\left(x-x_{eq}\right)^2\right]\right\} \quad . \tag{173}$$

Then it is straightforward to verify

$$\begin{aligned}
&\left(\left\langle\left(x-x_{eq}\right)^2\right\rangle_0^2, \left\langle\left(x-x_{eq}\right)^2\right\rangle_0 \left\langle\left(x-x_{eq}\right)p\right\rangle_0, \left\langle\left(x-x_{eq}\right)^2\right\rangle_0 \left\langle p^2\right\rangle_0\right)^T \\
&= \boldsymbol{\chi}_{\infty,0} \\
&= \left(\mathbf{1}-\bar{\bar{\mathbf{G}}}\right)^{-1}\bar{\bar{\mathbf{g}}} \\
&= \left(\frac{1}{\beta^2 M^2\omega^4}, 0, \frac{1-\frac{\omega^2\Delta t^2}{4}}{\beta^2\omega^2}\right)^T
\end{aligned} \quad , \tag{174}$$

and

$$\begin{aligned}
&\boldsymbol{\chi}_{0,0} - \left(\mathbf{1}-\bar{\bar{\mathbf{G}}}\right)^{-1}\bar{\bar{\mathbf{g}}} \\
&= \boldsymbol{\chi}_{0,0} - \boldsymbol{\chi}_{\infty,0} \\
&= \left(\left\langle\left(x-x_{eq}\right)^4\right\rangle_0 - \left\langle\left(x-x_{eq}\right)^2\right\rangle_0^2, \ \left\langle\left(x-x_{eq}\right)^3 p\right\rangle_0 - \left\langle\left(x-x_{eq}\right)^2\right\rangle_0 \left\langle\left(x-x_{eq}\right)p\right\rangle_0, \right. \\
&\qquad \left. \left\langle\left(x-x_{eq}\right)^2 p^2\right\rangle_0 - \left\langle\left(x-x_{eq}\right)^2\right\rangle_0 \left\langle p^2\right\rangle_0\right)^T \\
&= \left(\frac{2}{\beta^2 M^2\omega^4}, 0, 0\right)^T
\end{aligned} \tag{175}$$

Rearranging Eq. (170) leads to

$$\boldsymbol{\chi}_{n+1,0}-\left(\mathbf{1}-\bar{\bar{\mathbf{G}}}\right)^{-1}\bar{\bar{\mathbf{g}}}=\bar{\bar{\mathbf{G}}}\left[\boldsymbol{\chi}_{n,0}-\left(\mathbf{1}-\bar{\bar{\mathbf{G}}}\right)^{-1}\bar{\bar{\mathbf{g}}}\right] \quad . \tag{176}$$

The recursion formula Eq. (176) leads to

$$\boldsymbol{\chi}_{n,0}-\left(\mathbf{1}-\bar{\bar{\mathbf{G}}}\right)^{-1}\bar{\bar{\mathbf{g}}}=\bar{\bar{\mathbf{G}}}^{n}\left[\boldsymbol{\chi}_{0,0}-\left(\mathbf{1}-\bar{\bar{\mathbf{G}}}\right)^{-1}\bar{\bar{\mathbf{g}}}\right] \quad . \tag{177}$$

Summing over $n$ from 0 to infinity in both sides of Eq. (177) produces

$$\sum_{n=0}^{\infty}\left[\boldsymbol{\chi}_{n,0}-\left(\mathbf{1}-\bar{\bar{\mathbf{G}}}\right)^{-1}\bar{\bar{\mathbf{g}}}\right]=\left(\mathbf{1}-\bar{\bar{\mathbf{G}}}\right)^{-1}\left[\boldsymbol{\chi}_{0,0}-\left(\mathbf{1}-\bar{\bar{\mathbf{G}}}\right)^{-1}\bar{\bar{\mathbf{g}}}\right] \quad . \tag{178}$$

Substituting Eqs. (174)-(175) into Eq. (178), we obtain

$$\begin{pmatrix} \sum_{n=0}^{\infty}\left(\left\langle\left(x_0-x_{eq}\right)^2\left(x_n-x_{eq}\right)^2\right\rangle_0-\left\langle\left(x-x_{eq}\right)^2\right\rangle_0^2\right) \\ \sum_{n=0}^{\infty}\left(\left\langle\left(x_0-x_{eq}\right)^2\left(x_n-x_{eq}\right)p_n\right\rangle_0-\left\langle\left(x-x_{eq}\right)^2\right\rangle_0\left\langle\left(x-x_{eq}\right)p\right\rangle_0\right) \\ \sum_{n=0}^{\infty}\left(\left\langle\left(x_0-x_{eq}\right)^2 p_n^2\right\rangle_0-\left\langle\left(x-x_{eq}\right)^2\right\rangle_0\left\langle p^2\right\rangle_0\right) \end{pmatrix}=\left(\mathbf{1}-\bar{\bar{\mathbf{G}}}\right)^{-1}\begin{pmatrix} \dfrac{2}{\beta^2 M^2\omega^4} \\ 0 \\ 0 \end{pmatrix} \quad . \tag{179}$$

The characteristic correlation time of the potential for a finite time interval $\Delta t$ [Eqs. (150)-(151)] is

$$\tau_{pot}=\Delta t\sum_{n=0}^{\infty}\frac{\left\langle\left(x_0-x_{eq}\right)^2\left(x_n-x_{eq}\right)^2\right\rangle_0-\left\langle\left(x-x_{eq}\right)^2\right\rangle_0^2}{\left\langle\left(x-x_{eq}\right)^4\right\rangle_0-\left\langle\left(x-x_{eq}\right)^2\right\rangle_0^2} \tag{180}$$

for the "middle" scheme. Eqs. (174), (179), and (180) lead to

$$\tau_{pot}^{\text{Middle}}=\left[\left(\mathbf{1}-\bar{\bar{\mathbf{G}}}\right)^{-1}\right]_{11}\Delta t \quad . \tag{181}$$

Here $\left[\left(\mathbf{1}-\bar{\bar{\mathbf{G}}}\right)^{-1}\right]_{11}$ represents the element in the 1st row and 1st column of the matrix $\left(\mathbf{1}-\bar{\bar{\mathbf{G}}}\right)^{-1}$. Substituting Eqs. (165)-(167) and (171) into Eq. (181) yields the explicit form

$$\tau_{pot}^{\text{Middle}}=\frac{\left(1-e^{-\gamma\Delta t}\right)^2+\left(1+e^{-\gamma\Delta t}\right)\left(3-e^{-\gamma\Delta t}\right)\left(\dfrac{\omega\Delta t}{2}\right)^2}{\omega^2\Delta t\left(1+e^{-\gamma\Delta t}\right)\left(1-e^{-\gamma\Delta t}\right)} \quad . \tag{182}$$

Interestingly, Eq. (182) indicates

$$\tau_{pot}^{\text{Middle}} \xrightarrow{\gamma \to 0} \infty \tag{183}$$

$$\tau_{pot}^{\text{Middle}} \xrightarrow{\gamma \to \infty} \frac{1+3\left(\frac{\omega \Delta t}{2}\right)^2}{\omega^2 \Delta t} \quad . \tag{184}$$

Eq. (183) holds for both an infinitesimal time interval and a finite one. While in the limit $\gamma \to \infty$ for an infinitesimal time interval the characteristic correlation time of the potential is infinite, that for a finite time interval is, however, a constant!

The optimal friction coefficient for Eq. (182) is

$$\gamma_{pot}^{\text{Middle, opt}} = \frac{1}{\Delta t} \ln\left(\frac{2+\omega \Delta t}{2-\omega \Delta t}\right) \tag{185}$$

such that the characteristic correlation time reaches the minimum value

$$\tau_{pot}^{\text{Middle, min}} = \frac{2+\omega \Delta t}{2\omega} \quad . \tag{186}$$

As $\Delta t \to 0$, Eq. (182), Eq. (185), and Eq. (186) approach Eq. (144), Eq. (145) and Eq. (146), respectively.

Similarly, the characteristic correlation time of the Hamiltonian for a finite time interval $\Delta t$ for the "middle" scheme may be shown as

$$\tau_{Ham}^{\text{Middle}} = \frac{\left(1-e^{-\gamma \Delta t}\right)^2 + \left(3+e^{-\gamma \Delta t}\right)^2 \left(\frac{\omega \Delta t}{2}\right)^2 - \left(3+e^{-\gamma \Delta t}\right)^2 \left(\frac{\omega \Delta t}{2}\right)^4 + \left(3-e^{-\gamma \Delta t}\right)\left(1+e^{-\gamma \Delta t}\right)\left(\frac{\omega \Delta t}{2}\right)^6}{\omega^2 \Delta t \left(1+e^{-\gamma \Delta t}\right)\left(1-e^{-\gamma \Delta t}\right)\left\{\left[1-\left(\frac{\omega \Delta t}{2}\right)^2\right]^2 + 1\right\}} . \tag{187}$$

Eq. (187) leads to

$$\tau_{Ham}^{\text{Middle}} \xrightarrow{\gamma \to 0} \infty \tag{188}$$

$$\tau_{Ham}^{\text{Middle}} \overset{\gamma\to\infty}{\to} \frac{1+9\left(\frac{\omega\Delta t}{2}\right)^2-9\left(\frac{\omega\Delta t}{2}\right)^4+3\left(\frac{\omega\Delta t}{2}\right)^6}{\omega^2\Delta t\left\{\left[1-\left(\frac{\omega\Delta t}{2}\right)^2\right]^2+1\right\}} . \tag{189}$$

The characteristic correlation time of the Hamiltonian in the limit $\gamma\to\infty$ for a finite time interval is also a constant.

The optimal friction coefficient for Eq. (187) is

$$\gamma_{Ham}^{\text{Middle, opt}} = \frac{1}{\Delta t}\ln\left\{\frac{1+5\left(\frac{\omega\Delta t}{2}\right)^2-5\left(\frac{\omega\Delta t}{2}\right)^4+\left(\frac{\omega\Delta t}{2}\right)^6+\omega\Delta t\left[2-\left(\frac{\omega\Delta t}{2}\right)^2\right]\sqrt{1+\left(\frac{\omega\Delta t}{2}\right)^2-\left(\frac{\omega\Delta t}{2}\right)^4}}{\left[1-\left(\frac{\omega\Delta t}{2}\right)^2\right]^3}\right\} \tag{190}$$

such that the characteristic correlation time reaches the minimum value

$$\tau_{Ham}^{\text{Middle, min}} = -\frac{\phi_1\Delta t}{\phi_2} \tag{191}$$

with

$$\begin{aligned}
\phi_1 &= 6\omega^{14}\Delta t^{14}-192\omega^{12}\Delta t^{12}+2400\omega^{10}\Delta t^{10}-13696\omega^{8}\Delta t^{8}\\
&\quad+23552\omega^{6}\Delta t^{6}+88064\omega^{4}\Delta t^{4}-294912\omega^{2}\Delta t^{2}-131072\\
&\quad+\left(-\omega^{10}\Delta t^{10}+36\omega^{8}\Delta t^{8}-432\omega^{6}\Delta t^{6}+2112\omega^{4}\Delta t^{4}-2048\omega^{2}\Delta t^{2}-10240\right)\\
&\qquad\times\sqrt{\omega^2\Delta t^2\left(8-\omega^2\Delta t^2\right)^2\left(-\omega^4\Delta t^4+4\omega^2\Delta t^2+16\right)}\\
\phi_2 &= 32\left\{\left[1-\left(\frac{\omega\Delta t}{2}\right)^2\right]^2+1\right\}\left[-2\omega^4\Delta t^4+8\omega^2\Delta t^2+32+\sqrt{\omega^2\Delta t^2\left(8-\omega^2\Delta t^2\right)^2\left(-\omega^4\Delta t^4+4\omega^2\Delta t^2+16\right)}\right]\\
&\quad\times\left[\omega^6\Delta t^6-16\omega^4\Delta t^4+64\omega^2\Delta t^2+2\sqrt{\omega^2\Delta t^2\left(8-\omega^2\Delta t^2\right)^2\left(-\omega^4\Delta t^4+4\omega^2\Delta t^2+16\right)}\right]
\end{aligned} \tag{192}$$

As $\Delta t\to 0$, Eq. (187), Eq. (190) and Eq. (191) approach Eq. (147), Eq. (148) and Eq. (149), respectively.

**2) Virtual dynamics case in the middle scheme**

Replacing the phase space propagator for the thermostat $e^{\mathcal{L}_T\Delta t}$ by its virtual dynamics version $e^{\mathcal{L}_T^{vir}\Delta t}$ in the “middle” scheme [Eq. (23)] leads to the “middle (vir)” scheme

$$e^{\mathcal{L}^{\text{Middle (vir)}}\Delta t} = e^{\mathcal{L}_p\Delta t/2} e^{\mathcal{L}_x\Delta t/2} e^{\mathcal{L}_T^{\text{vir}}\Delta t} e^{\mathcal{L}_x\Delta t/2} e^{\mathcal{L}_p\Delta t/2} \quad . \tag{193}$$

Note that Eq. (158) is also the stationary density distribution for the virtual dynamics case "middle (vir)" for the harmonic system.

Similar to Eq. (155), we have

$$\begin{aligned}
\rho_{n,0}(x,p) &\equiv \rho(x,p;n\Delta t|x_0,p_0;0) = \left(e^{\mathcal{L}^{\text{Middle (vir)}}\Delta t}\right)^n \delta(x-x_0)\delta(p-p_0) \\
\rho_{n,1}(x,p) &\equiv e^{\mathcal{L}_p\Delta t/2}\rho_{n,0}(x,p) \\
\rho_{n,2}(x,p) &\equiv e^{\mathcal{L}_x\Delta t/2}\rho_{n,1}(x,p) \\
\rho_{n,3}(x,p) &\equiv e^{\mathcal{L}_T^{vir}\Delta t}\rho_{n,2}(x,p) \\
\rho_{n,4}(x,p) &\equiv e^{\mathcal{L}_x\Delta t/2}\rho_{n,3}(x,p)
\end{aligned} \quad , \tag{194}$$

which lead to

$$\rho_{n+1,0}(x,p) = e^{\mathcal{L}_p\Delta t/2}\rho_{n,4}(x,p) \quad . \tag{195}$$

We define $\boldsymbol{\chi}_{i,n}$ in the same way as in the real dynamics case [Eq. (164)]. Analogously, we may verify

$$\begin{aligned}
\boldsymbol{\chi}_{n,1} &= \tilde{\mathbf{G}}_1\boldsymbol{\chi}_{n,0} \\
\boldsymbol{\chi}_{n,2} &= \tilde{\mathbf{G}}_2\boldsymbol{\chi}_{n,1} \\
\boldsymbol{\chi}_{n,3} &= \tilde{\mathbf{G}}_3'\boldsymbol{\chi}_{n,2} + \tilde{\tilde{\mathbf{g}}} \\
\boldsymbol{\chi}_{n,4} &= \tilde{\mathbf{G}}_2\boldsymbol{\chi}_{n,3} \\
\boldsymbol{\chi}_{n+1,0} &= \tilde{\mathbf{G}}_1\boldsymbol{\chi}_{n,4}
\end{aligned} \tag{196}$$

with $\tilde{\mathbf{G}}_1$, $\tilde{\mathbf{G}}_2$ and $\tilde{\tilde{\mathbf{g}}}$ defined in Eq. (165), Eq. (166), and Eq. (168), respectively, and

$$\tilde{\mathbf{G}}_3' = \begin{pmatrix} 1 & 0 & 0 \\ 0 & -e^{-\gamma\Delta t} & 0 \\ 0 & 0 & e^{-2\gamma\Delta t} \end{pmatrix} \quad . \tag{197}$$

Eq. (196) leads to

$$\boldsymbol{\chi}_{n+1,0} = \bar{\bar{\mathbf{G}}}'\boldsymbol{\chi}_{n,0} + \bar{\bar{\mathbf{g}}} \tag{198}$$

with $\bar{\bar{\mathbf{G}}}' = \tilde{\mathbf{G}}_1\tilde{\mathbf{G}}_2\tilde{\mathbf{G}}_3'\tilde{\mathbf{G}}_2\tilde{\mathbf{G}}_1$ and $\bar{\bar{\mathbf{g}}} = \tilde{\mathbf{G}}_1\tilde{\mathbf{G}}_2\tilde{\tilde{\mathbf{g}}}$. Following the same procedure as in the real dynamics case, the characteristic correlation time of the potential for "middle (vir)" may be

shown as

$$\tau_{pot}^{\text{Middle (vir)}} = \left[ \left( \mathbf{1} - \bar{\bar{\mathbf{G}}}' \right)^{-1} \right]_{11} \Delta t = \frac{\left(1+e^{-\gamma\Delta t}\right)^2 + \left(1-e^{-\gamma\Delta t}\right)\left(3+e^{-\gamma\Delta t}\right)\left(\frac{\omega\Delta t}{2}\right)^2}{\omega^2 \Delta t \left(1+e^{-\gamma\Delta t}\right)\left(1-e^{-\gamma\Delta t}\right)} \quad . \tag{199}$$

Similarly, we obtain the characteristic correlation time of the Hamiltonian for "middle (vir)"

$$\tau_{Ham}^{\text{Middle (vir)}} = \frac{\left(1+e^{-\gamma\Delta t}\right)^2 + \left(3-e^{-\gamma\Delta t}\right)^2\left(\frac{\omega\Delta t}{2}\right)^2 - \left(3-e^{-\gamma\Delta t}\right)^2\left(\frac{\omega\Delta t}{2}\right)^4 + \left(3+e^{-\gamma\Delta t}\right)\left(1-e^{-\gamma\Delta t}\right)\left(\frac{\omega\Delta t}{2}\right)^6}{\omega^2 \Delta t \left(1+e^{-\gamma\Delta t}\right)\left(1-e^{-\gamma\Delta t}\right)\left\{\left[1-\left(\frac{\omega\Delta t}{2}\right)^2\right]^2 + 1\right\}} \ . \tag{200}$$

In the virtual dynamics case of the "middle" scheme, the characteristic correlation length of either the potential or the Hamiltonian monotonically decreases as the friction $\gamma$ increases.

It is easy to show

$$\tau_{pot}^{\text{Middle (vir)}} - \tau_{pot}^{\text{Middle}} = \frac{4e^{-\gamma\Delta t}\left[1-\left(\frac{\omega\Delta t}{2}\right)^2\right]}{\omega^2 \Delta t \left(1+e^{-\gamma\Delta t}\right)\left(1-e^{-\gamma\Delta t}\right)} > 0 \tag{201}$$

$$\tau_{Ham}^{\text{Middle (vir)}} - \tau_{Ham}^{\text{Middle}} = \frac{4e^{-\gamma\Delta t}\left[1-\left(\frac{\omega\Delta t}{2}\right)^2\right]^3}{\omega^2 \Delta t \left(1+e^{-\gamma\Delta t}\right)\left(1-e^{-\gamma\Delta t}\right)\left\{\left[1-\left(\frac{\omega\Delta t}{2}\right)^2\right]^2 + 1\right\}} > 0 \quad , \tag{202}$$

i.e., $\tau_{pot}^{\text{Middle (vir)}} > \tau_{pot}^{\text{Middle}}$ and $\tau_{Ham}^{\text{Middle (vir)}} > \tau_{Ham}^{\text{Middle}}$ when the friction coefficient $\gamma$ is finite. The characteristic correlation time for the virtual dynamics case is *always* larger than that for the real dynamics case.

Interestingly, for a finite time interval $\Delta t$ we have

$$\tau_{pot}^{\text{Middle (vir)}} \xrightarrow{\gamma\to\infty} \frac{1+3\left(\frac{\omega\Delta t}{2}\right)^2}{\omega^2 \Delta t} \tag{203}$$

$$\tau_{Ham}^{\text{Middle (vir)}} \xrightarrow{\gamma\to\infty} \frac{1+9\left(\frac{\omega\Delta t}{2}\right)^2 - 9\left(\frac{\omega\Delta t}{2}\right)^4 + 3\left(\frac{\omega\Delta t}{2}\right)^6}{\omega^2\Delta t\left\{\left[1-\left(\frac{\omega\Delta t}{2}\right)^2\right]^2+1\right\}} \quad . \tag{204}$$

That is, as $\gamma\to\infty$ the characteristic correlation time for the virtual dynamics case approaches the same limit as that for the real dynamics case does.

**2B. Trajectory-based approach**

**1) Real dynamics case in the middle scheme**

Eq. (80) leads to

$$\begin{aligned}
&\mathbf{R}_n - \bar{\mathbf{R}} \\
&= \tilde{\mathbf{M}}^n\mathbf{R}_0 + \sum_{j=0}^{n-1}\tilde{\mathbf{M}}^j\left(\mathbf{F}_0 + \tilde{\mathbf{\Omega}}_{n-1-j}\right) - \left(\lim_{m\to\infty}\tilde{\mathbf{M}}^m\mathbf{R}_0 + \sum_{j=0}^{\infty}\tilde{\mathbf{M}}^j\mathbf{F}_0\right) \\
&= \tilde{\mathbf{M}}^n\left(\mathbf{R}_0 - \lim_{m\to\infty}\tilde{\mathbf{M}}^m\mathbf{R}_0 - \sum_{j=0}^{\infty}\tilde{\mathbf{M}}^j\mathbf{F}_0\right) + \sum_{j=0}^{n-1}\tilde{\mathbf{M}}^j\tilde{\mathbf{\Omega}}_{n-1-j} \\
&= \tilde{\mathbf{M}}^n\left(\mathbf{R}_0 - \bar{\mathbf{R}}\right) + \sum_{j=0}^{n-1}\tilde{\mathbf{M}}^j\tilde{\mathbf{\Omega}}_{n-1-j}
\end{aligned} \quad . \tag{205}$$

Substituting Eqs. (105) and (106) into Eq. (205) produces

$$x_n - x_{eq} = \left(v_{11}u_{11}\varepsilon_1^n + v_{12}u_{21}\varepsilon_2^n\right)\left(x_0 - x_{eq}\right) + \left(v_{11}u_{12}\varepsilon_1^n + v_{12}u_{22}\varepsilon_2^n\right)p_0 + \sum_{j=0}^{n-1}\left(c_{x1}\varepsilon_1^j + c_{x2}\varepsilon_2^j\right)\Omega_{n-1-j} \quad . \tag{206}$$

Because the stationary density distribution of the "middle" scheme [Eq. (55)] is a product of the position distribution and the momentum distribution, then it is straightforward to show

$$\begin{aligned}
&\left\langle\left(x_n - x_{eq}\right)^2\left(x_0 - x_{eq}\right)^2\right\rangle - \left\langle\left(x_n - x_{eq}\right)^2\right\rangle\left\langle\left(x_0 - x_{eq}\right)^2\right\rangle \\
&= \left(v_{11}u_{11}\varepsilon_1^n + v_{12}u_{21}\varepsilon_2^n\right)^2\left[\left\langle\left(x_0 - x_{eq}\right)^4\right\rangle - \left\langle\left(x_0 - x_{eq}\right)^2\right\rangle^2\right]
\end{aligned} \quad . \tag{207}$$

Eq. (150) for the real dynamics case then becomes

$$C_{pot}^{\text{Middle}}(n\Delta t) = \left(v_{11}u_{11}\varepsilon_1^n + v_{12}u_{21}\varepsilon_2^n\right)^2 \quad . \tag{208}$$

The characteristic correlation time of the potential energy is

$$\begin{aligned}\tau_{pot}^{\text{Middle}} &= \sum_{n=0}^{\infty}\left(v_{11}u_{11}\varepsilon_1^n + v_{12}u_{21}\varepsilon_2^n\right)^2 \Delta t \\ &= \sum_{n=0}^{\infty}\left[\left(v_{11}u_{11}\right)^2 \varepsilon_1^{2n} + \left(v_{12}u_{21}\right)^2 \varepsilon_2^{2n} + 2v_{11}u_{11}v_{12}u_{21}\varepsilon_1^n\varepsilon_2^n\right]\Delta t\end{aligned} \quad . \tag{209}$$

The summation over $n$ in Eq. (209) produces

$$\tau_{pot}^{\text{Middle}} = \left[\left(v_{11}u_{11}\right)^2 \frac{1}{1-\varepsilon_1^2} + \left(v_{12}u_{21}\right)^2 \frac{1}{1-\varepsilon_2^2} + 2v_{11}u_{11}v_{12}u_{21}\frac{1}{1-\varepsilon_1\varepsilon_2}\right]\Delta t \quad . \tag{210}$$

Substituting the elements of Eq. (101), of Eq. (113), and of Eq. (114) into Eq. (210), we obtain

$$\begin{aligned}\tau_{pot}^{\text{Middle}} = \frac{1}{4D\sinh^2\lambda}\Bigg[&\left(\sqrt{D}\sinh\lambda + \frac{1-D}{2}\right)^2 \frac{1}{1-\varepsilon_1^2} + \left(\sqrt{D}\sinh\lambda - \frac{1-D}{2}\right)^2 \frac{1}{1-\varepsilon_2^2} \\ &+2\left(\sqrt{D}\sinh\lambda + \frac{1-D}{2}\right)\left(\sqrt{D}\sinh\lambda - \frac{1-D}{2}\right)\frac{1}{1-\varepsilon_1\varepsilon_2}\Bigg]\Delta t\end{aligned} \quad . \tag{211}$$

Using Eq. (101), Eq. (102), and Eq. (111), we may express Eq. (211) in a more simplified form

$$\begin{aligned}\tau_{pot}^{\text{Middle}} &= \frac{\left(5-2D+D^2-3T+DT\right)\Delta t}{4\left(D+1-T\right)\left(1-D\right)} \\ &= \frac{\left(1-e^{-\gamma\Delta t}\right)^2 + \left(1+e^{-\gamma\Delta t}\right)\left(3-e^{-\gamma\Delta t}\right)\left(\frac{\omega\Delta t}{2}\right)^2}{\omega^2\Delta t\left(1+e^{-\gamma\Delta t}\right)\left(1-e^{-\gamma\Delta t}\right)}\end{aligned} \quad . \tag{212}$$

That is, the trajectory-based approach also leads to the same result as Eq. (182).

We may follow the same procedure to prove that the characteristic correlation time of the Hamiltonian for a finite time interval $\Delta t$ is the same as Eq. (187).

**2) Virtual dynamics case in the middle scheme**

Analogously, we may obtain the characteristic correlation time of the potential and that of the Hamiltonian for the virtual dynamics case of the "middle" scheme, which are the same as Eq. (199) and Eq. (200), respectively.

**2C. Other schemes**

It is straightforward to use either the phase space propagator approach or the trajectory-based one to obtain the characteristic correlation time for other schemes. While Table 2 presents the characteristic correlation time of the potential for each scheme, Table 3 shows that of the Hamiltonian.

Since the characteristic correlation time reaches a plateau in the limit $\gamma \to \infty$ in any one of the eight schemes that employ the first type of repartition, Tables 4 and 5 present the plateau value of the characteristic correlation time of the potential and that of the Hamiltonian, respectively. It should be stressed that for the harmonic system the real and virtual dynamics cases in the "side" scheme share effectively same behaviors (in Tables 2-5). This is also true for the "PV-side" scheme.

Finally we note that Leimkuhler and Matthews employed a different approach to obtain the mean and second-order moments of the stationary state distribution for a few Langevin dynamics algorithms for a one-dimensional harmonic potential when the time interval is finite[21], but they did not show the form of the stationary state distribution, neither did they study the characteristic correlation time.

## VI. Examples and discussions

### 1. Comparison of the schemes for one-dimensional models

We first test two standard 1-dimensional models—a harmonic oscillator

$$U(x) = \frac{1}{2} M \omega^2 x^2 \tag{213}$$

[i.e., $A = M\omega^2$ and $x_{eq} = 0$ in Eq. (100)] and a quartic potential

$$U(x) = x^4/4 \quad . \tag{214}$$

Since the quartic potential has no harmonic terms, it presents a challenging model to verify the conclusions drawn from the analytical analysis for the harmonic system.

**1A. Thermal fluctuations**

We consider thermal fluctuations of the potential, kinetic energy, and Hamiltonian, which indicate the accuracy of an underlying algorithm for sampling configurational space, the momentum space, and the whole phase space, respectively. Figs. 1 and 2 demonstrate the thermal fluctuation as a function of the time interval for the harmonic oscillator and for the quartic model. The numerical results in Figs. 1 and 2 show that the "middle/middle (vir)" scheme is the most accurate in sampling the configurational space, while the "PV-end/PV-end (vir)" scheme is the best in sampling the momentum space. It is easy to follow the proof in Section IV of our earlier work[25] to prove that the "PV-end"/ "PV-side"/ "PV-beginning" schemes lead the same stationary state marginal distribution of the coordinate for a general system while the "end"/ "side"/ "beginning" schemes do so as well, regardless of whether real or virtual dynamics is implemented. Analogously, we may follow Ref. [25] to prove that the ascending order for the error of the momentum distribution for a general system is

"PV-end" ≤ "PV-side" ≤ "PV-beginning"

and

"end" ≤ "side" ≤ "beginning",

irrespective of whether the real or virtual dynamics case is employed. Figs. 1 and 2 also demonstrate that when the friction coefficient $\gamma$ takes a reasonable value, the "middle-xT" scheme is the most accurate in sampling the phase space among the twelve schemes.

In addition to thermal fluctuations, we also calculate the average values for the potential and kinetic energy. (See supplementary material.) The results agree well with the conclusion that we draw.

**1B. Characteristic correlation time**

We then study the characteristic correlation time as a function of the friction coefficient $\gamma$. The twelve schemes may be divided into three categories. The stationary distribution in the harmonic limit is independent of $\gamma$ for any one of the first eight schemes. The first category consists of the "side" and "PV-side" schemes, since in either scheme results of the real dynamics case and those of the virtual dynamics case share effectively same behaviors (at least for the two one-dimensional model systems). The rest six schemes fall into the second category. In each of these schemes virtual dynamics yields a substantially different algorithm from what real dynamics leads to. For demonstration we choose the "side" scheme in the first category and the "middle" scheme in the second category. The third category include the schemes of which the stationary distribution in the harmonic limit depends on $\gamma$. The last four schemes fall into this category. We choose the "middle-xT" scheme for demonstration. While Figs. 3 and 4 demonstrate the analytic results on the characteristic correlation time of the potential and on that of the Hamiltonian, respectively, for the "middle"/"side"/"middle-xT" schemes for the harmonic system [Eq. (213)], Figs. 5 and 6 show the numerical results for three schemes for the quartic model [Eq. (214)].

Consider the real dynamics case in the first eight schemes. While the characteristic correlation time goes to infinity as the friction coefficient approaches zero, it gradually reaches a plateau as the friction coefficient approaches infinity. [See Panels (a)-(b) of Figs. 3-6.] As shown in Fig. 7 for the harmonic system, the product of the plateau value and the frequency $\tau^{\gamma\to\infty}\omega$ is a function of $\omega\Delta t$ [the explicit form of which is given in Table 4 or Table 5]. In the "middle" scheme the plateau value $\tau^{\gamma\to\infty}$ (overall) decreases as the time interval $\Delta t$ increases, regardless of whether the characteristic correlation time of the potential or that of the Hamiltonian is considered.

When the real dynamics case in any one of these schemes is employed, the optimal friction coefficient that produces the minimum characteristic correlation time is often a function of the time interval $\Delta t$. The optimal friction coefficient is often finite. That is, while the characteristic correlation time decays as the friction coefficient increases from zero to the optimal friction, it increases to reach a plateau as the friction coefficient increases from the optimal value to infinity. For the schemes other than "middle", the optimal friction coefficient is not always finite (i.e., the characteristic correlation time may monotonically decay as the friction coefficient increases). [See Panel (b) of Figs. 3-4.] For the harmonic system, the range of $\omega\Delta t$ where the optimal friction coefficient is finite is listed for each of the first eight schemes in Table 6. While Panels (b) and (d) of Fig. 8 show the optimal friction coefficient as a function of the time interval $\Delta t$, Panels (a) and (c) of Fig. 8 do so for the corresponding minimum characteristic correlation time. In the "middle" scheme (for the harmonic system) the optimal friction coefficient is always finite for either of the characteristic correlation time of the potential and that of the Hamiltonian, as long as the dynamics is in the stable region $\omega\Delta t < 2$.

When $\Delta t$ is reasonably large without loss of much accuracy, both the minimum and the plateau (in the high friction limit) of the characteristic correlation time are considerably small in such a scheme as “middle”, which indicates that it may be efficient and robust within a broad range of the friction coefficient. [See Panels (a)-(b) of Figs. 3-6, Fig. 7, and Panels (a) and (c) of Fig. 8.] So in terms of sampling efficiency it is often more favorable to choose a relatively and reasonably large friction coefficient rather than a small one in the “middle” scheme, when no knowledge of the optimal friction coefficient for a system is available.

The virtual dynamics case of any one scheme in the first category (i.e., “side” or ”PV-side”) is effectively identical to the real dynamics case of the same scheme. [See Panel (b) of Figs. 3-6.] In contrast to the characteristic correlation time in the real dynamics case of each scheme of the second category, the corresponding virtual dynamics case always monotonically decreases as the friction coefficient increases. When the friction coefficient approaches infinity, the characteristic correlation time in the virtual dynamics case reaches the same plateau as that in the real dynamics case of the same scheme does. [See Panel (a) of Figs. 3-6.]

We then consider the last four schemes (“middle-xT”/“side-xT”/“middle-pT”/ “side-xT”). The characteristic correlation time of the potential (or the Hamiltonian) also has a minimum value. The optimal value of the friction coefficient that produces the minimum characteristic correlation time is also a function of the time interval $\Delta t$ . The characteristic correlation time goes to infinity as the friction coefficient approaches either zero or infinity. Unlike the first eight schemes, the last four schemes do not lead to a plateau of the characteristic correlation time in the high friction limit. The characteristic correlation time is much more sensitive to the friction coefficient in the last four schemes. [See Panel (c) of Figs. 3-6] Fig. 8 compares "middle-xT" to "middle" and "side" on the minimum value of the characteristic correlation time and the corresponding optimal friction coefficient for the harmonic system. Such as the "middle-xT" scheme in the third category may offer an alternative good algorithm if the optimal friction region is easy to obtain for anharmonic systems.

In conclusion, the curves in Figs. 3-4 represent the analytic results of the characteristic correlation time for the harmonic system[36] presented in Section V and Tables 2-6, the numerical results in Figs. 5-6 for the quartic potential suggest the similar behavior of the characteristic correlation time also exists for anharmonic systems.

**2. Numerical performance of the "middle" scheme**

Since the configurational sampling is often more important in molecular simulations, the "middle/middle(vir)" scheme is then recommended for Langevin dynamics simulations. In addition to the two one-dimensional models, two typical "real" systems are investigated for studying the dependence of the accuracy and efficiency of the "middle/middle(vir)" scheme on the friction coefficient. The first molecular system is the $H_2O$ molecule with the accurate potential energy surface developed by Partridge and Schwenke from extensive *ab initio* calculations and experimental data[37]. As the explicit form of the PES is available, that of the force can be expressed. The simulations are performed for $T = 100\ \mathrm{K}$. The second "real" system is $(Ne)_{13}$, a Lennard-Jones (LJ) cluster. The parameters of the system are described in Ref.[38]. The simulations are performed for $T = 14\ \mathrm{K}$.

First consider the dependence of accuracy (of numerical results) on the friction coefficient.

Two coordinate-dependent properties are studied, which include the average potential energy and the fluctuation of the potential. For the harmonic system the "middle/middle(vir)" scheme leads to the exact configurational distribution regardless of the value of the friction coefficient. This indicates that coordinate-dependent properties obtained from the "middle/middle(vir)" schemes are exact and independent of the friction coefficient in the harmonic limit. The numerical results in Panels (a)-(b) of Fig. 9 are in good agreement (within statistical error bars) with the analytical analysis. Similarly, it is shown in Panels (c)-(d) of Fig. 9 and in Fig. 10 that for anharmonic and/or "real" systems the results produced by the "middle/middle(vir)" scheme are also relatively insensitive to the friction coefficient in a broad region when the time interval is fixed. The region ranges from around the optimal value to infinity.

Since the dependence of accuracy on the friction coefficient is weak in such a wide range, the characteristic correlation time is the next important factor to consider. Fig. 11 shows that the characteristic correlation time gradually approaches a plateau in the high friction limit for the two "real" systems. This is consistent with the behavior of the characteristic correlation time depicted in Figs. 3-6 for the two one-dimensional systems. It is then suggested that a reasonable value for the friction should be around the optimal value [Eq. (185)] or larger.

In summary, the results of Figs. 3-6 and Figs. 9-11 suggest that both the accuracy and efficiency of the "middle/middle(vir)" scheme are insensitive to the friction coefficient in a broad range even for anharmonic and/or "real" systems.

## VII. Conclusion remarks

To present a unified theoretical framework for understanding the Boltzmann thermostat based on Langevin dynamics, we give a comprehensive study on the performance of different numerical schemes for solving the Langevin equation. Three types of repartition of the Langevin equation describing the finite-time change of the phase-space point are thus proposed and investigated. While the first type of repartition [Eq. (10)] of the Langevin equation involves the exact realization of the Ornstein-Uhlenbeck noise [Eq. (13)], of which numerical algorithms include both real and virtual dynamics cases. This type of repartition includes the first eight schemes ("middle"/"end"/"beginning"/"side"/"PV-middle"/"PV-end"/ "PV-beginning"/"PV-side"). The second and third types of repartition [Eq. (31) and Eq. (36)] do not "obviously" result in the virtual dynamics analogue giving the correct Boltzmann distribution[39]. They lead to the last four schemes ("middle-pT"/ "side-pT"/ "middle-xT"/"side-xT"). Other types of repartition may also be introduced. (See more discussion in supplementary material.)

By either directly solving the discrete equations of motion or using the related phase space propagators, we introduce two different theoretical approaches to obtain both the stationary state distribution and characteristic correlation time for the harmonic system when the time interval $\Delta t$ is finite. It is shown that the stationary distribution of each of the first eight schemes is independent of the friction coefficient, regardless of whether the real or virtual dynamics case is employed. In contrast the stationary distribution of each of the last four schemes depends on the friction coefficient. When the friction coefficient approaches infinity, the stationary distribution (in each of the last four schemes) is even not well-defined.

The characteristic correlation time may be used as a measure for the simulation efficiency.

It turns out that the real dynamics algorithm in any one of the schemes often has a minimum correlation time when the time interval $\Delta t$ is in a certain range. The optimal value of the friction coefficient that produces the minimum correlation time is often a function of the time interval $\Delta t$. In any one of the schemes the characteristic correlation time goes to infinity as the friction coefficient approaches zero. As the friction coefficient approaches infinity, the characteristic correlation time in each of the last four schemes goes to infinity while that in each of the first eight schemes gradually reaches a plateau. It is then expected that simulation results produced by the first eight schemes are much less sensitive to the friction coefficient than those obtained by the last four schemes.

The real and virtual dynamics cases of the "side" scheme share same/similar behaviors, so are those of the "PV-side" scheme. More interesting is the virtual dynamics case in each of the "middle"/"end"/"beginning"/"PV-middle"/"PV-end"/"PV-beginning" schemes, which is substantially different from the corresponding real dynamics case. The characteristic correlation time in the virtual dynamics case monotonically decreases as the friction coefficient increases, which eventually approaches the same plateau in the high friction limit as that in the real dynamics case of the same scheme does.

The numerical examples show that the conclusions drawn from the analytical analysis for the harmonic system are applicable to anharmonic models and "real" molecular systems. The real dynamics case of the "middle" scheme is recommended as the best (second-order) algorithm for performing Langevin dynamics. It has two important properties:

1) It produces the most accurate configurational sampling. Its numerical performance in accuracy is relatively insensitive to the friction coefficient in a wide range from around the optimal value to the high friction value.

2) Its characteristic correlation time always has a plateau as an upper bound in the high friction limit when the time interval is finite. This guarantees the sampling efficiency when the friction coefficient is chosen in a wide range from around the optimal value to the high friction value for a reasonably large time interval.

Note that the Boltzmann distribution $e^{-\beta H(\mathbf{x},\mathbf{p})} = \exp\left[-\beta \mathbf{p}^T \mathbf{M}^{-1} \mathbf{p}/2\right] \exp\left[-\beta U(\mathbf{x})\right]$ is a product of the configurational distribution $e^{-\beta U(\mathbf{x})}$ and the Maxwell momentum distribution. Because the latter is simply a Gaussian function, it is trivial to employ Monte Carlo (techniques) for accurate momentum sampling. We may use the "middle" scheme to obtain the marginal distribution of the coordinate while sampling the momentum space separately by Monte Carlo. Such an approach will offer accurate and efficient phase space sampling for the Boltzmann distribution.

Finally, we note that our investigation has implications to other types of thermostats. Our recent study[25] indicates that the conclusions for numerical schemes with the first type of repartition of the Langevin equation in the present work may in principle apply to other thermostats. Further investigation along the direction is certainly warranted.

## SUPPLEMENTARY MATERIAL

See supplementary material for more results of the numerical examples in part 1 of Section VI and for more discussion on numerical algorithms on Langevin dynamics.

## Acknowledgements

This work was supported by the Ministry of Science and Technology of China (MOST)

Grant No. 2016YFC0202803, by the 973 program of MOST No. 2013CB834606, by the National Natural Science Foundation of China (NSFC) Grants No. 21373018, No. 21573007, and No. 21421003, by the Recruitment Program of Global Experts, by Specialized Research Fund for the Doctoral Program of Higher Education No. 20130001110009, and by Special Program for Applied Research on Super Computation of the NSFC-Guangdong Joint Fund (the second phase) under Grant No.U1501501. We acknowledge the Beijing and Tianjin supercomputer centers and the High-performance Computing Platform of Peking University for providing computational resources. This research also used resources of the National Energy Research Scientific Computing Center, a DOE Office of Science User Facility supported by the Office of Science of the U.S. Department of Energy under Contract No. DE-AC02-05CH11231. We thank Lin Lin for informing Ref. [10] when the manuscript was prepared.

## Appendix A. Phase space propagator approach for deriving the stationary state distribution for the harmonic system

We may employ phase space propagators to do the analysis for all the twelve schemes. Here we adopt a strategy similar to what we have recently developed for the analysis for the Andersen thermostat[25].

### 1. Schemes that employ the second or third type of repartition

Since it is more difficult to obtain the stationary state distribution for the harmonic system for the last four schemes [Eq. (34), Eq. (35), Eq. (48), and Eq. (49)] that employ the second or third type of repartition, we choose “middle-xT” (of these four schemes) for demonstration of the derivation procedure.

Consider a one-dimensional harmonic system where Eq. (50) becomes Eq. (100). Use the "middle-xT" scheme [Eq. (48)] as the example. The relevant Kolmogorov operators in Eq. (21) and Eq. (47) become

$$\mathcal{L}_p \rho = A\left(x - x_{eq}\right)\frac{\partial \rho}{\partial p} \tag{A1}$$

$$\mathcal{L}_{x-T}\rho = -\frac{p}{M}\frac{\partial \rho}{\partial x} + \frac{\partial}{\partial p}\left(\gamma p \cdot \rho\right) + \frac{\gamma M}{\beta}\frac{\partial^2 \rho}{\partial p^2} \quad . \tag{A2}$$

We define the following density distributions

$$\begin{aligned} \rho_{n,0}\left(x,p\right) &\equiv \left(e^{\mathcal{L}^{\text{Middle-xT}}\Delta t}\right)^n \rho_0\left(x,p\right) \\ \rho_{n,1}\left(x,p\right) &\equiv e^{\mathcal{L}_p \Delta t/2}\rho_{n,0}\left(x,p\right) \qquad , \\ \rho_{n,2}\left(x,p\right) &\equiv e^{\mathcal{L}_{x-T}\Delta t}\rho_{n,1}\left(x,p\right) \end{aligned} \tag{A3}$$

which lead to

$$\rho_{n+1,0}\left(x,p\right) = e^{\mathcal{L}_p \Delta t/2}\rho_{n,2}\left(x,p\right) \quad . \tag{A4}$$

Then we may define $\boldsymbol{\upsilon}_{n,i}$ as

$$\boldsymbol{\upsilon}_{n,i} = \begin{pmatrix} \langle x \rangle_{n,i} \\ \langle p \rangle_{n,i} \end{pmatrix} \quad \left(i = \overline{0,2}\right) \quad , \tag{A5}$$

with

$$\langle x \rangle_{n,i} = \int \rho_{n,i}\left(x,p\right) x \, dxdp \quad \left(i = \overline{0,2}\right) \tag{A6}$$

$$\langle p \rangle_{n,i} = \int \rho_{n,i}\left(x,p\right) p \, dxdp \quad \left(i = \overline{0,2}\right) \quad . \tag{A7}$$

It is easy to verify

$$\begin{aligned} \langle x \rangle_{n,1} &= \int \rho_{n,1}\left(x,p\right) x \, dxdp \\ &= \int e^{\mathcal{L}_p \Delta t/2}\rho_{n,0}\left(x,p\right) x \, dxdp \\ &= \int \rho_{n,0}\left(x,p\right) x \, dxdp \\ &= \langle x \rangle_{n,0} \end{aligned} \tag{A8}$$

and

$$\begin{aligned}\langle p\rangle_{n,1} &= \int \rho_{n,1}(x,p)\, p\; dxdp \\ &= \int e^{\mathcal{L}_p \Delta t/2} \rho_{n,0}(x,p)\, p\; dxdp \\ &= \int \rho_{n,0}(x,p)\left[ p - \frac{\Delta t}{2} A\left(x - x_{eq}\right)\right] dxdp \quad , \\ &= \langle p\rangle_{n,0} - \frac{\Delta t}{2} A \langle x\rangle_{n,0} + \frac{\Delta t}{2} A x_{eq}\end{aligned} \tag{A9}$$

which can be expressed as

$$\boldsymbol{\upsilon}_{n,1} = \mathbf{J}_1 \boldsymbol{\upsilon}_{n,0} + \mathbf{j} \tag{A10}$$

with

$$\mathbf{J}_1 = \begin{pmatrix} 1 & 0 \\ -\frac{\Delta t}{2} A & 1 \end{pmatrix} \quad , \tag{A11}$$

$$\mathbf{j} = \begin{pmatrix} 0 \\ \frac{\Delta t}{2} A x_{eq} \end{pmatrix} \quad . \tag{A12}$$

Analogously, we may obtain

$$\boldsymbol{\upsilon}_{n,2} = \mathbf{J}_2 \boldsymbol{\upsilon}_{n,1} \quad , \tag{A13}$$

$$\boldsymbol{\upsilon}_{n+1,0} = \mathbf{J}_1 \boldsymbol{\upsilon}_{n,2} + \mathbf{j} \quad , \tag{A14}$$

with

$$\mathbf{J}_2 = \begin{pmatrix} 1 & \frac{1 - e^{-\gamma \Delta t}}{\gamma M} \\ 0 & e^{-\gamma \Delta t} \end{pmatrix} \quad . \tag{A15}$$

Substituting Eq. (A10) and Eq. (A13) into Eq. (A14), we have

$$\boldsymbol{\upsilon}_{n+1,0} = \overline{\mathbf{J}} \boldsymbol{\upsilon}_{n,0} + \overline{\mathbf{j}} \quad , \tag{A16}$$

with $\overline{\mathbf{J}} = \mathbf{J}_1 \mathbf{J}_2 \mathbf{J}_1$ and $\overline{\mathbf{j}} = \left(\mathbf{1} + \mathbf{J}_1 \mathbf{J}_2\right)\mathbf{j}$. Then the averaged phased-space point can be obtained by

$$\begin{pmatrix} \bar{x} \\ \bar{p} \end{pmatrix} = \lim_{n\to\infty} \begin{pmatrix} \langle x \rangle_{n,0} \\ \langle p \rangle_{n,0} \end{pmatrix} = \boldsymbol{\upsilon}_{\infty,0} = \left(\mathbf{1} - \bar{\mathbf{J}}\right)^{-1} \bar{\mathbf{j}} = \begin{pmatrix} x_{eq} \\ 0 \end{pmatrix} . \tag{A17}$$

Consider a series of vectors $\boldsymbol{\xi}_{n,i}$ defined as

$$\boldsymbol{\xi}_{n,i} = \left( \left\langle \left(x - x_{eq}\right)^2 \right\rangle_{n,i} ,\ \left\langle \left(x - x_{eq}\right) p \right\rangle_{n,i} ,\ \left\langle p^2 \right\rangle_{n,i} \right)^T \quad \left(i = \overline{0,2}\right) , \tag{A18}$$

with

$$\left\langle \left(x - x_{eq}\right)^2 \right\rangle_{n,i} = \int \rho_{n,i}\left(x, p\right)\left(x - x_{eq}\right)^2 dxdp \quad \left(i = \overline{0,2}\right) , \tag{A19}$$

$$\left\langle \left(x - x_{eq}\right) p \right\rangle_{n,i} = \int \rho_{n,i}\left(x, p\right)\left(x - x_{eq}\right) p \ dxdp \quad \left(i = \overline{0,2}\right) , \tag{A20}$$

and

$$\left\langle p^2 \right\rangle_{n,i} = \int \rho_{n,i}\left(x, p\right) p^2 dxdp \quad \left(i = \overline{0,2}\right) . \tag{A21}$$

Following the same strategy in Eqs. (A10), (A13), and (A14), we may show

$$\boldsymbol{\xi}_{n,1} = \mathbf{G}_1 \boldsymbol{\xi}_{n,0} , \tag{A22}$$

$$\boldsymbol{\xi}_{n,2} = \mathbf{G}_2 \boldsymbol{\xi}_{n,1} + \mathbf{g} , \tag{A23}$$

$$\boldsymbol{\xi}_{n+1,0} = \mathbf{G}_1 \boldsymbol{\xi}_{n,2} , \tag{A24}$$

with

$$\mathbf{G}_1 = \begin{pmatrix} 1 & 0 & 0 \\ -A\dfrac{\Delta t}{2} & 1 & 0 \\ A^2 \dfrac{\Delta t^2}{4} & -A\Delta t & 1 \end{pmatrix} , \tag{A25}$$

$$
\mathbf{G}_2 = \begin{pmatrix} 1 & \dfrac{2\left(1-e^{-\gamma\Delta t}\right)}{\gamma M} & \dfrac{\left(1-e^{-\gamma\Delta t}\right)^2}{\gamma^2 M^2} \\ 0 & e^{-\gamma\Delta t} & \dfrac{e^{-\gamma\Delta t}\left(1-e^{-\gamma\Delta t}\right)}{\gamma M} \\ 0 & 0 & e^{-2\gamma\Delta t} \end{pmatrix} , \tag{A26}
$$

and

$$
\mathbf{g} = \begin{pmatrix} \dfrac{2\gamma\Delta t - 3 + 4e^{-\gamma\Delta t} - e^{-2\gamma\Delta t}}{\beta\gamma^2 M} \\ \dfrac{\left(1-e^{-\gamma\Delta t}\right)^2}{\beta\gamma} \\ \dfrac{M\left(1-e^{-2\gamma\Delta t}\right)}{\beta} \end{pmatrix} . \tag{A27}
$$

Substituting Eqs. (A22)-(A23) into Eq. (A24), we have

$$
\boldsymbol{\xi}_{n+1,0} = \bar{\mathbf{G}}\boldsymbol{\xi}_{n,0} + \bar{\mathbf{g}} \tag{A28}
$$

with $\bar{\mathbf{G}} = \mathbf{G}_1\mathbf{G}_2\mathbf{G}_1$ and $\bar{\mathbf{g}} = \mathbf{G}_1\mathbf{g}$. It is straightforward to verify

$$
\begin{aligned}
&\left(\left\langle\left(x-x_{eq}\right)^2\right\rangle, \ \left\langle\left(x-x_{eq}\right)p\right\rangle, \ \left\langle p^2\right\rangle\right)^T \\
&= \lim_{n\to\infty}\left(\left\langle\left(x-x_{eq}\right)^2\right\rangle_{n,0}, \ \left\langle\left(x-x_{eq}\right)p\right\rangle_{n,0}, \ \left\langle p^2\right\rangle_{n,0}\right)^T \\
&= \boldsymbol{\xi}_{\infty,0} \\
&= \left(\mathbf{1}-\bar{\mathbf{G}}\right)^{-1}\bar{\mathbf{g}}
\end{aligned} \quad . \tag{A29}
$$

We may then obtain

$$
\left(\left\langle\left(x-x_{eq}\right)^2\right\rangle,\ \left\langle\left(x-x_{eq}\right)p\right\rangle,\ \left\langle p^2\right\rangle\right)^T
$$

$$
=\begin{pmatrix}
\dfrac{1}{\beta A}\dfrac{2\left[\gamma^2\left(1-e^{-2\gamma\Delta t}\right)+\left(e^{-2\gamma\Delta t}+2\gamma\Delta t e^{-\gamma\Delta t}-1\right)\dfrac{A}{M}\right]}{\gamma\left(1-e^{-\gamma\Delta t}\right)\left[2\gamma\left(1+e^{-\gamma\Delta t}\right)-\Delta t\left(1-e^{-\gamma\Delta t}\right)\dfrac{A}{M}\right]} \\
\dfrac{1}{\beta}\dfrac{\left(1+e^{-\gamma\Delta t}\right)\left[\gamma\Delta t\left(1+e^{-\gamma\Delta t}\right)-2\left(1-e^{-\gamma\Delta t}\right)\right]}{\left(1-e^{-\gamma\Delta t}\right)\left[\Delta t\left(1-e^{-\gamma\Delta t}\right)\dfrac{A}{M}-2\gamma\left(1+e^{-\gamma\Delta t}\right)\right]} \\
\dfrac{M}{\beta}\dfrac{\theta_1}{\theta_2}
\end{pmatrix}
\tag{A30}
$$

with

$$
\begin{aligned}
\theta_1 = &\left\{\frac{1}{2}\left(1+e^{-\gamma\Delta t}\right)^3\frac{A}{M}\gamma^2\Delta t^2+\frac{1}{2}\left(1-e^{-\gamma\Delta t}\right)^2\left(1+e^{-\gamma\Delta t}\right)\left[4\gamma^2+\left(\frac{A}{M}\right)^2\Delta t^2\right]\right.\\
&\left.-\frac{A}{M}\gamma\Delta t\left(1-e^{-\gamma\Delta t}\right)\left[3\left(1+e^{-2\gamma\Delta t}\right)+e^{-\gamma\Delta t}\left(\frac{A}{M}\Delta t^2-2\right)\right]\right\} .\\
\theta_2 = &\ \gamma\left(1-e^{-\gamma\Delta t}\right)^2\left[2\gamma\left(1+e^{-\gamma\Delta t}\right)-\frac{A}{M}\Delta t\left(1-e^{-\gamma\Delta t}\right)\right]
\end{aligned}
\tag{A31}
$$

Define the fluctuation correlation matrix

$$
\mathbf{W}=\begin{pmatrix} W_{xx} & W_{xp} \\ W_{px} & W_{pp} \end{pmatrix}
\tag{A32}
$$

with

$$
\begin{aligned}
W_{xx} &= \left\langle\left(x-x_{eq}\right)^2\right\rangle \\
W_{xp} &= W_{px} = \left\langle\left(x-x_{eq}\right)p\right\rangle \\
W_{pp} &= \left\langle p^2\right\rangle
\end{aligned} .
\tag{A33}
$$

Substituting Eq. (A30) into Eq. (A32) generates the second-order moments for the "middle-xT" scheme. The mean [Eq. (A17)] and the second-order moments [Eq. (A32)] are not sufficient to obtain the stationary state distribution. Higher-order moments are required. It is straightforward (though tedious) to obtain all the higher-order moments and then the moment generating function, similar to our recent work on the Andersen thermostat (Appendix A of Ref. [25]).

Another method is to assume the stationary state distribution is of a Gaussian form, where the mean [Eq. (A17)] and the fluctuation correlation matrix [Eq. (A32)] are sufficient to determine the distribution. It is trivial to follow Section IV-1 to prove that the stationary state for the one-dimensional harmonic potential [Eq. (100)] produced by the "middle-xT" scheme is

$$\rho(x,p)=\frac{1}{\tilde{Z}}\exp\left[-\frac{1}{2}\left(\mathbf{R}-\bar{\mathbf{R}}\right)^T\mathbf{W}^{-1}\left(\mathbf{R}-\bar{\mathbf{R}}\right)\right] , \tag{A34}$$

where $\mathbf{R}=(x,p)^T$, $\bar{\mathbf{R}}=\left(x_{eq},0\right)^T$, and $\tilde{Z}$ is a normalization constant.

Similarly, we may prove that the stationary state of any one of the last four schemes [Eq. (34), Eq. (35), Eq. (48), and Eq. (49)] shares the same form as Eq. (A34) except that the corresponding fluctuation correlation matrix is different. The results are listed in Table 1.

### 2. Schemes that employ the first type of repartition

One may follow the derivation procedure in the first part of Appendix A to prove that the one-dimensional version of Eq. (55), Eq. (57), Eq. (58), or Eq. (59) is the stationary state for the corresponding scheme (for either real dynamics or virtual dynamics cases) for the one-dimensional harmonic system.

## Appendix B. Relation between two Langevin dynamics algorithms

The update of the position and momentum based on Grønbech-Jensen and Farago's algorithm[10] can be expressed as

$$\mathbf{x}_{n+1}=\mathbf{x}_n+\frac{\mathbf{1}+\mathbf{a}}{2}\Delta t\mathbf{M}^{-1}\mathbf{p}_n-\frac{\mathbf{1}+\mathbf{a}}{4}\Delta t^2\mathbf{M}^{-1}\nabla_{\mathbf{x}_n}U(\mathbf{x}_n)+\frac{\mathbf{1}+\mathbf{a}}{4}\Delta t\mathbf{M}^{-1}\sqrt{\frac{2\mathbf{M}\boldsymbol{\gamma}^{\mathrm{GF}}\Delta t}{\beta}}\boldsymbol{\mu}_n \tag{B1}$$

$$\mathbf{p}_{n+1}=\mathbf{a}\mathbf{p}_n-\frac{\Delta t}{2}\left(\mathbf{a}\nabla_{\mathbf{x}_n}U(\mathbf{x}_n)+\nabla_{\mathbf{x}_{n+1}}U(\mathbf{x}_{n+1})\right)+\frac{\mathbf{1}+\mathbf{a}}{2}\sqrt{\frac{2\mathbf{M}\boldsymbol{\gamma}^{\mathrm{GF}}\Delta t}{\beta}}\boldsymbol{\mu}_n . \tag{B2}$$

Here

$$\mathbf{a} = \left(\mathbf{1} + \frac{1}{2}\boldsymbol{\gamma}^{\mathrm{GF}}\Delta t\right)^{-1}\left(\mathbf{1} - \frac{1}{2}\boldsymbol{\gamma}^{\mathrm{GF}}\Delta t\right) \tag{B3}$$

with $\boldsymbol{\gamma}^{\mathrm{GF}}$ as the (diagonal) friction coefficient matrix defined in Grønbech-Jensen and Farago's algorithm, and

$$\boldsymbol{\mu}_n \equiv \boldsymbol{\mu}\left(n\Delta t, \Delta t\right) \tag{B4}$$

is a standard-Gaussian-random-number vector defined by Eqs. (17)-(18). Note that Eqs. (B1) and (B2) can be rewritten as

$$\mathbf{x}_{n+1} = \mathbf{x}_n + \frac{\mathbf{1}+\mathbf{a}}{2}\Delta t\mathbf{M}^{-1}\mathbf{p}_n - \frac{\mathbf{1}+\mathbf{a}}{4}\Delta t^2\mathbf{M}^{-1}\nabla_{\mathbf{x}_n}U(\mathbf{x}_n) + \frac{\Delta t}{2}\mathbf{M}^{-1}\sqrt{\frac{\mathbf{M}}{\beta}\left(\mathbf{1}-\mathbf{a}^2\right)}\boldsymbol{\mu}_n \tag{B5}$$

$$\mathbf{p}_{n+1} = \mathbf{a}\mathbf{p}_n - \frac{\Delta t}{2}\left(\mathbf{a}\nabla_{\mathbf{x}_n}U(\mathbf{x}_n) + \nabla_{\mathbf{x}_{n+1}}U(\mathbf{x}_{n+1})\right) + \sqrt{\frac{\mathbf{M}}{\beta}\left(\mathbf{1}-\mathbf{a}^2\right)}\boldsymbol{\mu}_n \quad . \tag{B6}$$

The update of the position and momentum in the "middle" scheme [Eq. (23)] is expressed as Eqs. (62)-(63). Note that the Gaussian random number vector $\boldsymbol{\Omega}_n$ in Eqs. (62)-(63) may be rewritten as

$$\boldsymbol{\Omega}_n \equiv \boldsymbol{\Omega}(n\Delta t, \Delta t) = \sqrt{\frac{1}{\beta}}\mathbf{M}^{1/2}\left(\mathbf{1} - e^{-2\gamma\Delta t}\right)^{1/2}\boldsymbol{\mu}_n(n\Delta t, \Delta t) \tag{B7}$$

with the standard-Gaussian-random-number vector $\boldsymbol{\mu}_n \equiv \boldsymbol{\mu}\left(n\Delta t, \Delta t\right)$ defined by Eqs. (17)-(18). Eqs. (62)-(63) then become

$$\mathbf{x}_{n+1} = \mathbf{x}_n + \frac{\Delta t}{2}\mathbf{M}^{-1}\left(\mathbf{1} + e^{-\gamma\Delta t}\right)\left[\mathbf{p}_n - \frac{\Delta t}{2}\nabla_{\mathbf{x}_n}U(\mathbf{x}_n)\right] + \frac{\Delta t}{2}\sqrt{\frac{1}{\beta}}\mathbf{M}^{-1/2}\left(\mathbf{1} - e^{-2\gamma\Delta t}\right)^{1/2}\boldsymbol{\mu}_n \tag{B8}$$

$$\mathbf{p}_{n+1} = e^{-\gamma\Delta t}\left[\mathbf{p}_n - \frac{\Delta t}{2}\nabla_{\mathbf{x}_n}U(\mathbf{x}_n)\right] - \frac{\Delta t}{2}\nabla_{\mathbf{x}_{n+1}}U(\mathbf{x}_{n+1}) + \sqrt{\frac{1}{\beta}}\mathbf{M}^{1/2}\left(\mathbf{1} - e^{-2\gamma\Delta t}\right)^{1/2}\boldsymbol{\mu}_n \quad . \tag{B9}$$

Grønbech-Jensen and Farago's algorithm[10] is equivalent to the "middle" scheme when

$$\mathbf{a} = e^{-\gamma\Delta t} \quad . \tag{B10}$$

Substituting Eq. (B10) into Eq. (B3), we obtain

$$\gamma^{\mathrm{GF}} = \frac{2}{\Delta t}\left(\mathbf{1}+e^{-\gamma\Delta t}\right)^{-1}\left(\mathbf{1}-e^{-\gamma\Delta t}\right) \quad . \tag{B11}$$

That is, the friction coefficient $\gamma^{\mathrm{GF}}$ defined in Grønbech-Jensen and Farago's algorithm is *not* the true friction coefficient $\gamma$ defined in the original Langevin equation Eqs. (2)-(3), but a function of both $\gamma$ and the time interval $\Delta t$. $\gamma^{\mathrm{GF}}$ is equivalent to $\gamma$ only in the limit $\Delta t \to 0$.

Analogously, it is straightforward to verify that when

$$\mathbf{a} = -e^{-\gamma\Delta t} \quad , \tag{B12}$$

Grønbech-Jensen and Farago's algorithm[10] is equivalent to the "middle (vir)" scheme. Substituting Eq. (B12) into Eq. (B3), we obtain

$$\gamma^{\mathrm{GF}} = \frac{2}{\Delta t}\left(\mathbf{1}-e^{-\gamma\Delta t}\right)^{-1}\left(\mathbf{1}+e^{-\gamma\Delta t}\right) \quad . \tag{B13}$$

## Tables and Figures

**Table 1.** Fluctuation correlation matrices in the stationary state distribution [Eq. (A34)] for the last four schemes for the one-dimensional harmonic system [Eq. (100)].

| Scheme | Fluctuation correlation matrix |
|---|---|
| Middle-pT | $W_{xx}=\frac{1}{\beta}\frac{1}{2}\gamma\Delta t\left(1-e^{-\gamma\Delta t}\right)^{-1}\left(1+e^{-\gamma\Delta t}\right)A^{-1}$<br>$W_{xp}=W_{px}=0$<br>$W_{pp}=\frac{M}{\beta}2\left(1+e^{-\gamma\Delta t}\right)\left[2\left(1+e^{-\gamma\Delta t}\right)-\Delta tAM^{-1}\gamma^{-1}\left(1-e^{-\gamma\Delta t}\right)\right]^{-1}$ |
| Side-pT | $W_{xx}=\frac{1}{\beta A}\frac{\gamma\Delta t\left(1+e^{-\gamma\Delta t}\right)^{2}}{\left(1-e^{-\gamma\Delta t}\right)\left[2\left(1+e^{-\gamma\Delta t}\right)-\Delta tAM^{-1}\gamma^{-1}\left(1-e^{-\gamma\Delta t}\right)\right]}$<br>$W_{xp}=-\frac{1}{\beta}\frac{\Delta t\left(1-e^{-2\gamma\Delta t}\right)}{(1+e^{-\frac{1}{2}\gamma\Delta t})^{2}\left[2\left(1+e^{-\gamma\Delta t}\right)-\Delta tAM^{-1}\gamma^{-1}\left(1-e^{-\gamma\Delta t}\right)\right]}$<br>$W_{pp}=\frac{M}{\beta}\frac{-4\Delta tAM^{-1}e^{-\frac{1}{2}\gamma\Delta t}(1-e^{-\frac{1}{2}\gamma\Delta t})+2\gamma\left(1+e^{-\gamma\Delta t}\right)(1+e^{-\frac{1}{2}\gamma\Delta t})}{(1+e^{-\frac{1}{2}\gamma\Delta t})\left[-\Delta tAM^{-1}\left(1-e^{-\gamma\Delta t}\right)+2\gamma\left(1+e^{-\gamma\Delta t}\right)\right]}$ |
| Middle-xT | $W_{xx}=\frac{1}{\beta A}\frac{2\left[\gamma^{2}\left(1-e^{-2\gamma\Delta t}\right)+\left(e^{-2\gamma\Delta t}+2\gamma\Delta te^{-\gamma\Delta t}-1\right)AM^{-1}\right]}{\gamma\left(1-e^{-\gamma\Delta t}\right)\left[2\gamma\left(1+e^{-\gamma\Delta t}\right)-\Delta t\left(1-e^{-\gamma\Delta t}\right)AM^{-1}\right]}$<br>$W_{xp}=W_{px}=\frac{1}{\beta}\frac{\left(1+e^{-\gamma\Delta t}\right)\left[\gamma\Delta t\left(1+e^{-\gamma\Delta t}\right)-2\left(1-e^{-\gamma\Delta t}\right)\right]}{\left(1-e^{-\gamma\Delta t}\right)\left[\Delta t\left(1-e^{-\gamma\Delta t}\right)AM^{-1}-2\gamma\left(1+e^{-\gamma\Delta t}\right)\right]}$<br>$W_{pp}=\frac{M}{\beta}\frac{\theta_1}{\theta_2}$<br>with<br>$\theta_1=\left\{\frac{1}{2}\left(1+e^{-\gamma\Delta t}\right)^{3}AM^{-1}\gamma^{2}\Delta t^{2}+\frac{1}{2}\left(1-e^{-\gamma\Delta t}\right)^{2}\left(1+e^{-\gamma\Delta t}\right)\left[4\gamma^{2}+\left(AM^{-1}\right)^{2}\Delta t^{2}\right]\right.$<br>$\left.-AM^{-1}\gamma\Delta t\left(1-e^{-\gamma\Delta t}\right)\left[3\left(1+e^{-2\gamma\Delta t}\right)+e^{-\gamma\Delta t}\left(AM^{-1}\Delta t^{2}-2\right)\right]\right\}$<br>$\theta_2=\gamma\left(1-e^{-\gamma\Delta t}\right)^{2}\left[2\gamma\left(1+e^{-\gamma\Delta t}\right)-AM^{-1}\Delta t\left(1-e^{-\gamma\Delta t}\right)\right]$ |

<table>
<tr>
<td>Side-xT</td>
<td>

$$W_{xx} = \frac{1}{\beta A}\frac{\bar{\theta}_1}{\bar{\theta}_2}$$

$$W_{xp} = W_{px} = \frac{1}{\beta}\frac{\bar{\theta}_3}{\bar{\theta}_4}$$

$$W_{pp} = \frac{M}{\beta}\frac{(1+e^{-\gamma\Delta t})\left[2\Delta t^2\gamma AM^{-1}e^{-\gamma\Delta t} + 2\gamma\left(1-e^{-\gamma\Delta t}\right)^2 - \Delta t AM^{-1}\left(1-e^{-2\gamma\Delta t}\right)\right]}{\left(1-e^{-\gamma\Delta t}\right)^2\left[-\Delta t\left(1-e^{-\gamma\Delta t}\right)AM^{-1} + 2\gamma\left(1+e^{-\gamma\Delta t}\right)\right]}$$

with

$$\begin{aligned}\bar{\theta}_1 = \Big\{ & -\Delta t^2\gamma\left(AM^{-1}\right)^2(-e^{-\frac{5}{2}\gamma\Delta t} + 3e^{-2\gamma\Delta t} - 3e^{-\frac{1}{2}\gamma\Delta t} + 1) \\ & -2\gamma\left(\gamma^2 - AM^{-1}\right)(1+e^{-\frac{1}{2}\gamma\Delta t})\left(1-e^{-2\gamma\Delta t}\right) \\ & +2\Delta t\left(AM^{-1}\right)(1+e^{-\frac{1}{2}\gamma\Delta t})\Big[4AM^{-1}e^{-\gamma\Delta t} + \gamma^2(1+e^{-2\gamma\Delta t}) \\ & \qquad -2\left(AM^{-1}+\gamma^2\right)e^{-\frac{1}{2}\gamma\Delta t}(1+e^{-\gamma\Delta t})\Big]\Big\}\end{aligned}$$

$$\bar{\theta}_2 = \gamma^2(1+e^{-\frac{1}{2}\gamma\Delta t})\left(1-e^{-\gamma\Delta t}\right)\left[\Delta t\left(1-e^{-\gamma\Delta t}\right)AM^{-1} - 2\gamma\left(1+e^{-\gamma\Delta t}\right)\right]$$

$$\begin{aligned}\bar{\theta}_3 = \Big\{ & -2\Delta t^2\gamma AM^{-1}e^{-\frac{1}{2}\gamma\Delta t}(1-e^{-\frac{3}{2}\gamma\Delta t}) - 2\gamma(1+e^{-\frac{1}{2}\gamma\Delta t})\left(1-e^{-2\gamma\Delta t}\right) \\ & +\Delta t(1+e^{-\frac{1}{2}\gamma\Delta t})\Big[AM^{-1}(1-e^{-\frac{1}{2}\gamma\Delta t})^2(1+4e^{-\frac{1}{2}\gamma\Delta t} + e^{-\gamma\Delta t}) \\ & \qquad +2\gamma^2 e^{-\frac{1}{2}\gamma\Delta t}\left(1+e^{-\gamma\Delta t}\right)\Big]\Big\}\end{aligned}$$

$$\bar{\theta}_4 = \gamma(1+e^{-\frac{1}{2}\gamma\Delta t})\left(1-e^{-\gamma\Delta t}\right)\left[\Delta t\left(1-e^{-\gamma\Delta t}\right)AM^{-1} - 2\gamma\left(1+e^{-\gamma\Delta t}\right)\right]$$

</td>
</tr>
</table>

**Table 2.** Characteristic correlation time of the potential energy

| Scheme | $\tau_{pot}$ |
|---|---|
| Middle(vir)<br>PV-middle(vir) | $\dfrac{\left(1+e^{-\gamma\Delta t}\right)^2+\left(1-e^{-\gamma\Delta t}\right)\left(3+e^{-\gamma\Delta t}\right)\left(\dfrac{\omega\Delta t}{2}\right)^2}{\omega^2\Delta t\left(1+e^{-\gamma\Delta t}\right)\left(1-e^{-\gamma\Delta t}\right)}$ |
| Middle<br>PV-middle | $\dfrac{\left(1-e^{-\gamma\Delta t}\right)^2+\left(1+e^{-\gamma\Delta t}\right)\left(3-e^{-\gamma\Delta t}\right)\left(\dfrac{\omega\Delta t}{2}\right)^2}{\omega^2\Delta t\left(1+e^{-\gamma\Delta t}\right)\left(1-e^{-\gamma\Delta t}\right)}$ |
| End(vir)<br>PV-end(vir)<br>Beginning(vir)<br>PV-beginning(vir) | $\dfrac{\left(1+e^{-\gamma\Delta t}\right)^2-e^{-\gamma\Delta t}\left(2+e^{-\gamma\Delta t}\right)\omega^2\Delta t^2\left(1-\dfrac{\omega^2\Delta t^2}{4}\right)}{\omega^2\Delta t\left(1+e^{-\gamma\Delta t}\right)\left(1-e^{-\gamma\Delta t}\right)\left(1-\dfrac{\omega^2\Delta t^2}{4}\right)}$ |
| Side(vir)<br>Side<br>PV-side(vir)<br>PV-side<br>End<br>PV-end<br>Beginning<br>PV-beginning | $\dfrac{\left(1-e^{-\gamma\Delta t}\right)^2+e^{-\gamma\Delta t}\left(2-e^{-\gamma\Delta t}\right)\omega^2\Delta t^2\left(1-\dfrac{\omega^2\Delta t^2}{4}\right)}{\omega^2\Delta t\left(1+e^{-\gamma\Delta t}\right)\left(1-e^{-\gamma\Delta t}\right)\left(1-\dfrac{\omega^2\Delta t^2}{4}\right)}$ |
| Middle-pT | $\dfrac{2\left(1-e^{-\gamma\Delta t}\right)+\omega^2\Delta t\gamma^{-1}\left(3-e^{-\gamma\Delta t}\right)}{4\omega^2\gamma^{-1}\left(1-e^{-\gamma\Delta t}\right)}$ |
| Side-pT | $\dfrac{\left[\omega^4\Delta t^2e^{-\gamma\Delta t}\left(1-e^{-\gamma\Delta t}\right)\left(2-e^{-\gamma\Delta t}\right) -2\omega^2\Delta t\gamma e^{-\gamma\Delta t}\left(1+e^{-\gamma\Delta t}\right)\left(2-e^{-\gamma\Delta t}\right)-\gamma^2\left(1-e^{-\gamma\Delta t}\right)\left(1+e^{-\gamma\Delta t}\right)^2\right]}{\omega^2\left(1+e^{-\gamma\Delta t}\right)\left(1-e^{-\gamma\Delta t}\right)\left[\omega^2\Delta t\left(1-e^{-\gamma\Delta t}\right)-2\gamma\left(1+e^{-\gamma\Delta t}\right)\right]}$ |
| Middle-xT<br>Side-xT | See supplementary material |

**Table 3.** Characteristic correlation time of the Hamiltonian

| Scheme | $\tau_{Ham}$ |
|---|---|
| Middle(vir) | $\dfrac{\left[\left(1+e^{-\gamma\Delta t}\right)^2+\left(3-e^{-\gamma\Delta t}\right)^2\left(\frac{\omega\Delta t}{2}\right)^2-\left(3-e^{-\gamma\Delta t}\right)^2\left(\frac{\omega\Delta t}{2}\right)^4+\left(3+e^{-\gamma\Delta t}\right)\left(1-e^{-\gamma\Delta t}\right)\left(\frac{\omega\Delta t}{2}\right)^6\right]}{\omega^2\Delta t\left(1+e^{-\gamma\Delta t}\right)\left(1-e^{-\gamma\Delta t}\right)\left[\left(1-\frac{\omega^2\Delta t^2}{4}\right)^2+1\right]}$ |
| Middle | $\dfrac{\left[\left(1-e^{-\gamma\Delta t}\right)^2+\left(3+e^{-\gamma\Delta t}\right)^2\left(\frac{\omega\Delta t}{2}\right)^2-\left(3+e^{-\gamma\Delta t}\right)^2\left(\frac{\omega\Delta t}{2}\right)^4+\left(3-e^{-\gamma\Delta t}\right)\left(1+e^{-\gamma\Delta t}\right)\left(\frac{\omega\Delta t}{2}\right)^6\right]}{\omega^2\Delta t\left(1+e^{-\gamma\Delta t}\right)\left(1-e^{-\gamma\Delta t}\right)\left[\left(1-\frac{\omega^2\Delta t^2}{4}\right)^2+1\right]}$ |
| PV-middle(vir) | $\dfrac{\left[256\left(1+e^{-\gamma\Delta t}\right)^2-128\left(e^{-2\gamma\Delta t}+6e^{-\gamma\Delta t}-3\right)\omega^2\Delta t^2+16\left(3e^{-2\gamma\Delta t}+22e^{-\gamma\Delta t}-13\right)\omega^4\Delta t^4-8\left(e^{-2\gamma\Delta t}+6e^{-\gamma\Delta t}-5\right)\omega^6\Delta t^6-\left(3+e^{-\gamma\Delta t}\right)\left(1-e^{-\gamma\Delta t}\right)\omega^8\Delta t^8\right]}{256\omega^2\Delta t\left(1+e^{-\gamma\Delta t}\right)\left(1-e^{-\gamma\Delta t}\right)\left(1-\frac{\omega^2\Delta t^2}{4}\right)\left[\left(1-\frac{\omega^2\Delta t^2}{4}\right)^2+1\right]}$ |
| PV-middle | $\dfrac{\left[256\left(1-e^{-\gamma\Delta t}\right)^2-128\left(e^{-2\gamma\Delta t}-6e^{-\gamma\Delta t}-3\right)\omega^2\Delta t^2+16\left(3e^{-2\gamma\Delta t}-22e^{-\gamma\Delta t}-13\right)\omega^4\Delta t^4-8\left(e^{-2\gamma\Delta t}-6e^{-\gamma\Delta t}-5\right)\omega^6\Delta t^6-\left(3-e^{-\gamma\Delta t}\right)\left(1+e^{-\gamma\Delta t}\right)\omega^8\Delta t^8\right]}{256\omega^2\Delta t\left(1+e^{-\gamma\Delta t}\right)\left(1-e^{-\gamma\Delta t}\right)\left(1-\frac{\omega^2\Delta t^2}{4}\right)\left[\left(1-\frac{\omega^2\Delta t^2}{4}\right)^2+1\right]}$ |
| End(vir)<br>Beginning(vir) | $\dfrac{\left[64\left(1+e^{-\gamma\Delta t}\right)^2+128\left(1-e^{-\gamma\Delta t}\right)\omega^2\Delta t^2-16\left(5-2e^{-\gamma\Delta t}+e^{-2\gamma\Delta t}\right)\omega^4\Delta t^4+4\left(4+e^{-2\gamma\Delta t}\right)\omega^6\Delta t^6-\omega^8\Delta t^8\right]}{64\omega^2\Delta t\left(1+e^{-\gamma\Delta t}\right)\left(1-e^{-\gamma\Delta t}\right)\left(1-\frac{\omega^2\Delta t^2}{4}\right)\left[\left(1-\frac{\omega^2\Delta t^2}{4}\right)^2+1\right]}$ |
| End<br>Beginning | $$\frac{\left[64\left(1-e^{-\gamma\Delta t}\right)^2+128\left(1+e^{-\gamma\Delta t}\right)\omega^2\Delta t^2-16\left(5+2e^{-\gamma\Delta t}+e^{-2\gamma\Delta t}\right)\omega^4\Delta t^4+4\left(4+e^{-2\gamma\Delta t}\right)\omega^6\Delta t^6-\omega^8\Delta t^8\right]}{64\omega^2\Delta t\left(1+e^{-\gamma\Delta t}\right)\left(1-e^{-\gamma\Delta t}\right)\left(1-\frac{\omega^2\Delta t^2}{4}\right)\left[\left(1-\frac{\omega^2\Delta t^2}{4}\right)^2+1\right]}$$ |
| PV-end(vir)<br>PV-<br>beginning(vir) | $$\frac{\left[16\left(1+e^{-\gamma\Delta t}\right)^2+\left(28-40e^{-\gamma\Delta t}-4e^{-2\gamma\Delta t}\right)\omega^2\Delta t^2-\left(4-16e^{-\gamma\Delta t}-4e^{-2\gamma\Delta t}\right)\omega^4\Delta t^4-e^{-\gamma\Delta t}\left(2+e^{-\gamma\Delta t}\right)\omega^6\Delta t^6\right]}{16\omega^2\Delta t\left(1+e^{-\gamma\Delta t}\right)\left(1-e^{-\gamma\Delta t}\right)\left[\left(1-\frac{\omega^2\Delta t^2}{4}\right)^2+1\right]}$$ |
| PV-end<br>PV-beginning | $$\frac{\left[16\left(1-e^{-\gamma\Delta t}\right)^2+\left(28+40e^{-\gamma\Delta t}-4e^{-2\gamma\Delta t}\right)\omega^2\Delta t^2-\left(4+16e^{-\gamma\Delta t}-4e^{-2\gamma\Delta t}\right)\omega^4\Delta t^4+e^{-\gamma\Delta t}\left(2-e^{-\gamma\Delta t}\right)\omega^6\Delta t^6\right]}{16\omega^2\Delta t\left(1+e^{-\gamma\Delta t}\right)\left(1-e^{-\gamma\Delta t}\right)\left[\left(1-\frac{\omega^2\Delta t^2}{4}\right)^2+1\right]}$$ |
| Side(vir)<br>Side | $$\frac{\left[\left(1-e^{-\gamma\Delta t}\right)^2+e^{-\gamma\Delta t}\left(2-e^{-\gamma\Delta t}\right)\omega^2\Delta t^2\left(1-\frac{\omega^2\Delta t^2}{4}\right)+2e^{-\gamma\Delta t}\omega^2\Delta t^2\left(1-\frac{\omega^2\Delta t^2}{4}\right)^2+\omega^2\Delta t^2\left(1-\frac{\omega^2\Delta t^2}{4}\right)^3\right]}{\omega^2\Delta t\left(1+e^{-\gamma\Delta t}\right)\left(1-e^{-\gamma\Delta t}\right)\left(1-\frac{\omega^2\Delta t^2}{4}\right)\left[\left(1-\frac{\omega^2\Delta t^2}{4}\right)^2+1\right]}$$ |
| PV-side(vir)<br>PV-side | $$\frac{\left[16\left(1-e^{-\gamma\Delta t}\right)^2+\left(-20e^{-2\gamma\Delta t}+72e^{-\gamma\Delta t}+12\right)\omega^2\Delta t^2+8e^{-\gamma\Delta t}\left(e^{-\gamma\Delta t}-3\right)\omega^4\Delta t^4+e^{-\gamma\Delta t}\left(2-e^{-\gamma\Delta t}\right)\omega^6\Delta t^6\right]}{16\omega^2\Delta t\left(1+e^{-\gamma\Delta t}\right)\left(1-e^{-\gamma\Delta t}\right)\left[\left(1-\frac{\omega^2\Delta t^2}{4}\right)^2+1\right]}$$ |
| Middle-pT<br>Side-pT<br>Middle-xT<br>Side-xT | See supplementary material |

**Table 4.** Characteristic correlation time of the potential energy in the limit $\gamma \to \infty$

| Scheme | $\tau_{pot}^{\gamma\to\infty}$ |
|---|---|
| Middle<br>PV-middle<br>Middle(vir)<br>PV-middle(vir) | $\dfrac{1+\frac{3}{4}\omega^2\Delta t^2}{\omega^2\Delta t}$ |
| End<br>Beginning<br>Side<br>PV-end PV-beginning<br>PV-side<br>End(vir)<br>Beginning(vir)<br>Side(vir)<br>PV-end(vir)<br>PV-beginning(vir)<br>PV-side(vir) | $\dfrac{1}{\omega^2\Delta t\left(1-\frac{\omega^2\Delta t^2}{4}\right)}$ |

**Table 5.** Characteristic correlation time of the Hamiltonian in the limit $\gamma \to \infty$

| Scheme | $\tau_{Ham}^{\gamma\to\infty}$ |
|---|---|
| Middle<br>Middle(vir) | $\dfrac{1+\frac{9}{4}\omega^2\Delta t^2-\frac{9}{16}\omega^4\Delta t^4+\frac{3}{64}\omega^6\Delta t^6}{\omega^2\Delta t\left[\left(1-\frac{\omega^2\Delta t^2}{4}\right)^2+1\right]}$ |
| PV-middle<br>PV-middle(vir) | $\dfrac{256+384\omega^2\Delta t^2-208\omega^4\Delta t^4+40\omega^6\Delta t^6-3\omega^8\Delta t^8}{256\omega^2\Delta t\left(1-\frac{\omega^2\Delta t^2}{4}\right)\left[\left(1-\frac{\omega^2\Delta t^2}{4}\right)^2+1\right]}$ |
| End<br>Beginning<br>End(vir)<br>Beginning(vir) | $\dfrac{64+128\omega^2\Delta t^2-80\omega^4\Delta t^4+16\omega^6\Delta t^6-\omega^8\Delta t^8}{64\omega^2\Delta t\left(1-\frac{\omega^2\Delta t^2}{4}\right)\left[\left(1-\frac{\omega^2\Delta t^2}{4}\right)^2+1\right]}$ |
| PV-end<br>PV-beginning<br>PV-end(vir)<br>PV-beginning(vir) | $\dfrac{4+7\omega^2\Delta t^2-\omega^4\Delta t^4}{4\omega^2\Delta t\left[\left(1-\frac{\omega^2\Delta t^2}{4}\right)^2+1\right]}$ |
| Side<br>Side(vir) | $\dfrac{1+\omega^2\Delta t^2\left(1-\frac{\omega^2\Delta t^2}{4}\right)^3}{\omega^2\Delta t\left(1-\frac{\omega^2\Delta t^2}{4}\right)\left[\left(1-\frac{\omega^2\Delta t^2}{4}\right)^2+1\right]}$ |
| PV-side<br>PV-side(vir) | $\dfrac{4+3\omega^2\Delta t^2}{4\omega^2\Delta t\left[\left(1-\frac{\omega^2\Delta t^2}{4}\right)^2+1\right]}$ |

**Table 6.** The range of $\omega\Delta t$ where the optimal friction coefficient is finite for the harmonic system $U(x)=\frac{1}{2}M\omega^2x^2$

| Scheme | | Range |
|---|---|---|
| Middle<br>PV-middle | $\tau_{pot}$ | (0,2) |
| End<br>Beginning<br>Side<br>Side(vir)<br>PV-end<br>PV-beginning<br>PV-side<br>PV-side(vir) | $\tau_{pot}$ | $(0,\sqrt{2})$ and $(\sqrt{2}, 2)$ |
| Middle | $\tau_{Ham}$ | (0,2) |
| End<br>Beginning | $\tau_{Ham}$ | $(0,\sqrt{2})$ and $(\sqrt{2}, 2)$ |
| Side<br>Side(vir) | $\tau_{Ham}$ | (0, 0.806064) and (1.709276, 2) |
| PV-middle | $\tau_{Ham}$ | $(0,\sqrt{5}-1)$ and $(\sqrt{5}-1, 2)$ |
| PV-end<br>PV-beginning | $\tau_{Ham}$ | $(0,\sqrt{2})$ and $(\sqrt{2}, 2)$ |
| PV-side<br>PV-side(vir) | $\tau_{Ham}$ | (0, 0.73205) |

**Fig. 1**. Results for the fluctuation of potential energy, kinetic energy and Hamiltonian using different time intervals for the harmonic system $U(x)=\frac{1}{2}x^2$. The friction coefficient $\gamma=1$, which is the optimal $\gamma$ for the characteristic correlation time of the potential energy for infinitesimal time interval. (a) The fluctuation of potential energy for all the schemes in the real dynamics case. (b) Same as Panel (a), but for the virtual dynamics case. (c) Same as Panel (a), but for the fluctuation of kinetic energy. (d) Same as Panel (c), but for the virtual dynamics case. (e) Same as Panel (a), but for the fluctuation of Hamiltonian. (f) Same as Panel (e), but for the virtual dynamics case.

**Fig. 2**. Same as **Fig. 1**, but for the quartic system $U(x)=\frac{1}{4}x^4$ and the friction coefficient $\gamma=1.2$, which is nearly the optimal $\gamma$ for the characteristic correlation time of the potential for infinitesimal time interval.

**Fig. 3**. Analytic results for the characteristic correlation time of the potential energy for the harmonic system $U(x)=\frac{1}{2}M\omega^2x^2$. The curves depict the equations that $\tau_{pot}\omega$ and $\gamma/\omega$ satisfy for the parameter $\omega\Delta t\to 0^+$, $\omega\Delta t=0.6, \sqrt{2}, \text{ and } 1.9$, respectively. (a) For the "(PV-)middle" schemes. (b) For the "(PV-)side/side(vir)/end/beginning" schemes. (c) For the "middle-xT" scheme.

**Fig. 4**. Analytic results of the characteristic correlation time of the Hamiltonian for the harmonic system $U(x)=\frac{1}{2}M\omega^2x^2$. The curves depict the equations that $\tau_{Ham}\omega$ and $\gamma/\omega$ satisfy for the parameter $\omega\Delta t\to 0^+$, $\omega\Delta t=0.6, \sqrt{2}, \text{ and } 1.9$, respectively. (a) For the "middle" scheme. (b) For the "side"/"side(vir)" schemes. (c) For the "middle-xT" scheme.

**Fig. 5**. Characteristic correlation time of the potential for the quartic system $U(x)=\frac{1}{4}x^4$. Three time intervals $\Delta t=0.1$, 0.3, 0.4 are used. The unit of all the parameters is atomic unit

(a.u.). Statistical error bars are included. (a) For the "middle" scheme. Hollow symbols with dotted line: numerical results for the virtual dynamics case. Solid symbols: numerical results for the real dynamics case. (b) Same as Panel (a), but for the "side" scheme. (c) For the "middle-xT" scheme. Solid symbols with dotted line: numerical results.

**Fig. 6**. Same as **Fig. 5**, but for the characteristic correlation time of the Hamiltonian.

**Fig. 7**. The characteristic correlation time of the potential and that of the Hamiltonian in the limit $\gamma \to \infty$ for the harmonic system $U(x) = \frac{1}{2} M\omega^2 x^2$. The eight schemes that employ the first type of repartition are considered. (a) The characteristic correlation time of the potential energy in the limit $\gamma \to \infty$ for all the eight schemes. (b) The characteristic correlation time of the Hamiltonian in the limit $\gamma \to \infty$ for the "middle", "side" and "end"/"begin" schemes. (c) Same as Panel (b), but for the "PV-middle", "PV-side" and "PV-end"/"PV-begin" schemes.

**Fig. 8**. The "middle", "side"/"side (vir)" and "middle-xT" schemes are considered as examples for the harmonic system $U(x) = \frac{1}{2} M\omega^2 x^2$. (a) $\tau_{\text{pot}}^{\text{min}}\omega$ as a function of $\omega\Delta t$, where $\tau_{\text{pot}}^{\text{min}}$ is the minimum value of the characteristic correlation time of the potential. (b) $\gamma_{\text{pot}}^{\text{opt}}/\omega$ as a function of $\omega\Delta t$, where $\gamma_{\text{pot}}^{\text{opt}}$ is the optimal friction coefficient for the characteristic correlation time of the potential. Panels (c) and (d) are the same as Panels (a) and (b), respectively, but for the characteristic correlation time of the Hamiltonian.

**Fig. 9**. Results for the averaged potential energy and the thermal fluctuation of the potential using different friction coefficients $\gamma$. Panels (a)-(b) for 1-dim harmonic model. Two time intervals $\Delta t = 1.0$, 1.8 are used. Panels (c)-(d) for 1-dim quartic model. Three time intervals $\Delta t = 0.3$, 0.4, 0.45 are used. The unit of all the parameters is atomic unit (au) in the figure. "real-1.0" represents the numerical results obtained by real dynamics for $\Delta t = 1.0$ au; "vir-1.0"

stands for those produced by virtual dynamics for $\Delta t = 1.0$ au; *etc*. Statistical error bars are included.

**Fig. 10**. Results for the averaged potential energy and the thermal fluctuation of the potential using different friction coefficients $\gamma$ (unit: $fs^{-1}$). Panels (a)-(b) for the $H_2O$ molecule. Three time intervals $\Delta t = 1.9, 2.2, 2.4$ (unit: fs) are used. Panels (c)-(d) for the $(Ne)_{13}$ cluster. Two time intervals $\Delta t = 20,\ 50$ (unit: fs) are used. [Panels (a) and (c) display averaged potential energy per atom $\langle U(\mathbf{x})\rangle / (N_{\mathrm{atom}} k_{\mathrm{B}})$ (unit: Kelvin). Panels (b) and (d) display the thermal fluctuation of the potential per atom $\sqrt{\langle U^2\rangle - \langle U\rangle^2} / (N_{\mathrm{atom}} k_{\mathrm{B}} T)$ .] "real-1.9" represents the numerical results obtained by real dynamics for $\Delta t = 1.9$ fs; "vir-1.9" stands for those produced by virtual dynamics for $\Delta t = 1.9$ fs; *etc*. Statistical error bars are included. [For comparison the converged results are obtained with the parameters: (a)-(b). $\gamma = 0.68\ fs^{-1}$, $\Delta t = 0.24$ fs; (c)-(d). $\gamma = 0.001\ fs^{-1}$, $\Delta t = 10$ fs.]

**Fig. 11**. Characteristic correlation time of the potential (Panel (a)) and that of the Hamiltonian (Panel (b)) for the "middle" scheme for the $H_2O$ molecule at 100 K. Panels (c) and (d) are the same as Panels (a) and (b), but for the $(Ne)_{13}$ cluster at 14 K. While two time intervals $\Delta t = 0.24$, 1.2 (unit: fs) are used for the $H_2O$ molecule in Panels (a)-(b), those used for the $(Ne)_{13}$ cluster at 14 K are $\Delta t = 20,\ 50$ (unit: fs). The unit of all the parameters is per femtosecond ($fs^{-1}$). Statistical error bars are included. Hollow squares/circles with dotted lines: numerical results for the virtual dynamics case. Solid squares/circles: numerical results for the real dynamics case. "vir-num-1.2" represents the numerical results from the virtual dynamics case for $\Delta t = 1.2$ fs; "real-num-1.2" stands for the numerical results from the real dynamics case for $\Delta t = 1.2$ fs; *etc*.

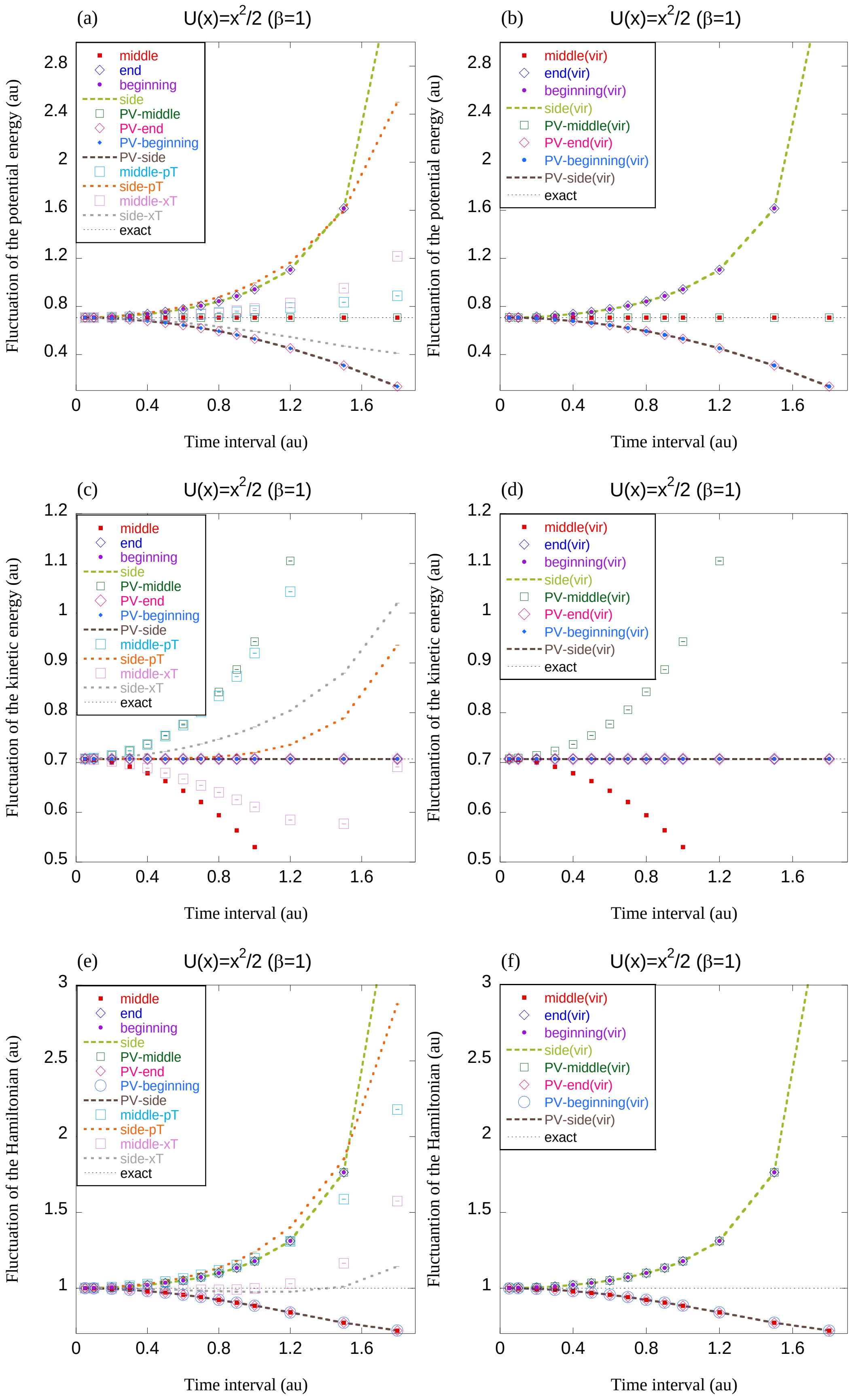

**Fig. 1**

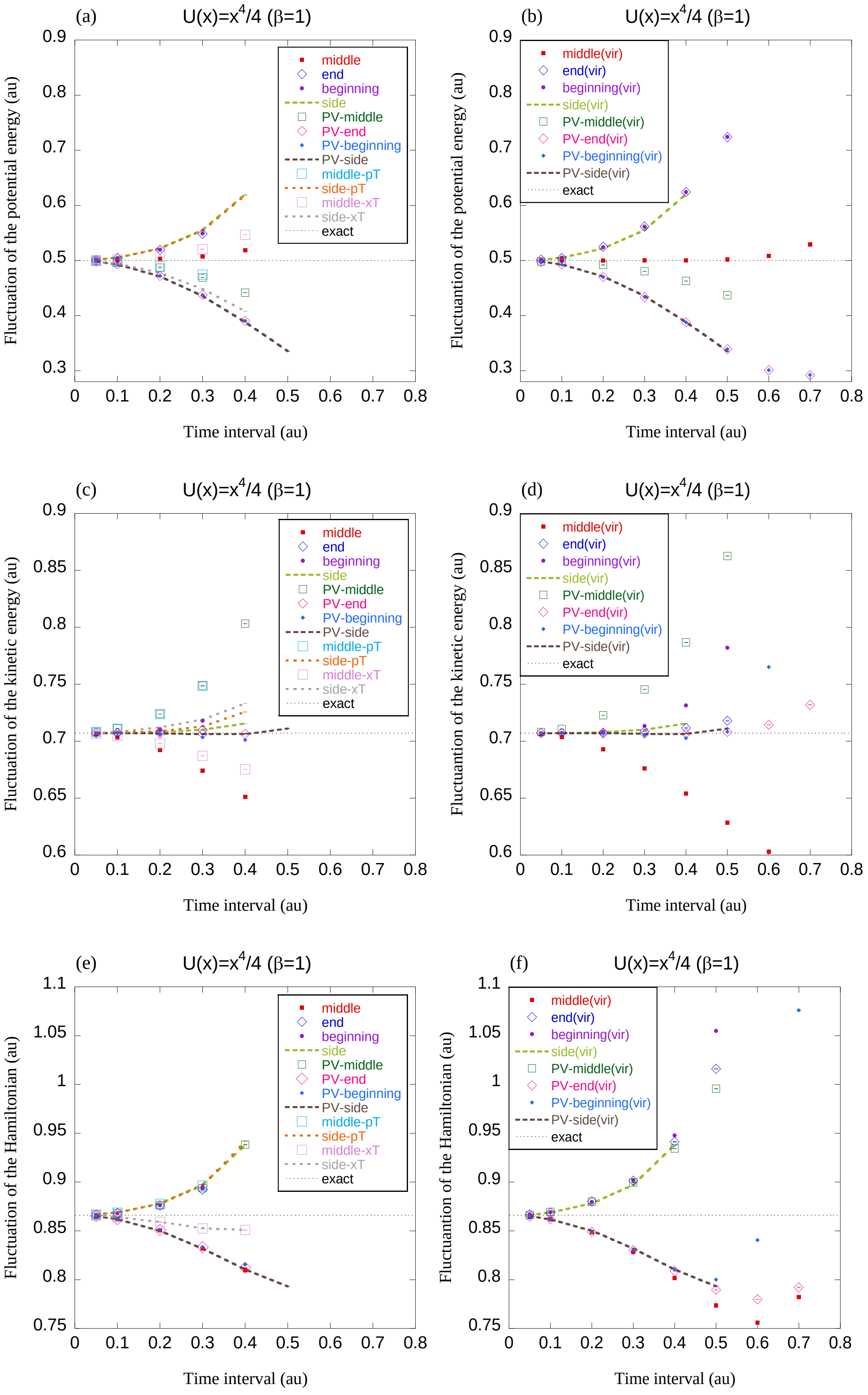

**Fig. 2**

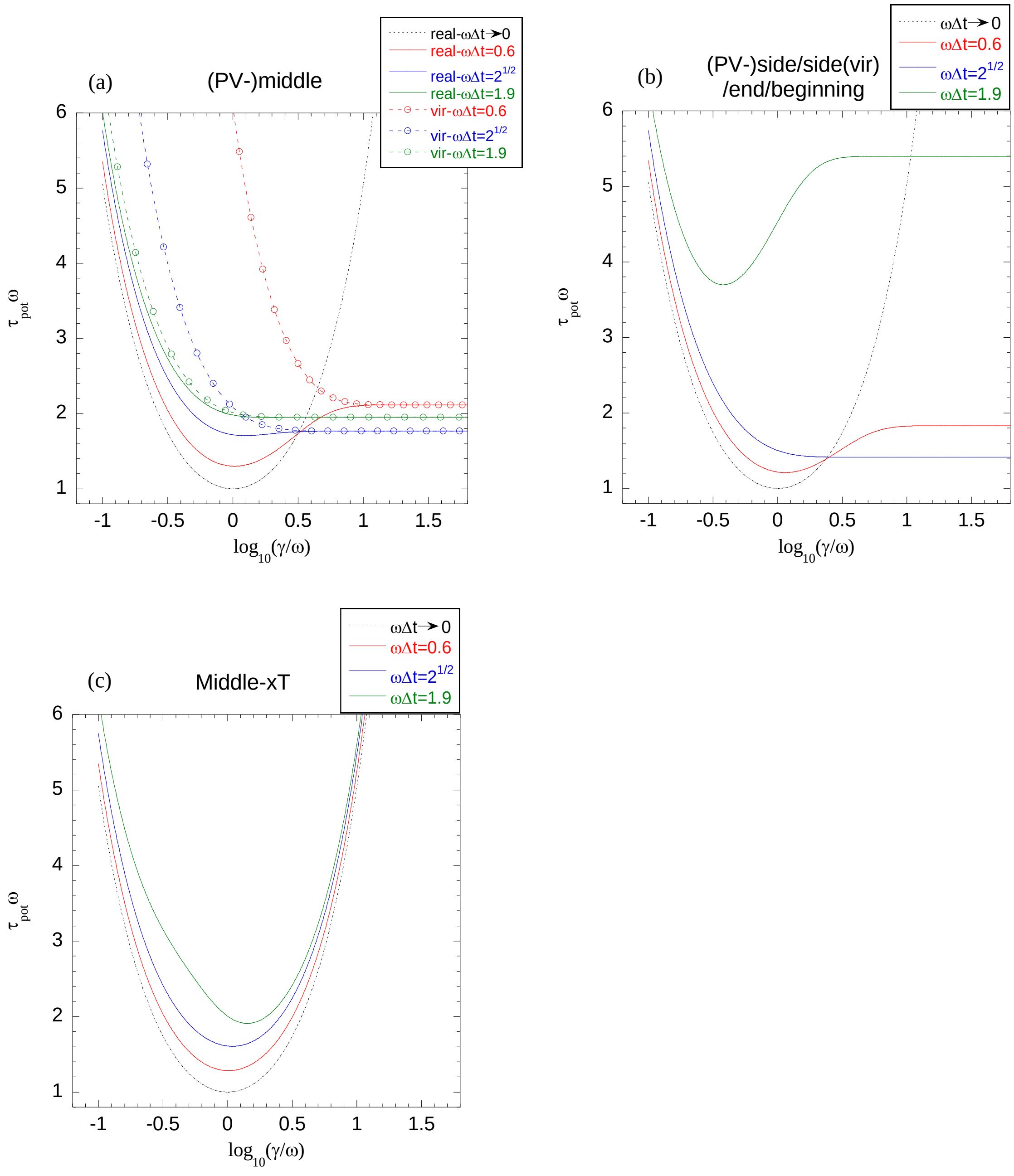

**Fig. 3**

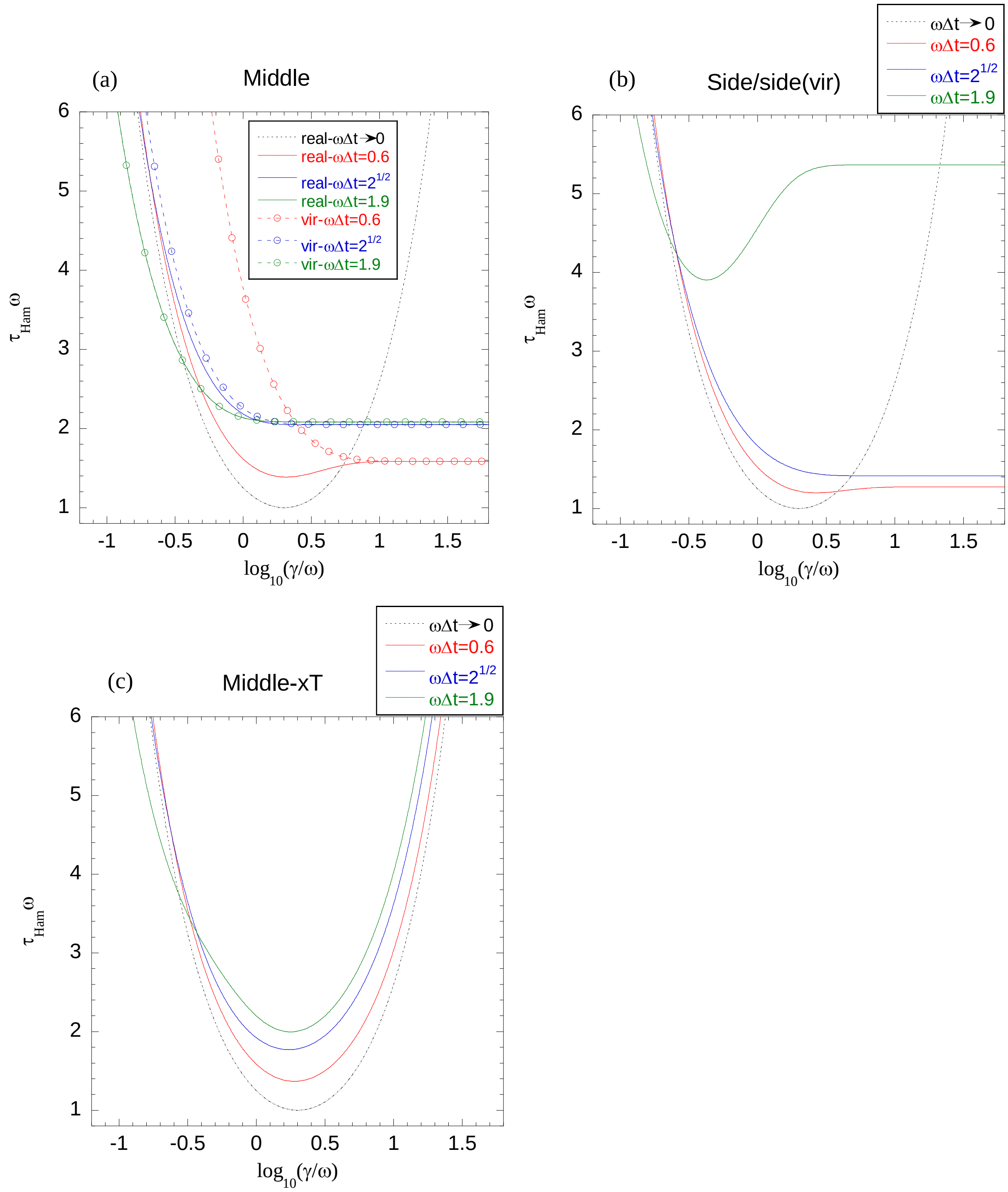

**Fig. 4**

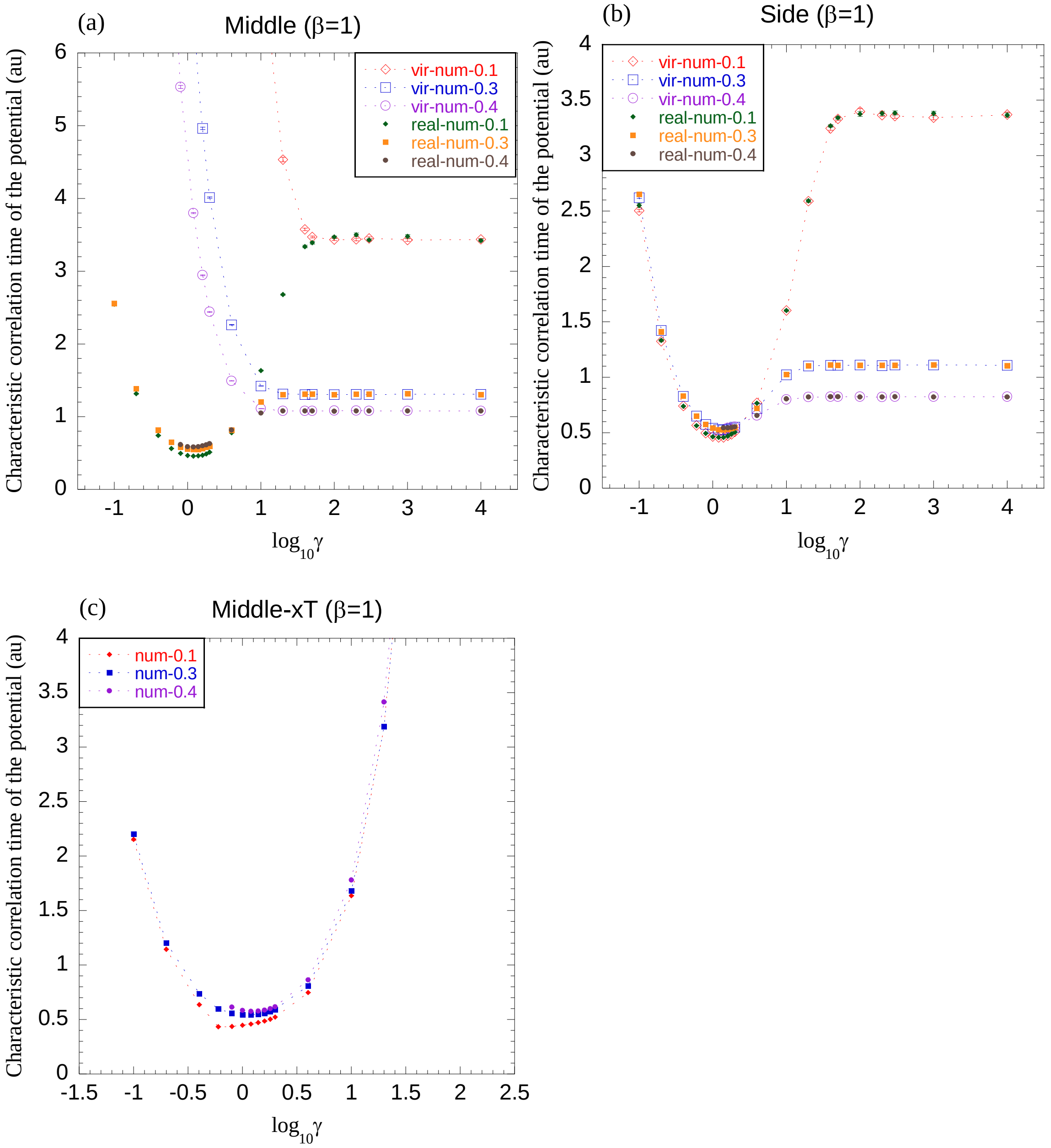

**Fig. 5**

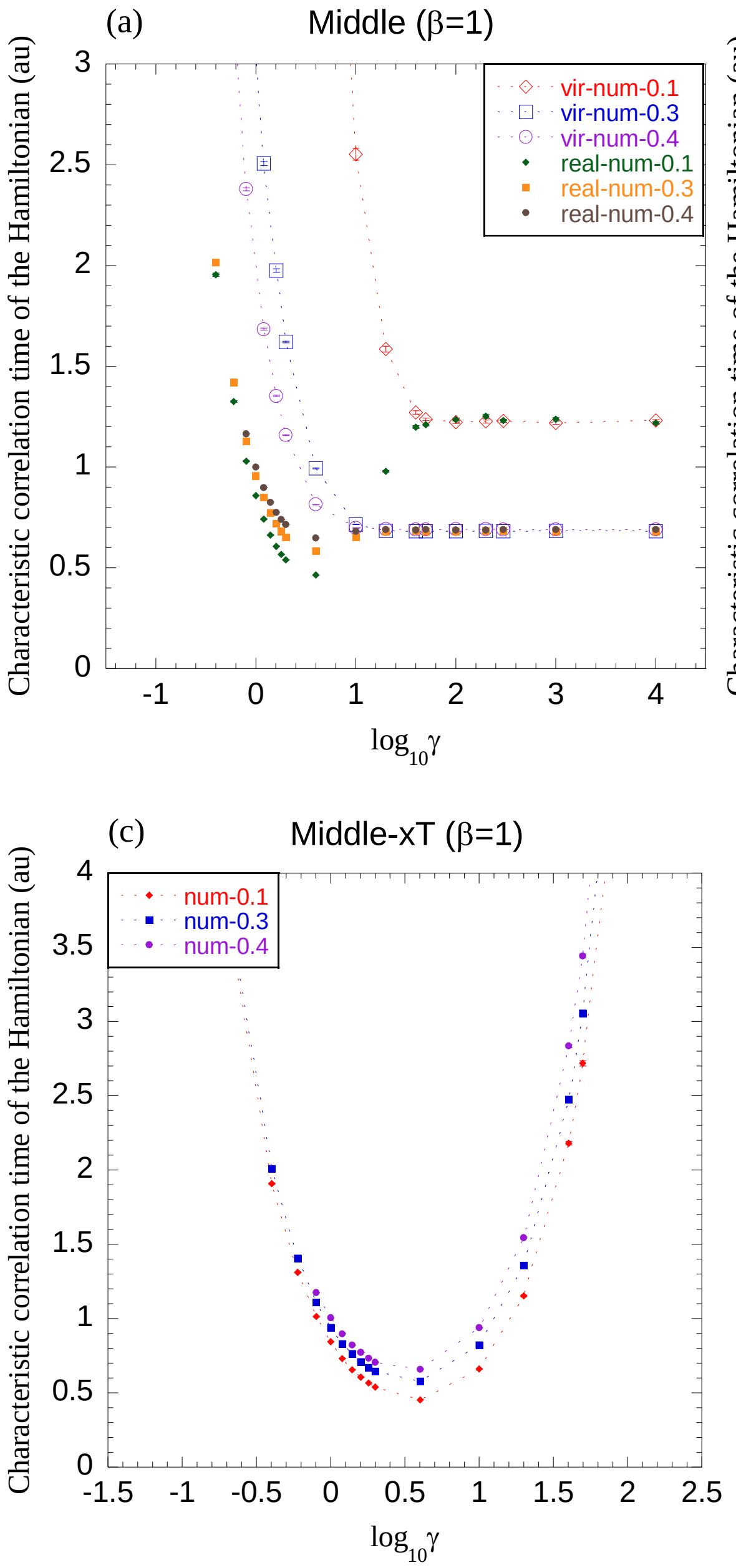

**Fig. 6**

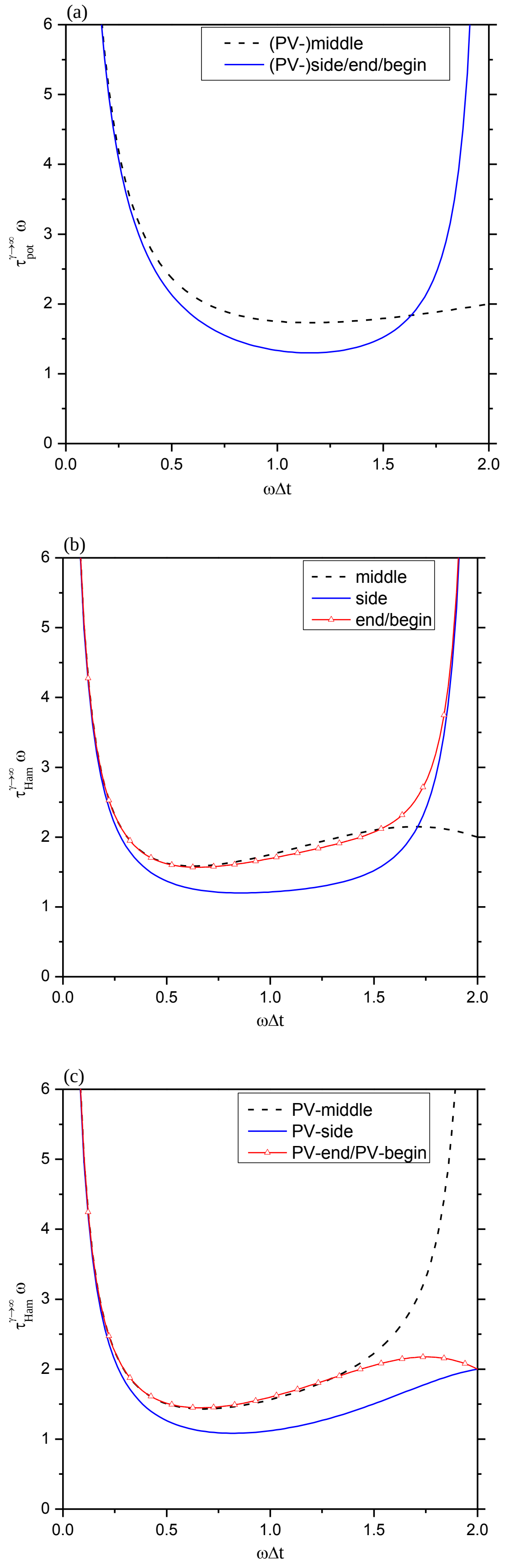

**Fig. 7**

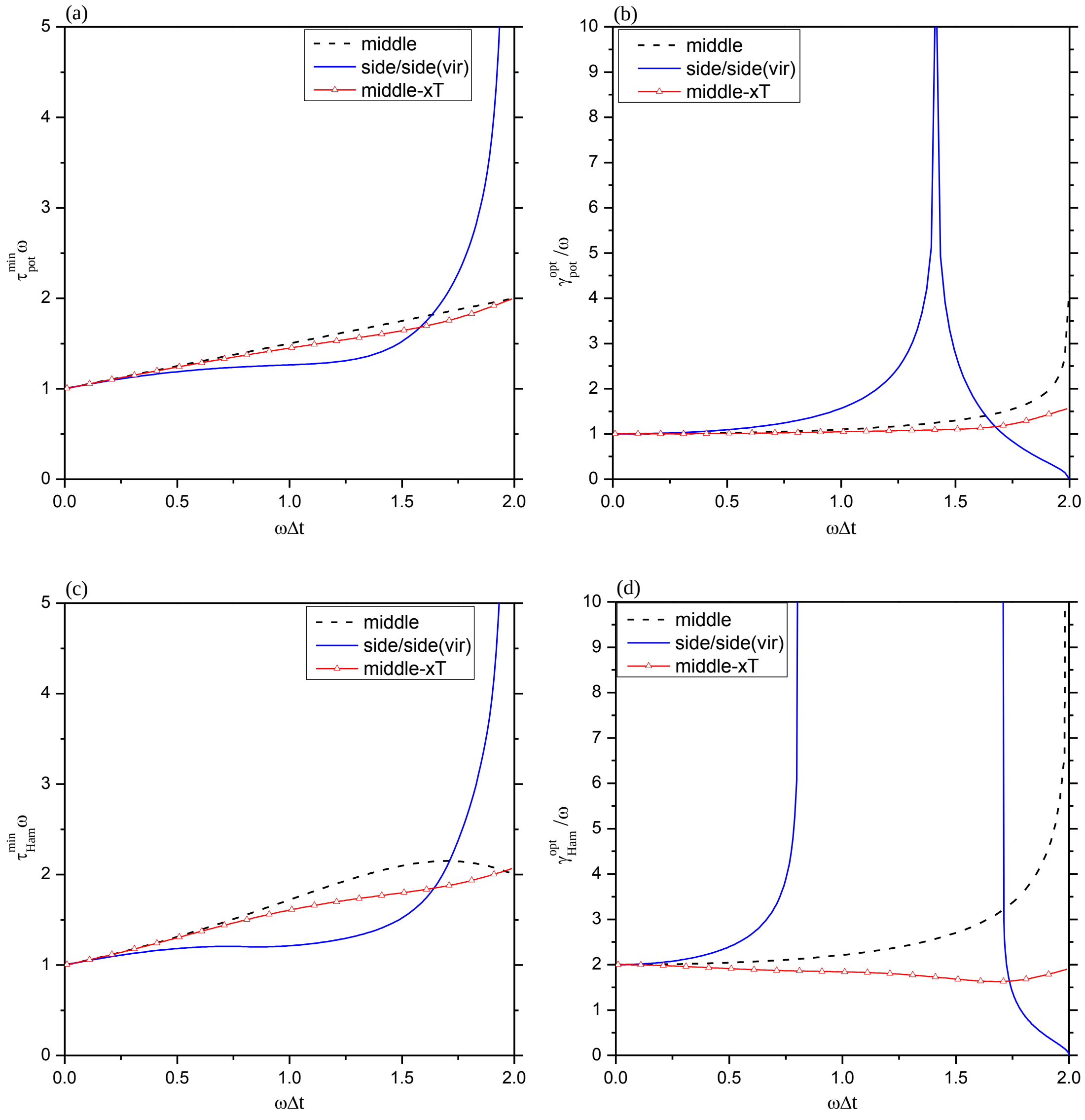

**Fig. 8**

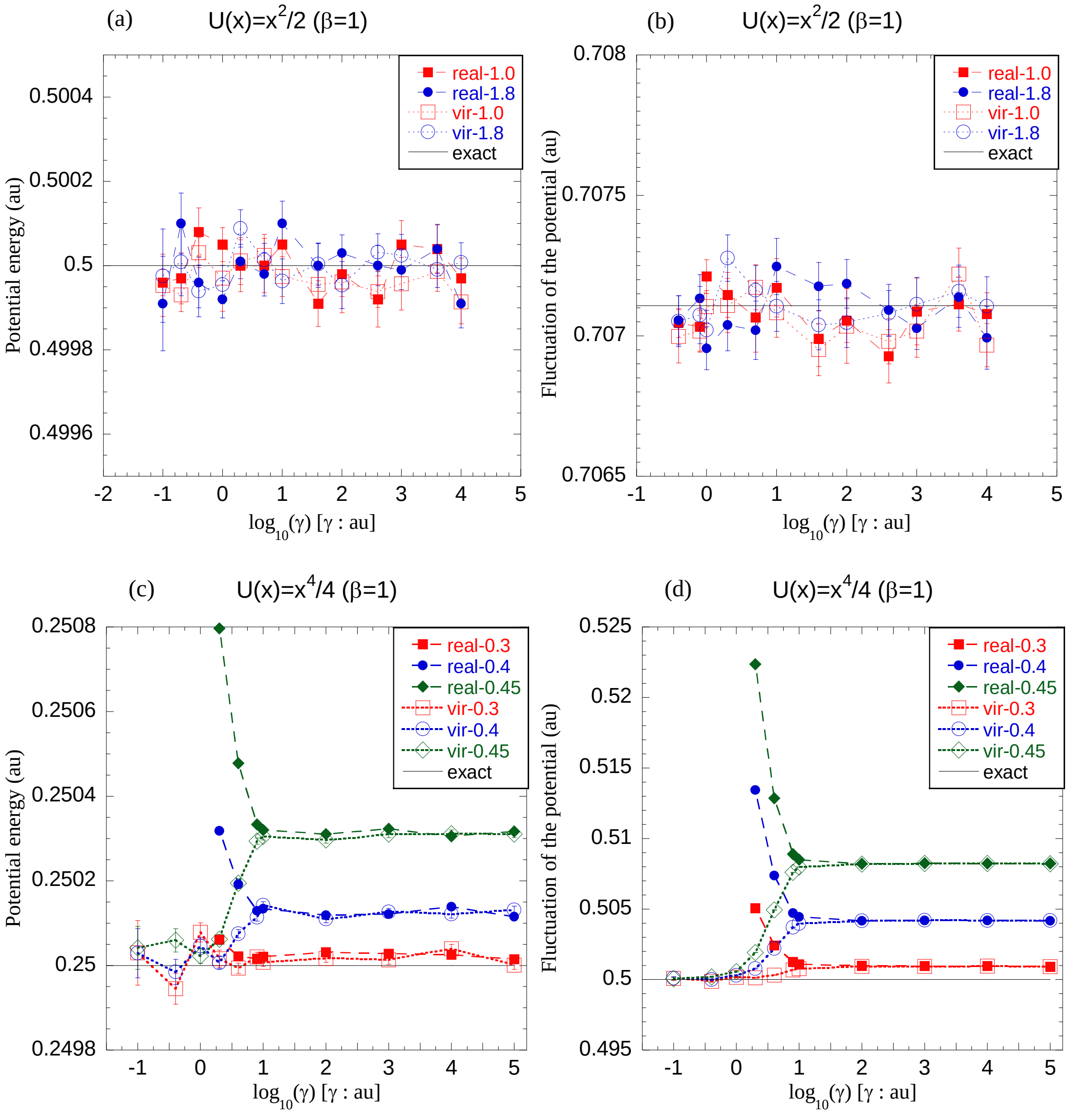

**Fig. 9**

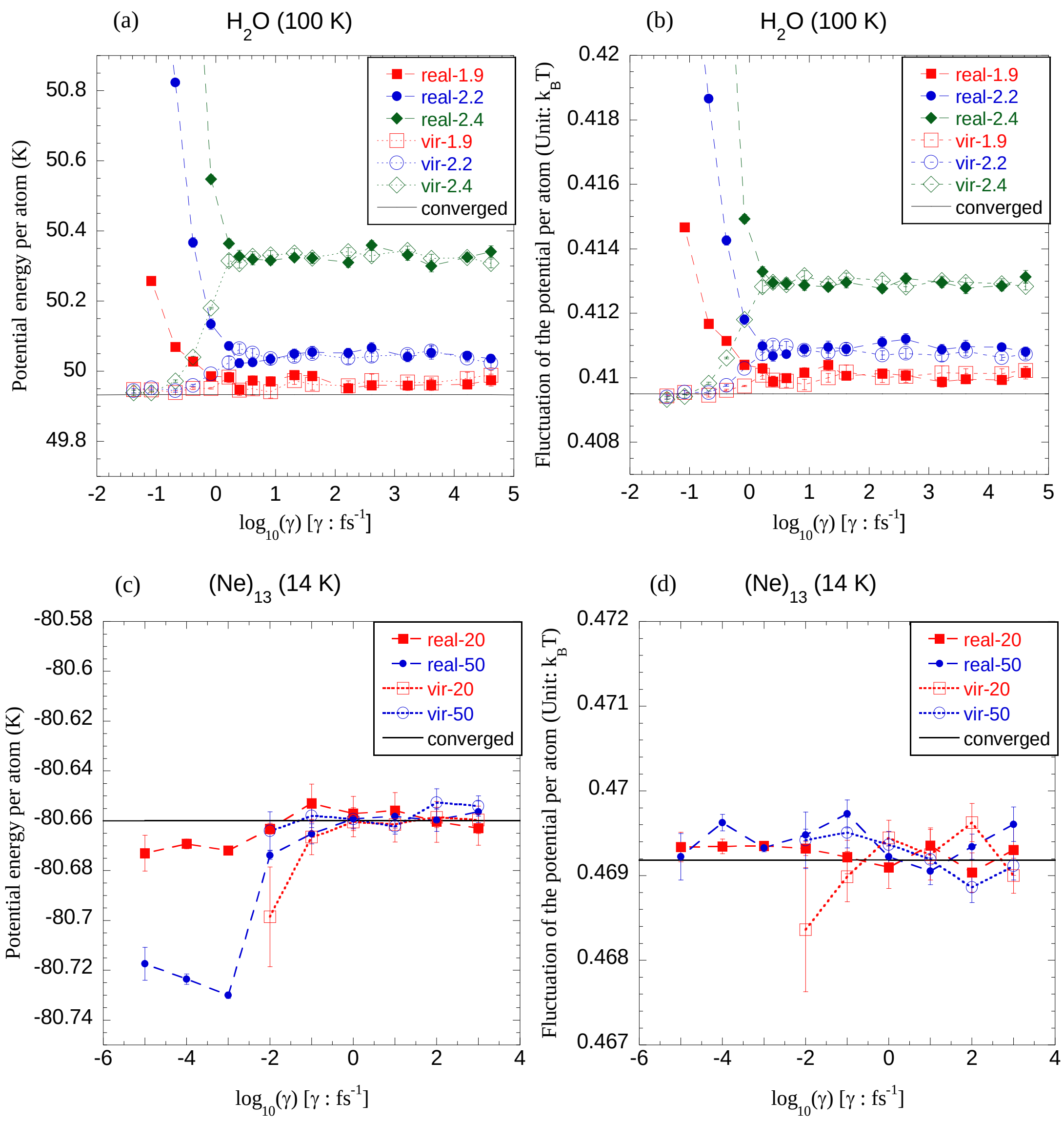

(a)
H2O (100 K)
Potential energy per atom (K)
log10(γ) [γ : fs-1]
real-1.9
real-2.2
real-2.4
vir-1.9
vir-2.2
vir-2.4
converged
(b)
H2O (100 K)
Fluctuation of the potential per atom (Unit: kBT)
log10(γ) [γ : fs-1]
(c)
(Ne)13 (14 K)
Potential energy per atom (K)
real-20
real-50
vir-20
vir-50
converged
(d)
(Ne)13 (14 K)
Fluctuation of the potential per atom (Unit: kBT)

**Fig. 10**

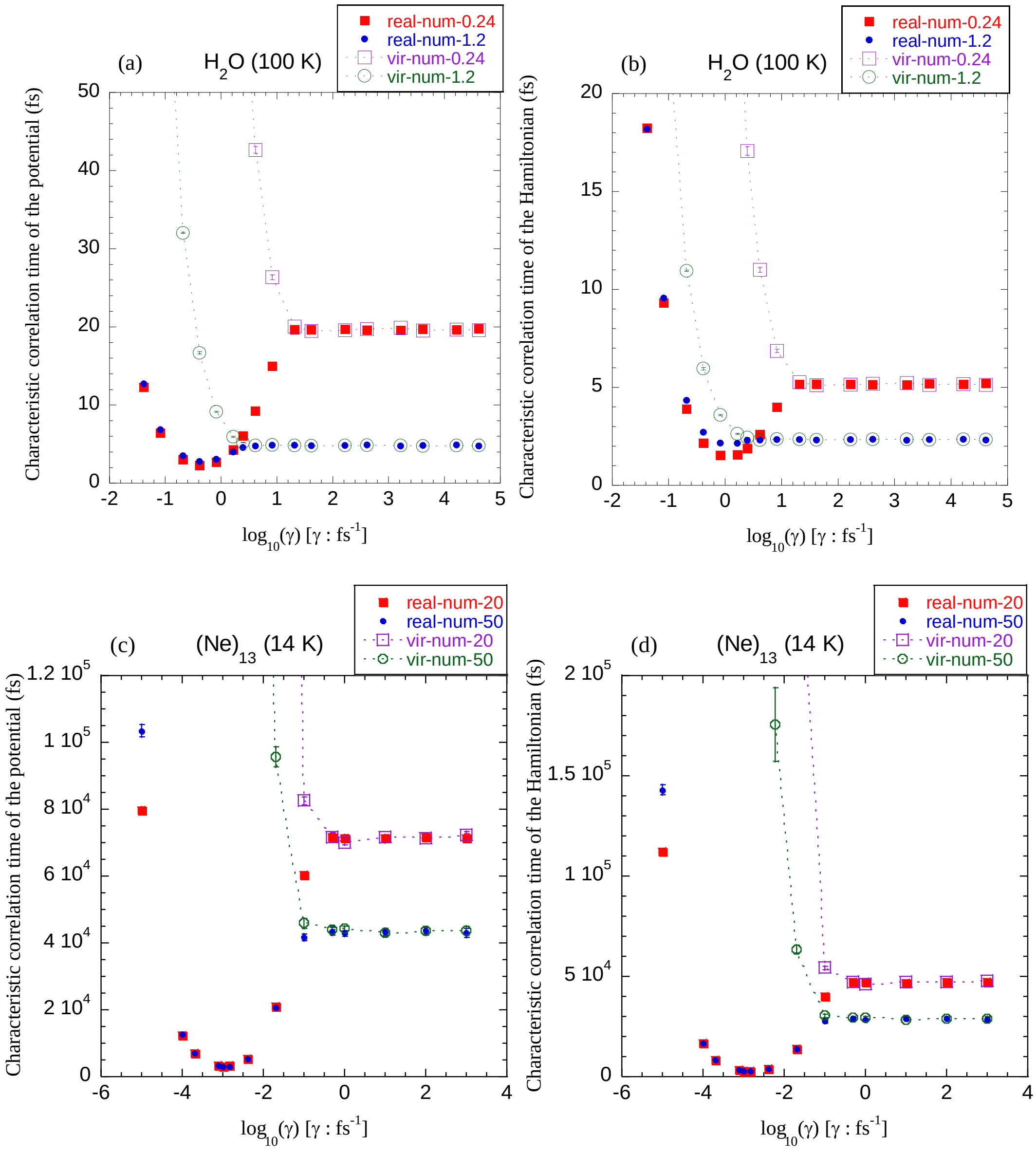
(a)
$H_2O$ (100 K)
real-num-0.24
real-num-1.2
vir-num-0.24
vir-num-1.2
Characteristic correlation time of the potential (fs)
$\log_{10}(\gamma)$ [$\gamma$ : fs$^{-1}$]
(b)
$H_2O$ (100 K)
real-num-0.24
real-num-1.2
vir-num-0.24
vir-num-1.2
Characteristic correlation time of the Hamiltonian (fs)
$\log_{10}(\gamma)$ [$\gamma$ : fs$^{-1}$]
(c)
$(Ne)_{13}$ (14 K)
real-num-20
real-num-50
vir-num-20
vir-num-50
Characteristic correlation time of the potential (fs)
$\log_{10}(\gamma)$ [$\gamma$ : fs$^{-1}$]
(d)
$(Ne)_{13}$ (14 K)
real-num-20
real-num-50
vir-num-20
vir-num-50
Characteristic correlation time of the Hamiltonian (fs)
$\log_{10}(\gamma)$ [$\gamma$ : fs$^{-1}$]

**Fig. 11**

# Supplementary material for
# "Stationary state distribution and efficiency analysis of the Langevin equation *via* real or virtual dynamics"

Dezhang Li[1], Xu Han[1], Yichen Chai[1], Cong Wang[2], Zhijun Zhang[1], Zifei Chen[1],

Jian Liu[1, a)], Jiushu Shao[2, b)]

1. Beijing National Laboratory for Molecular Sciences, Institute of Theoretical and Computational Chemistry, College of Chemistry and Molecular Engineering, Peking University, Beijing 100871, China
2. College of Chemistry and Center for Advanced Quantum Studies, Key Laboratory of Theoretical and Computational Photochemistry, Ministry of Education, Beijing Normal University, Beijing 100875, China

Corresponding Authors:

a) Electronic mail: jianliupku@pku.edu.cn

b) Electronic mail: jiushu@bnu.edu.cn

## S1. More results of the characteristic correlation time

### 1A. Characteristic correlation time of the potential energy

#### 1) "Middle-xT" scheme

$$\tau_{pot}^{\text{Middle-xT}} = \frac{\varphi_1 + \varphi_2}{\varphi_3} \tag{S1}$$

with

$$\begin{aligned}
\varphi_1 &= 4\left(-1+e^{2\gamma\Delta t}\right)^3 \gamma^6 + 8\Delta t\left(-1+2e^{\gamma\Delta t}\right)\left(-1+e^{2\gamma\Delta t}\right)^2 \gamma^5\omega^2 \\
&\quad + 4\Delta t^2\left(-1+6e^{\gamma\Delta t} - 8e^{2\gamma\Delta t} - 10e^{3\gamma\Delta t} + 9e^{4\gamma\Delta t} + 4e^{5\gamma\Delta t}\right)\gamma^4\omega^4 \\
&\quad - \Delta t^2\left(-1+e^{\gamma\Delta t}\right)^3\left(1+e^{\gamma\Delta t}\right)^2\left(-1+3e^{\gamma\Delta t}\right)\omega^8 \\
&\quad + 4\Delta t^2\left(-1+e^{\gamma\Delta t}\right)\gamma^2\omega^6\left[-1+3e^{\gamma\Delta t} - 7e^{4\gamma\Delta t} + e^{5\gamma\Delta t} + \Delta t^2 e^{2\gamma\Delta t}\omega^2 - 3e^{3\gamma\Delta t}\left(4+\Delta t^2\omega^2\right)\right] \\
\varphi_2 &= 4\Delta t e^{\gamma\Delta t}\left(-1+e^{\gamma\Delta t}\right)^2\left(1+e^{\gamma\Delta t}\right)\gamma\omega^6\left[2e^{2\gamma\Delta t} - \Delta t^2\omega^2 + e^{\gamma\Delta t}\left(2+3\Delta t^2\omega^2\right)\right] \\
&\quad - 4\Delta t\gamma^3\omega^4\left[-1+3e^{6\gamma\Delta t} + e^{\gamma\Delta t}\left(2-2\Delta t^2\omega^2\right) + 2e^{5\gamma\Delta t}\left(1+\Delta t^2\omega^2\right)\right. \\
&\qquad \left. - 4e^{3\gamma\Delta t}\left(1+2\Delta t^2\omega^2\right) + e^{2\gamma\Delta t}\left(5+8\Delta t^2\omega^2\right) - e^{4\gamma\Delta t}\left(7+8\Delta t^2\omega^2\right)\right] \\
\varphi_3 &= 4\left(-1+e^{\gamma\Delta t}\right)\omega^2\left[2\left(1+e^{\gamma\Delta t}\right)\gamma - \Delta t\left(-1+e^{\gamma\Delta t}\right)\omega^2\right] \\
&\quad \times\left[\left(-1+e^{2\gamma\Delta t}\right)\gamma^2 - \left(-1+e^{2\gamma\Delta t}\right)\omega^2 + 2\Delta t e^{\gamma\Delta t}\gamma\omega^2\right]^2
\end{aligned} \tag{S2}$$

#### 2) "Side-xT" scheme

$$\tau_{pot}^{\text{Side-xT}} = \frac{\left[-2\left(1+e^{\gamma\Delta t}\right)\gamma + \Delta t\left(-1+e^{\gamma\Delta t}\right)\omega^2\right]^2\left(\tilde{\varphi}_1+\tilde{\varphi}_2+\tilde{\varphi}_3+\tilde{\varphi}_4+\tilde{\varphi}_5+\tilde{\varphi}_6\right)}{\tilde{\varphi}_7} \tag{S3}$$

with

$$\begin{aligned}
\tilde{\varphi}_1 &= 4\left(1+e^{\frac{\gamma\Delta t}{2}}\right)^5\left(-1+e^{\frac{\gamma\Delta t}{2}} - e^{\gamma\Delta t} + e^{\frac{3\gamma\Delta t}{2}}\right)^3\gamma^8 \\
&\quad - 8\Delta t\left(1+e^{\frac{\gamma\Delta t}{2}}\right)^4\left(-1+e^{\frac{\gamma\Delta t}{2}} - e^{\gamma\Delta t} + e^{\frac{3\gamma\Delta t}{2}}\right)^2\left(2-2e^{\frac{\gamma\Delta t}{2}} - 2e^{\frac{3\gamma\Delta t}{2}} + e^{2\gamma\Delta t}\right)\gamma^7\omega^2 \\
&\quad + 4\Delta t^2\left(1+e^{\frac{\gamma\Delta t}{2}}\right)^3\left[-7+21e^{\frac{\gamma\Delta t}{2}} - 27e^{\gamma\Delta t} + 37e^{\frac{3\gamma\Delta t}{2}} - 47e^{2\gamma\Delta t} + 35e^{\frac{5\gamma\Delta t}{2}} - 37e^{3\gamma\Delta t}\right. \\
&\qquad \left. + 33e^{\frac{7\gamma\Delta t}{2}} - 18e^{4\gamma\Delta t} + 16e^{\frac{9\gamma\Delta t}{2}} - 8e^{5\gamma\Delta t} + 2e^{\frac{11\gamma\Delta t}{2}}\right]\gamma^6\omega^4
\end{aligned} \tag{S4}$$

$$\tilde{\varphi}_2 = \Delta t^4\left(-1+e^{\frac{\gamma\Delta t}{2}}\right)^5\left(1+e^{\frac{\gamma\Delta t}{2}}\right)^3\left(1-4e^{\frac{\gamma\Delta t}{2}}+3e^{\gamma\Delta t}+8e^{\frac{3\gamma\Delta t}{2}}-13e^{2\gamma\Delta t}-4e^{\frac{5\gamma\Delta t}{2}}+e^{3\gamma\Delta t}\right)\omega^{12}$$

$$\begin{aligned}&-2\Delta t^3\left(-1+e^{\frac{\gamma\Delta t}{2}}\right)^4\left(1+e^{\frac{\gamma\Delta t}{2}}\right)^2\gamma\omega^{10}\Big[1+4e^{\gamma\Delta t}+\Delta t^2\omega^2+e^{\frac{7\gamma\Delta t}{2}}\left(-12+\Delta t^2\omega^2\right)\\&+e^{4\gamma\Delta t}\left(1+\Delta t^2\omega^2\right)-2e^{2\gamma\Delta t}\left(5+\Delta t^2\omega^2\right)-4e^{3\gamma\Delta t}\left(7+2\Delta t^2\omega^2\right)\\&-e^{\frac{\gamma\Delta t}{2}}\left(4+3\Delta t^2\omega^2\right)-e^{\frac{5\gamma\Delta t}{2}}\left(4+5\Delta t^2\omega^2\right)+e^{\frac{3\gamma\Delta t}{2}}\left(4+7\Delta t^2\omega^2\right)\Big]\end{aligned} \tag{S5}$$

$$\begin{aligned}\tilde{\varphi}_3 = &-4\Delta t\left(1+e^{\frac{\gamma\Delta t}{2}}\right)^2\gamma^5\omega^4\Big[1-12e^{3\gamma\Delta t}+7\Delta t^2\omega^2+e^{\frac{7\gamma\Delta t}{2}}\left(8-18\Delta t^2\omega^2\right)\\&+e^{\frac{5\gamma\Delta t}{2}}\left(8-10\Delta t^2\omega^2\right)+6e^{5\gamma\Delta t}\left(1+\Delta t^2\omega^2\right)-3e^{4\gamma\Delta t}\left(3+\Delta t^2\omega^2\right)\\&-e^{\frac{9\gamma\Delta t}{2}}\left(4+\Delta t^2\omega^2\right)-e^{\frac{11\gamma\Delta t}{2}}\left(4+\Delta t^2\omega^2\right)+e^{6\gamma\Delta t}\left(5+\Delta t^2\omega^2\right)+6e^{\gamma\Delta t}\left(1+3\Delta t^2\omega^2\right)\\&+3e^{2\gamma\Delta t}\left(1+9\Delta t^2\omega^2\right)-e^{\frac{3\gamma\Delta t}{2}}\left(4+13\Delta t^2\omega^2\right)-e^{\frac{\gamma\Delta t}{2}}\left(4+21\Delta t^2\omega^2\right)\Big]\end{aligned} \tag{S6}$$

$$\begin{aligned}\tilde{\varphi}_4 = &\Delta t^2\left(-1+e^{\gamma\Delta t}\right)\gamma^4\omega^6\Big[12+17\Delta t^2\omega^2+e^{\gamma\Delta t}\left(8-14\Delta t^2\omega^2\right)+2e^{5\gamma\Delta t}\left(-12+\Delta t^2\omega^2\right)\\&+e^{6\gamma\Delta t}\left(36+\Delta t^2\omega^2\right)+12e^{\frac{5\gamma\Delta t}{2}}\left(4+3\Delta t^2\omega^2\right)-12e^{3\gamma\Delta t}\left(4+5\Delta t^2\omega^2\right)+e^{\frac{9\gamma\Delta t}{2}}\left(-80+6\Delta t^2\omega^2\right)\\&+2e^{\frac{11\gamma\Delta t}{2}}\left(8+7\Delta t^2\omega^2\right)-4e^{\frac{7\gamma\Delta t}{2}}\left(28+15\Delta t^2\omega^2\right)+3e^{2\gamma\Delta t}\left(4+17\Delta t^2\omega^2\right)\\&-2e^{\frac{\gamma\Delta t}{2}}\left(16+17\Delta t^2\omega^2\right)+e^{\frac{3\gamma\Delta t}{2}}\left(32+22\Delta t^2\omega^2\right)-e^{4\gamma\Delta t}\left(124+77\Delta t^2\omega^2\right)\Big]\end{aligned} \tag{S7}$$

$$\begin{aligned}\tilde{\varphi}_5 = &\Delta t^2\left(-1+e^{\frac{\gamma\Delta t}{2}}\right)^3\left(1+e^{\frac{\gamma\Delta t}{2}}\right)\gamma^2\omega^8\Big[1+8\Delta t^2\omega^2+\Delta t^4\omega^4+4\Delta t^2e^{\frac{3\gamma\Delta t}{2}}\omega^2\left(10+\Delta t^2\omega^2\right)\\&-4e^{\frac{5\gamma\Delta t}{2}}\left(10+3\Delta t^2\omega^2\right)+3e^{5\gamma\Delta t}\left(-1+4\Delta t^2\omega^2\right)+e^{\gamma\Delta t}\left(5-12\Delta t^2\omega^2-2\Delta t^4\omega^4\right)\\&+2e^{\frac{9\gamma\Delta t}{2}}\left(-26+7\Delta t^2\omega^2+\Delta t^4\omega^4\right)-2e^{\frac{\gamma\Delta t}{2}}\left(2+9\Delta t^2\omega^2+\Delta t^4\omega^4\right)\\&-4e^{\frac{7\gamma\Delta t}{2}}\left(24+22\Delta t^2\omega^2+\Delta t^4\omega^4\right)-e^{4\gamma\Delta t}\left(87+64\Delta t^2\omega^2+\Delta t^4\omega^4\right)\\&+2e^{2\gamma\Delta t}\left(-5+12\Delta t^2\omega^2+2\Delta t^4\omega^4\right)-2e^{3\gamma\Delta t}\left(49+48\Delta t^2\omega^2+5\Delta t^4\omega^4\right)\Big]\end{aligned} \tag{S8}$$

$$\tilde{\varphi}_6 = -2\Delta t\left(-1+e^{\frac{\gamma\Delta t}{2}}\right)^2 \gamma^3\omega^6\Big[-10\Delta t^2 e^{\gamma\Delta t}\omega^2\left(2+\Delta t^2\omega^2\right)+\Delta t^2\omega^2\left(7+3\Delta t^2\omega^2\right)$$

$$-\Delta t^2 e^{\frac{\gamma\Delta t}{2}}\omega^2\left(10+3\Delta t^2\omega^2\right)+\Delta t^2 e^{\frac{3\gamma\Delta t}{2}}\omega^2\left(22+3\Delta t^2\omega^2\right)+e^{6\gamma\Delta t}\left(-4+15\Delta t^2\omega^2\right)$$

$$+2e^{5\gamma\Delta t}\left(-16-14\Delta t^2\omega^2+\Delta t^4\omega^4\right)-16e^{3\gamma\Delta t}\left(2+3\Delta t^2\omega^2+\Delta t^4\omega^4\right) \quad \text{(S9)}$$

$$+4e^{\frac{5\gamma\Delta t}{2}}\left(-4+5\Delta t^2\omega^2+4\Delta t^4\omega^4\right)+e^{\frac{11\gamma\Delta t}{2}}\left(-16+22\Delta t^2\omega^2+5\Delta t^4\omega^4\right)$$

$$-e^{\frac{9\gamma\Delta t}{2}}\left(48+106\Delta t^2\omega^2+5\Delta t^4\omega^4\right)-4e^{\frac{7\gamma\Delta t}{2}}\left(12+35\Delta t^2\omega^2+8\Delta t^4\omega^4\right)$$

$$+e^{2\gamma\Delta t}\left(-4+41\Delta t^2\omega^2+18\Delta t^4\omega^4\right)-e^{4\gamma\Delta t}\left(56+159\Delta t^2\omega^2+29\Delta t^4\omega^4\right)\Big]$$

$$\tilde{\varphi}_7 = \left(-1+e^{\gamma\Delta t}\right)\omega^2\left[2\left(1+e^{\gamma\Delta t}\right)\gamma-\Delta t\left(-1+e^{\gamma\Delta t}\right)\omega^2\right]^3$$

$$\times\left\{\Delta t^2\left(-1+3e^{\frac{\gamma\Delta t}{2}}-3e^{2\gamma\Delta t}+e^{\frac{5\gamma\Delta t}{2}}\right)\gamma\omega^4+2\left(1+e^{\frac{\gamma\Delta t}{2}}\right)^2\left(-1+e^{\frac{\gamma\Delta t}{2}}-e^{\gamma\Delta t}+e^{\frac{3\gamma\Delta t}{2}}\right)\gamma\left(\gamma^2-\omega^2\right)\right. \quad \text{(S10)}$$

$$\left.-2\Delta t\left(1+e^{\frac{\gamma\Delta t}{2}}\right)\omega^2\left[\gamma^2+e^{2\gamma\Delta t}\gamma^2+4e^{\gamma\Delta t}\omega^2-2e^{\frac{\gamma\Delta t}{2}}\left(\gamma^2+\omega^2\right)-2e^{\frac{3\gamma\Delta t}{2}}\left(\gamma^2+\omega^2\right)\right]\right)^2\right\}$$

**1B. Characteristic correlation time of the Hamiltonian**

**1) "Middle-pT" scheme**

$$\tau_{Ham}^{\text{Middle-pT}} = \frac{\psi_1}{\psi_2} \quad \text{(S11)}$$

with

$$\psi_1 = \Delta t\Big\{-128e^{2\gamma\Delta t}\left(-1+e^{\gamma\Delta t}\right)^2\gamma\omega^2-8\Delta t^4\left(-1+e^{\gamma\Delta t}\right)^2\left(-1+2e^{\gamma\Delta t}+2e^{2\gamma\Delta t}\right)\gamma\omega^6$$

$$+\Delta t^5\left(-1+e^{\gamma\Delta t}\right)^3\left(-1+3e^{\gamma\Delta t}\right)\omega^8+32\Delta t^2\left(1+e^{\gamma\Delta t}\right)^2\gamma\omega^2\left[\left(1-2e^{\gamma\Delta t}\right)\gamma^2+\left(-1+e^{\gamma\Delta t}\right)^2\omega^2\right]$$

$$-8\Delta t^3\left(-1+e^{2\gamma\Delta t}\right)\omega^4\left[-3\left(-1+2e^{\gamma\Delta t}+e^{2\gamma\Delta t}\right)\gamma^2+\left(-1+e^{\gamma\Delta t}\right)^2\omega^2\right]$$

$$-16\Delta t\left(-1+e^{\gamma\Delta t}\right)\left[\left(1+e^{\gamma\Delta t}\right)^3\gamma^4+2\left(1+e^{\gamma\Delta t}\right)^3\gamma^2\omega^2-\left(-1+e^{\gamma\Delta t}\right)^2\left(-1+3e^{\gamma\Delta t}\right)\omega^4\right]\Big\} \quad \text{(S12)}$$

$$\psi_2 = 4\left(-1+e^{\gamma\Delta t}\right)\omega^2\left[-2\left(1+e^{\gamma\Delta t}\right)\gamma+\Delta t\left(-1+e^{\gamma\Delta t}\right)\omega^2\right]$$

$$\times\Big[16+4\Delta t^2\gamma^2+4\Delta t^3\gamma\omega^2+\Delta t^4\omega^4-2e^{\gamma\Delta t}\left(16-4\Delta t^2\gamma^2+\Delta t^4\omega^4\right)$$

$$+e^{2\gamma\Delta t}\left(16+4\Delta t^2\gamma^2-4\Delta t^3\gamma\omega^2+\Delta t^4\omega^4\right)\Big]$$

**2) "Side-pT" scheme**

$$\tau_{Ham}^{\text{Side-pT}} = \frac{\tilde{\psi}_1 + \tilde{\psi}_2 + \tilde{\psi}_3}{\tilde{\psi}_4} \tag{S13}$$

with

$$\tilde{\psi}_1 = \Delta t \Bigg\{ -8e^{2\gamma\Delta t}\left(-1+e^{2\gamma\Delta t}\right)^2 \gamma^3\omega^2 - \Delta t^3\left(-1+e^{\gamma\Delta t}\right)\omega^4 \Bigg[ -\left(1+e^{\gamma\Delta t}\right)^3\left(-1+2e^{\gamma\Delta t}\right)\gamma^4$$
$$-2\left(-1+e^{\frac{\gamma\Delta t}{2}}\right)^2\left(-1+2e^{\frac{\gamma\Delta t}{2}} - e^{\gamma\Delta t} - 4e^{\frac{3\gamma\Delta t}{2}} + e^{2\gamma\Delta t} - 6e^{\frac{5\gamma\Delta t}{2}} + e^{3\gamma\Delta t}\right)\gamma^2\omega^2$$
$$+\left(-1+e^{\frac{\gamma\Delta t}{2}}\right)^4\left(1-4e^{\frac{\gamma\Delta t}{2}} + 3e^{\gamma\Delta t} + 8e^{\frac{3\gamma\Delta t}{2}} - 13e^{2\gamma\Delta t} - 4e^{\frac{5\gamma\Delta t}{2}} + e^{3\gamma\Delta t}\right)\omega^4 \Bigg]\Bigg\}$$

(S14)

$$\tilde{\psi}_2 = -\Delta t^2\left(-1+e^{2\gamma\Delta t}\right)\gamma^2\Bigg[\left(1+e^{\gamma\Delta t}\right)^4\gamma^4 + 2\left(1+e^{\gamma\Delta t}\right)^2\left(1-4e^{\frac{\gamma\Delta t}{2}} + 10e^{\gamma\Delta t} - 4e^{\frac{3\gamma\Delta t}{2}} + e^{2\gamma\Delta t}\right)\gamma^2\omega^2$$
$$-\left(-1+e^{\frac{\gamma\Delta t}{2}}\right)^2\left(-1+6e^{\frac{\gamma\Delta t}{2}} - 15e^{\gamma\Delta t} + 20e^{\frac{3\gamma\Delta t}{2}} - 11e^{2\gamma\Delta t} + 46e^{\frac{5\gamma\Delta t}{2}} + 3e^{3\gamma\Delta t}\right)\omega^4\Bigg] \tag{S15}$$

$$\tilde{\psi}_3 = -2\Delta t^3\gamma\omega^2\Bigg[\left(1+e^{\gamma\Delta t}\right)^4\left(-1+2e^{\gamma\Delta t}\right)\gamma^4$$
$$-2\left(-1+e^{\frac{\gamma\Delta t}{2}} - e^{\gamma\Delta t} + e^{\frac{3\gamma\Delta t}{2}}\right)^2\left(1-2e^{\frac{\gamma\Delta t}{2}} + e^{\gamma\Delta t} + 8e^{\frac{3\gamma\Delta t}{2}}\right)\gamma^2\omega^2 \tag{S16}$$
$$-\left(-1+e^{\frac{\gamma\Delta t}{2}}\right)^4\left(1-4e^{\frac{\gamma\Delta t}{2}} + 4e^{\gamma\Delta t} + 4e^{\frac{3\gamma\Delta t}{2}} - 10e^{2\gamma\Delta t} - 4e^{\frac{5\gamma\Delta t}{2}} - 28e^{3\gamma\Delta t} - 12e^{\frac{7\gamma\Delta t}{2}} + e^{4\gamma\Delta t}\right)\omega^4\Bigg]$$

$$\tilde{\psi}_4 = \left(-1+e^{\frac{\gamma\Delta t}{2}}\right)\left(1+e^{\frac{\gamma\Delta t}{2}}\right)\omega^2\left[-2\left(1+e^{\gamma\Delta t}\right)\gamma + \Delta t\left(-1+e^{\gamma\Delta t}\right)\omega^2\right]$$
$$\times\Bigg\{\Delta t^2\left(1+e^{\gamma\Delta t}\right)^4\gamma^4 - 16\Delta t e^{\frac{\gamma\Delta t}{2}}\left(-1+e^{\frac{\gamma\Delta t}{2}}\right)^3\left(1+e^{\frac{\gamma\Delta t}{2}} + e^{\gamma\Delta t} + e^{\frac{3\gamma\Delta t}{2}}\right)\gamma\omega^2$$
$$+16\Delta t^2 e^{\gamma\Delta t}\left(-1+e^{\frac{\gamma\Delta t}{2}}\right)^4\omega^4 + 2\left(-1+e^{\frac{\gamma\Delta t}{2}} - e^{\gamma\Delta t} + e^{\frac{3\gamma\Delta t}{2}}\right)^2\gamma^2 \tag{S17}$$
$$\times\left[2+\Delta t^2\omega^2 + e^{\frac{\gamma\Delta t}{2}}\left(4-2\Delta t^2\omega^2\right) + e^{\gamma\Delta t}\left(2+\Delta t^2\omega^2\right)\right]\Bigg\}$$

**3) "Middle-xT" scheme**

$$\tau_{Ham}^{\text{Middle-xT}} = \frac{\bar{\psi}_1 + \bar{\psi}_2 + \bar{\psi}_3 + \bar{\psi}_4 + \bar{\psi}_5 + \bar{\psi}_6}{4\left(-1+e^{\gamma\Delta t}\right)\omega^2\left[-2\left(1+e^{\gamma\Delta t}\right)\gamma + \Delta t\left(-1+e^{\gamma\Delta t}\right)\omega^2\right]\left(\bar{\psi}_7 + \bar{\psi}_8 + \bar{\psi}_9 + \bar{\psi}_{10}\right)} \quad \text{(S18)}$$

with

$$\begin{aligned}\bar{\psi}_1 = &-4\left(-1+e^{\gamma\Delta t}\right)\left(1+e^{\gamma\Delta t}\right)^3\gamma^6\left[4+\Delta t^2\omega^2+2e^{\gamma\Delta t}\left(-4+\Delta t^2\omega^2\right)+e^{2\gamma\Delta t}\left(4+\Delta t^2\omega^2\right)\right]^2 \\ &+\Delta t^2\left(-1+e^{\gamma\Delta t}\right)^5\left(1+e^{\gamma\Delta t}\right)^2\omega^8\left[-\left(4+\Delta t^2\omega^2\right)^2+e^{\gamma\Delta t}\left(48-8\Delta t^2\omega^2+3\Delta t^4\omega^4\right)\right]\end{aligned} \quad \text{(S19)}$$

$$\begin{aligned}\bar{\psi}_2 = &-8\Delta t\left(1+e^{\gamma\Delta t}\right)^2\gamma^5\omega^2\left[-32-12\Delta t^2\omega^2+4\Delta t^2 e^{6\gamma\Delta t}\omega^2-\Delta t^4\omega^4+2\Delta t^2 e^{5\gamma\Delta t}\omega^2\left(8+\Delta t^2\omega^2\right)\right. \\ &-2e^{\gamma\Delta t}\left(-64-8\Delta t^2\omega^2+\Delta t^4\omega^4\right)+8e^{3\gamma\Delta t}\left(16-4\Delta t^2\omega^2+\Delta t^4\omega^4\right) \\ &\left.+2e^{2\gamma\Delta t}\left(-96+14\Delta t^2\omega^2+\Delta t^4\omega^4\right)+e^{4\gamma\Delta t}\left(-32-20\Delta t^2\omega^2+7\Delta t^4\omega^4\right)\right]\end{aligned} \quad \text{(S20)}$$

$$\begin{aligned}\bar{\psi}_3 = &-4\Delta t\left(-1+e^{\gamma\Delta t}\right)^4\left(1+e^{\gamma\Delta t}\right)\gamma\omega^6\left[-4\Delta t^2\omega^2\left(4+\Delta t^2\omega^2\right)\right. \\ &-\Delta t^2 e^{\gamma\Delta t}\omega^2\left(32-4\Delta t^2\omega^2+\Delta t^4\omega^4\right)+2e^{3\gamma\Delta t}\left(16-8\Delta t^2\omega^2+5\Delta t^4\omega^4\right) \\ &\left.+e^{2\gamma\Delta t}\left(32+32\Delta t^2\omega^2-14\Delta t^4\omega^4+3\Delta t^6\omega^6\right)\right]\end{aligned} \quad \text{(S21)}$$

$$\begin{aligned}\bar{\psi}_4 = &\,4\Delta t^2\left(-1+e^{\gamma\Delta t}\right)^3\gamma^2\omega^6\left[-64-20\Delta t^2\omega^2-\Delta t^4\omega^4+e^{\gamma\Delta t}\left(4\Delta t^2\omega^2-9\Delta t^4\omega^4\right)\right. \\ &+e^{5\gamma\Delta t}\left(-64+48\Delta t^2\omega^2+\Delta t^4\omega^4\right)+e^{4\gamma\Delta t}\left(128-80\Delta t^2\omega^2+29\Delta t^4\omega^4\right) \\ &\left.-e^{2\gamma\Delta t}\left(64+124\Delta t^2\omega^2-20\Delta t^4\omega^4+\Delta t^6\omega^6\right)+e^{3\gamma\Delta t}\left(64+44\Delta t^2\omega^2-8\Delta t^4\omega^4+3\Delta t^6\omega^6\right)\right]\end{aligned} \quad \text{(S22)}$$

$$\begin{aligned}\bar{\psi}_5 = &-4\Delta t\left(-1+e^{\gamma\Delta t}\right)^2\gamma^3\omega^4\left[-112-64\Delta t^2\omega^2-9\Delta t^4\omega^4+e^{6\gamma\Delta t}\left(-112+104\Delta t^2\omega^2+7\Delta t^4\omega^4\right)\right. \\ &+e^{2\gamma\Delta t}\left(112-152\Delta t^2\omega^2+9\Delta t^4\omega^4\right)-2e^{\gamma\Delta t}\left(-48-32\Delta t^2\omega^2+5\Delta t^4\omega^4+\Delta t^6\omega^6\right) \\ &+2e^{5\gamma\Delta t}\left(48-112\Delta t^2\omega^2+43\Delta t^4\omega^4+\Delta t^6\omega^6\right)+4e^{3\gamma\Delta t}\left(-48-40\Delta t^2\omega^2+5\Delta t^4\omega^4+2\Delta t^6\omega^6\right) \\ &\left.+e^{4\gamma\Delta t}\left(112+176\Delta t^2\omega^2-39\Delta t^4\omega^4+16\Delta t^6\omega^6\right)\right]\end{aligned} \quad \text{(S23)}$$

$$\begin{aligned}\bar{\psi}_6 = &\,4\left(-1+e^{2\gamma\Delta t}\right)\gamma^4\omega^2\left[-64-96\Delta t^2\omega^2-24\Delta t^4\omega^4-\Delta t^6\omega^6\right. \\ &+16e^{6\gamma\Delta t}\left(-4+6\Delta t^2\omega^2+\Delta t^4\omega^4\right)-4e^{\gamma\Delta t}\left(-32-56\Delta t^2\omega^2-4\Delta t^4\omega^4+\Delta t^6\omega^6\right) \\ &+8e^{5\gamma\Delta t}\left(16-36\Delta t^2\omega^2+10\Delta t^4\omega^4+\Delta t^6\omega^6\right)+2e^{2\gamma\Delta t}\left(32-80\Delta t^2\omega^2+16\Delta t^4\omega^4+\Delta t^6\omega^6\right) \\ &\left.+4e^{3\gamma\Delta t}\left(-64-16\Delta t^2\omega^2-8\Delta t^4\omega^4+5\Delta t^6\omega^6\right)+e^{4\gamma\Delta t}\left(64+288\Delta t^2\omega^2-88\Delta t^4\omega^4+23\Delta t^6\omega^6\right)\right]\end{aligned} \quad \text{(S24)}$$

$$\begin{aligned}\bar{\psi}_7 = &\left(-1+e^{\gamma\Delta t}\right)^4\left(1+e^{\gamma\Delta t}\right)^2\omega^4\left(16+\Delta t^4\omega^4\right)-4\Delta t\left(-1+e^{\gamma\Delta t}\right)^3\left(1+e^{\gamma\Delta t}\right)\gamma\omega^4 \\ &\times\left[3\Delta t^2\omega^2+3\Delta t^2 e^{2\gamma\Delta t}\omega^2+e^{\gamma\Delta t}\left(16-2\Delta t^2\omega^2+\Delta t^4\omega^4\right)\right]\end{aligned} \quad \text{(S25)}$$

$$
\begin{aligned}
\overline{\psi}_8 = 2\Delta t^2\left(-1+e^{\gamma\Delta t}\right)^2\gamma^2\omega^4\Big[&22+\Delta t^2\omega^2+e^{4\gamma\Delta t}\left(22+\Delta t^2\omega^2\right)\\
&+8e^{\gamma\Delta t}\left(-3+2\Delta t^2\omega^2\right)+8e^{3\gamma\Delta t}\left(-3+2\Delta t^2\omega^2\right)\\
&+2e^{2\gamma\Delta t}\left(34-\Delta t^2\omega^2+\Delta t^4\omega^4\right)\Big]
\end{aligned}
\tag{S26}
$$

$$
\begin{aligned}
\overline{\psi}_9 = -4\Delta t\left(-1+e^{2\gamma\Delta t}\right)\gamma^3\omega^2\Big[&20+3\Delta t^2\omega^2+e^{4\gamma\Delta t}\left(20+3\Delta t^2\omega^2\right)\\
&+2e^{2\gamma\Delta t}\left(28-3\Delta t^2\omega^2+\Delta t^4\omega^4\right)+e^{\gamma\Delta t}\left(-48+8\Delta t^2\omega^2+\Delta t^4\omega^4\right)\\
&+e^{3\gamma\Delta t}\left(-48+8\Delta t^2\omega^2+\Delta t^4\omega^4\right)\Big]
\end{aligned}
\tag{S27}
$$

$$
\begin{aligned}
\overline{\psi}_{10} = \left(1+e^{\gamma\Delta t}\right)^2\gamma^4\Big[&32+16\Delta t^2\omega^2+\Delta t^4\omega^4+4e^{\gamma\Delta t}\left(-32+\Delta t^4\omega^4\right)\\
&+4e^{3\gamma\Delta t}\left(-32+\Delta t^4\omega^4\right)+e^{4\gamma\Delta t}\left(32+16\Delta t^2\omega^2+\Delta t^4\omega^4\right)\\
&+2e^{2\gamma\Delta t}\left(96-16\Delta t^2\omega^2+3\Delta t^4\omega^4\right)\Big]
\end{aligned}
\tag{S28}
$$

## 4) "Side-xT" scheme

$$
\tau_{Ham}^{\text{Side-xT}} = \frac{\widehat{\psi}_1+\widehat{\psi}_2+\widehat{\psi}_3+\widehat{\psi}_4+\widehat{\psi}_5+\widehat{\psi}_6+\widehat{\psi}_7+\widehat{\psi}_8}{\left\{\left(-1+e^{\frac{\gamma\Delta t}{2}}\right)\left(1+e^{\frac{\gamma\Delta t}{2}}\right)\omega^2\left[-2\left(1+e^{\gamma\Delta t}\right)\gamma+\Delta t\left(-1+e^{\gamma\Delta t}\right)\omega^2\right]\left(\widehat{\psi}_9+\widehat{\psi}_{10}+\widehat{\psi}_{11}+\widehat{\psi}_{12}+\widehat{\psi}_{13}+\widehat{\psi}_{14}\right)\right\}}
\tag{S29}
$$

with

$$
\begin{aligned}
\widehat{\psi}_1 = &-\Delta t^4\left(-1+e^{\frac{\gamma\Delta t}{2}}\right)^7\left(1+e^{\frac{\gamma\Delta t}{2}}\right)^3\left(1-4e^{\frac{\gamma\Delta t}{2}}+3e^{\gamma\Delta t}+8e^{\frac{3\gamma\Delta t}{2}}-13e^{2\gamma\Delta t}-4e^{\frac{5\gamma\Delta t}{2}}+e^{3\gamma\Delta t}\right)\omega^{12}\\
&+4\left(-1+e^{\gamma\Delta t}\right)\left(1+e^{\gamma\Delta t}\right)^3\gamma^8\left[1+e^{2\gamma\Delta t}+e^{\gamma\Delta t}\left(-2+\Delta t^2\omega^2\right)\right]^2
\end{aligned}
\tag{S30}
$$

$$
\begin{aligned}
\widehat{\psi}_2 = 2\Delta t^3\left(-1+e^{\frac{\gamma\Delta t}{2}}\right)^6\left(1+e^{\frac{\gamma\Delta t}{2}}\right)^2\gamma\omega^{10}\Bigg[&1+4e^{\gamma\Delta t}+\Delta t^2\omega^2+e^{\frac{7\gamma\Delta t}{2}}\left(-12+\Delta t^2\omega^2\right)\\
&+e^{4\gamma\Delta t}\left(1+\Delta t^2\omega^2\right)-2e^{2\gamma\Delta t}\left(5+\Delta t^2\omega^2\right)-4e^{3\gamma\Delta t}\left(7+2\Delta t^2\omega^2\right)\\
&-e^{\frac{\gamma\Delta t}{2}}\left(4+3\Delta t^2\omega^2\right)-e^{\frac{5\gamma\Delta t}{2}}\left(4+5\Delta t^2\omega^2\right)+e^{\frac{3\gamma\Delta t}{2}}\left(4+7\Delta t^2\omega^2\right)\Bigg]
\end{aligned}
\tag{S31}
$$

$$
\begin{aligned}
\widehat{\psi}_3 = -8\Delta t\left(1+e^{\gamma\Delta t}\right)^2\gamma^7\omega^2\Big[&-2+\Delta t^2e^{5\gamma\Delta t}\omega^2+e^{\gamma\Delta t}\left(8-3\Delta t^2\omega^2\right)\\
&+2e^{4\gamma\Delta t}\left(-1+\Delta t^2\omega^2\right)-e^{2\gamma\Delta t}\left(12-10\Delta t^2\omega^2+\Delta t^4\omega^4\right)\\
&+2e^{3\gamma\Delta t}\left(4-5\Delta t^2\omega^2+\Delta t^4\omega^4\right)\Big]
\end{aligned}
\tag{S32}
$$

$$\widehat{\psi}_4 = -\Delta t^2\left(-1+e^{\frac{\gamma\Delta t}{2}}\right)^5\left(1+e^{\frac{\gamma\Delta t}{2}}\right)\gamma^2\omega^8\left[1+12\Delta t^2\omega^2+\Delta t^4\omega^4+4\Delta t^2 e^{\frac{3\gamma\Delta t}{2}}\omega^2\left(4+\Delta t^2\omega^2\right)\right.$$
$$+4e^{\frac{5\gamma\Delta t}{2}}\left(-10+7\Delta t^2\omega^2\right)+e^{5\gamma\Delta t}\left(-3+14\Delta t^2\omega^2\right)+e^{\gamma\Delta t}\left(5-6\Delta t^2\omega^2-2\Delta t^4\omega^4\right)$$
$$-2e^{\frac{\gamma\Delta t}{2}}\left(2+3\Delta t^2\omega^2+\Delta t^4\omega^4\right)+2e^{\frac{9\gamma\Delta t}{2}}\left(-26+5\Delta t^2\omega^2+\Delta t^4\omega^4\right) \quad \text{(S33)}$$
$$-4e^{\frac{7\gamma\Delta t}{2}}\left(24+20\Delta t^2\omega^2+\Delta t^4\omega^4\right)-e^{4\gamma\Delta t}\left(87+88\Delta t^2\omega^2+\Delta t^4\omega^4\right)$$
$$\left.+2e^{2\gamma\Delta t}\left(-5-2\Delta t^2\omega^2+2\Delta t^4\omega^4\right)-2e^{3\gamma\Delta t}\left(49+12\Delta t^2\omega^2+5\Delta t^4\omega^4\right)\right]$$

$$\widehat{\psi}_5 = -2\Delta t\left(-1+e^{\frac{\gamma\Delta t}{2}}\right)^2\gamma^5\omega^4\left[-28e^{\frac{13\gamma\Delta t}{2}}+e^{7\gamma\Delta t}\left(-26+\Delta t^2\omega^2\right)-2\left(13+7\Delta t^2\omega^2\right)\right.$$
$$-2e^{\frac{\gamma\Delta t}{2}}\left(14+11\Delta t^2\omega^2\right)-6e^{\frac{3\gamma\Delta t}{2}}\left(-4-2\Delta t^2\omega^2+\Delta t^4\omega^4\right)-10e^{\gamma\Delta t}\left(-1+\Delta t^2\omega^2+\Delta t^4\omega^4\right)$$
$$+4e^{\frac{5\gamma\Delta t}{2}}\left(7-5\Delta t^2\omega^2+2\Delta t^4\omega^4\right)-8e^{\frac{7\gamma\Delta t}{2}}\left(6+4\Delta t^2\omega^2+3\Delta t^4\omega^4\right)$$
$$+e^{5\gamma\Delta t}\left(62+8\Delta t^2\omega^2+4\Delta t^4\omega^4\right)+e^{3\gamma\Delta t}\left(-46-31\Delta t^2\omega^2+6\Delta t^4\omega^4\right) \quad \text{(S34)}$$
$$+e^{\frac{11\gamma\Delta t}{2}}\left(24+4\Delta t^2\omega^2+6\Delta t^4\omega^4\right)+e^{2\gamma\Delta t}\left(62+33\Delta t^2\omega^2+6\Delta t^4\omega^4\right)$$
$$-2e^{4\gamma\Delta t}\left(23+34\Delta t^2\omega^2+7\Delta t^4\omega^4\right)-2e^{\frac{9\gamma\Delta t}{2}}\left(-14+3\Delta t^2\omega^2+8\Delta t^4\omega^4\right)$$
$$\left.+e^{6\gamma\Delta t}\left(10+17\Delta t^2\omega^2+8\Delta t^4\omega^4\right)\right]$$

$$\widehat{\psi}_6 = -\Delta t^2\left(-1+e^{\frac{\gamma\Delta t}{2}}\right)^3\left(1+e^{\frac{\gamma\Delta t}{2}}\right)\gamma^4\omega^6\left[64+64e^{6\gamma\Delta t}+17\Delta t^2\omega^2+4e^{\frac{11\gamma\Delta t}{2}}\left(-4+3\Delta t^2\omega^2\right)\right.$$
$$+2e^{\frac{\gamma\Delta t}{2}}\left(8+3\Delta t^2\omega^2\right)-e^{4\Delta t\gamma}\left(96+19\Delta t^2\omega^2\right)-2e^{\frac{9\gamma\Delta t}{2}}\left(40-16\Delta t^2\omega^2+\Delta t^4\omega^4\right)$$
$$-2e^{\frac{3\gamma\Delta t}{2}}\left(24+9\Delta t^2\omega^2+\Delta t^4\omega^4\right)-3e^{5\gamma\Delta t}\left(16+22\Delta t^2\omega^2+\Delta t^4\omega^4\right) \quad \text{(S35)}$$
$$+2e^{\frac{5\gamma\Delta t}{2}}\left(32+29\Delta t^2\omega^2+\Delta t^4\omega^4\right)-2e^{2\gamma\Delta t}\left(16+7\Delta t^2\omega^2+4\Delta t^4\omega^4\right)$$
$$+e^{\gamma\Delta t}\left(-16+18\Delta t^2\omega^2+5\Delta t^4\omega^4\right)+e^{3\gamma\Delta t}\left(64+32\Delta t^2\omega^2+6\Delta t^4\omega^4\right)$$
$$\left.+2e^{\frac{7\gamma\Delta t}{2}}\left(32+35\Delta t^2\omega^2+9\Delta t^4\omega^4\right)\right]$$

$$\hat{\psi}_7 = 2\Delta t\left(-1+e^{\frac{\gamma\Delta t}{2}}\right)^4 \gamma^3\omega^6\Big[\Delta t^4 e^{\gamma\Delta t}\omega^4 + \Delta t^2 e^{\frac{\gamma\Delta t}{2}}\omega^2\left(24+\Delta t^2\omega^2\right)$$

$$+\Delta t^2 e^{\frac{3\gamma\Delta t}{2}}\omega^2\left(-16+3\Delta t^2\omega^2\right)+\Delta t^2\omega^2\left(19+3\Delta t^2\omega^2\right)+e^{6\gamma\Delta t}\left(-4+21\Delta t^2\omega^2\right)$$

$$+16e^{\frac{7\gamma\Delta t}{2}}\left(-3-2\Delta t^2\omega^2+\Delta t^4\omega^4\right)+8e^{\frac{5\gamma\Delta t}{2}}\left(-2+6\Delta t^2\omega^2+\Delta t^4\omega^4\right) \quad \text{(S36)}$$

$$+e^{\frac{11\gamma\Delta t}{2}}\left(-16+16\Delta t^2\omega^2+5\Delta t^4\omega^4\right)-e^{2\gamma\Delta t}\left(4+9\Delta t^2\omega^2+6\Delta t^4\omega^4\right)$$

$$-e^{5\gamma\Delta t}\left(32+72\Delta t^2\omega^2+7\Delta t^4\omega^4\right)-e^{4\gamma\Delta t}\left(56+63\Delta t^2\omega^2+13\Delta t^4\omega^4\right)$$

$$-e^{\frac{9\gamma\Delta t}{2}}\left(48+104\Delta t^2\omega^2+17\Delta t^4\omega^4\right)+e^{3\gamma\Delta t}\left(-32+40\Delta t^2\omega^2+22\Delta t^4\omega^4\right)\Big]$$

$$\hat{\psi}_8 = 4\left(-1+e^{2\gamma\Delta t}\right)\gamma^6\omega^2\Big[-4-7\Delta t^2\omega^2+e^{6\gamma\Delta t}\left(-4+\Delta t^2\omega^2\right)+4\Delta t^2 e^{\frac{5\gamma\Delta t}{2}}\omega^2\left(-2+\Delta t^2\omega^2\right)$$

$$-4\Delta t^2 e^{\frac{9\gamma\Delta t}{2}}\omega^2\left(-2+\Delta t^2\omega^2\right)+4\Delta t^2 e^{\frac{3\gamma\Delta t}{2}}\omega^2\left(2+\Delta t^2\omega^2\right)$$

$$-4\Delta t^2 e^{\frac{7\gamma\Delta t}{2}}\omega^2\left(2+\Delta t^2\omega^2\right)+e^{\gamma\Delta t}\left(8+12\Delta t^2\omega^2-8\Delta t^4\omega^4\right) \quad \text{(S37)}$$

$$+e^{4\gamma\Delta t}\left(4-5\Delta t^2\omega^2+6\Delta t^4\omega^4\right)+e^{5\gamma\Delta t}\left(8-4\Delta t^2\omega^2+6\Delta t^4\omega^4\right)$$

$$+e^{2\gamma\Delta t}\left(4-13\Delta t^2\omega^2+10\Delta t^4\omega^4-\Delta t^6\omega^6\right)+2e^{3\gamma\Delta t}\left(-8+8\Delta t^2\omega^2-7\Delta t^4\omega^4+\Delta t^6\omega^6\right)\Big]$$

$$\hat{\psi}_9 = 16\Delta t^2\left(-1+e^{\frac{\gamma\Delta t}{2}}\right)^6\left(e^{\frac{\gamma\Delta t}{2}}+e^{\gamma\Delta t}\right)^2\omega^8+8\Delta t e^{\frac{\gamma\Delta t}{2}}\left(-1+e^{\frac{\gamma\Delta t}{2}}\right)^5\left(1+e^{\frac{\gamma\Delta t}{2}}\right)\gamma\omega^6$$

$$\times\Big[-2+\Delta t^2\omega^2+e^{2\gamma\Delta t}\left(-2+\Delta t^2\omega^2\right)-2e^{\frac{\gamma\Delta t}{2}}\left(2+\Delta t^2\omega^2\right) \quad \text{(S38)}$$

$$-2e^{\gamma\Delta t}\left(2+\Delta t^2\omega^2\right)-2e^{\frac{3\gamma\Delta t}{2}}\left(2+\Delta t^2\omega^2\right)\Big]$$

$$\hat{\psi}_{10} = -4\Delta t\left(-1+e^{2\gamma\Delta t}\right)\gamma^5\omega^2\Big[3+3e^{4\gamma\Delta t}-4\Delta t^2 e^{\frac{3\gamma\Delta t}{2}}\omega^2-4\Delta t^2 e^{\frac{5\gamma\Delta t}{2}}\omega^2 \quad \text{(S39)}$$

$$+2e^{2\gamma\Delta t}\left(1+\Delta t^2\omega^2\right)+e^{\gamma\Delta t}\left(-4+5\Delta t^2\omega^2\right)+e^{3\gamma\Delta t}\left(-4+5\Delta t^2\omega^2\right)\Big]$$

$$\hat{\psi}_{11} = -4\Delta t\left(-1+e^{\frac{\gamma\Delta t}{2}}\right)^3\left(1+e^{\frac{\gamma\Delta t}{2}}\right)\gamma^3\omega^4\Big[\Delta t^2\omega^2+\Delta t^2 e^{4\gamma\Delta t}\omega^2+e^{\frac{3\gamma\Delta t}{2}}\left(4-6\Delta t^2\omega^2\right)$$

$$+e^{\frac{5\gamma\Delta t}{2}}\left(4-6\Delta t^2\omega^2\right)+e^{\frac{\gamma\Delta t}{2}}\left(4-2\Delta t^2\omega^2\right)+e^{\frac{7\gamma\Delta t}{2}}\left(4-2\Delta t^2\omega^2\right) \quad \text{(S40)}$$

$$+e^{2\gamma\Delta t}\left(8+6\Delta t^2\omega^2\right)+e^{\gamma\Delta t}\left(4+8\Delta t^2\omega^2\right)+e^{3\gamma\Delta t}\left(4+8\Delta t^2\omega^2\right)\Big]$$

$$\widehat{\psi}_{12} = \Delta t^2 \left(-1+e^{\frac{\gamma\Delta t}{2}}\right)^2 \gamma^4\omega^4 \left[9+10e^{\frac{\gamma\Delta t}{2}}+8e^{\frac{3\gamma\Delta t}{2}}-6e^{2\gamma\Delta t}-6e^{3\gamma\Delta t}+8e^{\frac{7\gamma\Delta t}{2}}+10e^{\frac{9\gamma\Delta t}{2}}\right.$$
$$\left.+9e^{5\gamma\Delta t}-4e^{\frac{5\gamma\Delta t}{2}}\left(-7+4\Delta t^2\omega^2\right)+e^{\gamma\Delta t}\left(29+8\Delta t^2\omega^2\right)+e^{4\gamma\Delta t}\left(29+8\Delta t^2\omega^2\right)\right] \tag{S41}$$

$$\widehat{\psi}_{13} = 4\left(1+e^{\gamma\Delta t}\right)^2 \gamma^6 \left[2+2e^{4\gamma\Delta t}+4e^{\gamma\Delta t}\left(-2+\Delta t^2\omega^2\right)+4e^{3\gamma\Delta t}\left(-2+\Delta t^2\omega^2\right)\right.$$
$$\left.+e^{2\gamma\Delta t}\left(12-8\Delta t^2\omega^2+\Delta t^4\omega^4\right)\right] \tag{S42}$$

$$\widehat{\psi}_{14} = \left(-1+e^{\frac{\gamma\Delta t}{2}}\right)^4 \gamma^2\omega^4 \left[4-2\Delta t^2\omega^2+\Delta t^4\omega^4+16e^{\gamma\Delta t}\left(2+3\Delta t^2\omega^2\right)+16e^{3\gamma\Delta t}\left(2+3\Delta t^2\omega^2\right)\right.$$
$$-4e^{\frac{\gamma\Delta t}{2}}\left(-4+\Delta t^4\omega^4\right)-4e^{\frac{7\gamma\Delta t}{2}}\left(-4+\Delta t^4\omega^4\right)+e^{4\gamma\Delta t}\left(4-2\Delta t^2\omega^2+\Delta t^4\omega^4\right) \tag{S43}$$
$$+4e^{\frac{3\gamma\Delta t}{2}}\left(12+16\Delta t^2\omega^2+\Delta t^4\omega^4\right)+4e^{\frac{5\gamma\Delta t}{2}}\left(12+16\Delta t^2\omega^2+\Delta t^4\omega^4\right)$$
$$\left.+2e^{2\gamma\Delta t}\left(28+18\Delta t^2\omega^2+7\Delta t^4\omega^4\right)\right]$$

## S2. More results of numerical examples

### 2A. Potential and kinetic energy

We show the average potential and the average kinetic energy of the two one-dimensional models. While Figs. S1-S4 demonstrate the results for the harmonic system, Figs. S5-S6 do so for the quartic model. The results show that the "middle" scheme is the most accurate in the configurational sampling, while the "PV-end" scheme is the most accurate in the momentum sampling.

**Fig. S1.** Results for the potential energy and the kinetic energy using different time intervals for the harmonic system $U(x)=\frac{1}{2}x^2$. The friction coefficient $\gamma=1$, which is the optimal $\gamma$ for the characteristic correlation time of the potential energy for infinitesimal

time interval. (a) The potential energy for the eight schemes that employ the first type of repartition, in the real dynamics case. (b) Same as Panel (a), but for the virtual dynamics case. (c) The potential energy for the four schemes that employ the second and third type of repartition. (d) Same as Panel (a), but for the kinetic energy. (e) Same as Panel (b), but for the kinetic energy. (f) Same as Panel (c), but for the kinetic energy.

**Fig. S2.** Results for the potential energy and the kinetic energy using different time intervals for harmonic system $U(x)=\frac{1}{2}x^2$. The friction coefficient $\gamma=2$, which is the optimal $\gamma$ for the characteristic correlation time of the Hamiltonian for infinitesimal time interval. (a) The potential energy for the four schemes that employ the second and third type of repartition. (b) Same as Panel (a), but for the kinetic energy.

**Fig. S3.** Same as **Fig. S2**, but the friction coefficient $\gamma$ is chosen to be the optimal $\gamma$ for the characteristic correlation time of the potential energy for each time interval for each scheme.

**Fig. S4.** Same as **Fig. S2**, but the friction coefficient $\gamma$ is chosen to be the optimal $\gamma$ for the characteristic correlation time of the Hamiltonian for each time interval for each scheme.

**Fig. S5.** Same as **Fig. S1**, but for the quartic system $U(x)=\frac{1}{4}x^4$ and the friction coefficient $\gamma=1.2$, which is nearly the optimal $\gamma$ for the characteristic correlation time of the potential energy for infinitesimal time interval.

**Fig. S6.** Same as **Fig. S5**, but for the friction coefficient $\gamma=4$, which is nearly the optimal $\gamma$ for the characteristic correlation time of the Hamiltonian for infinitesimal time interval.

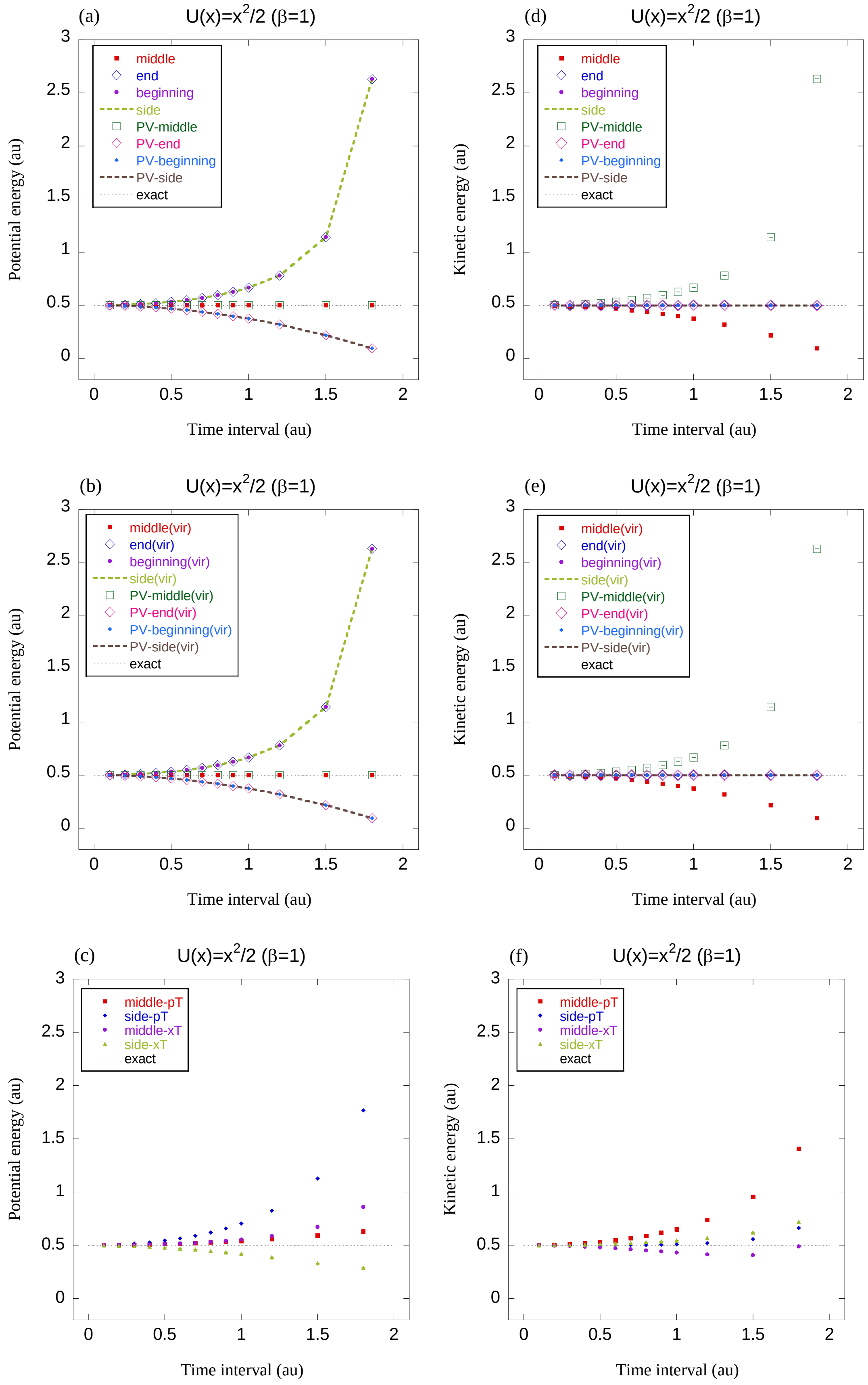

**Fig. S1**

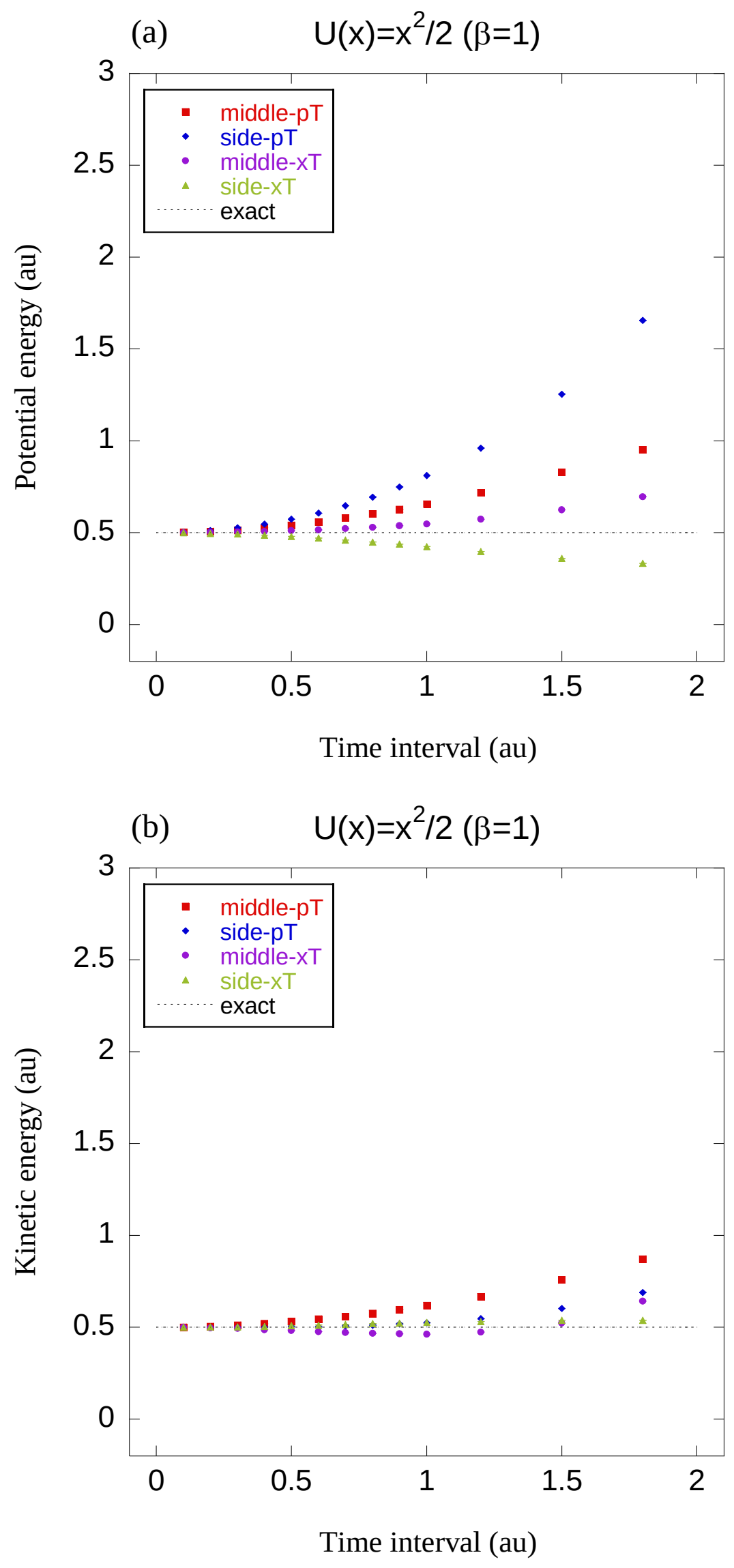

**Fig. S2**

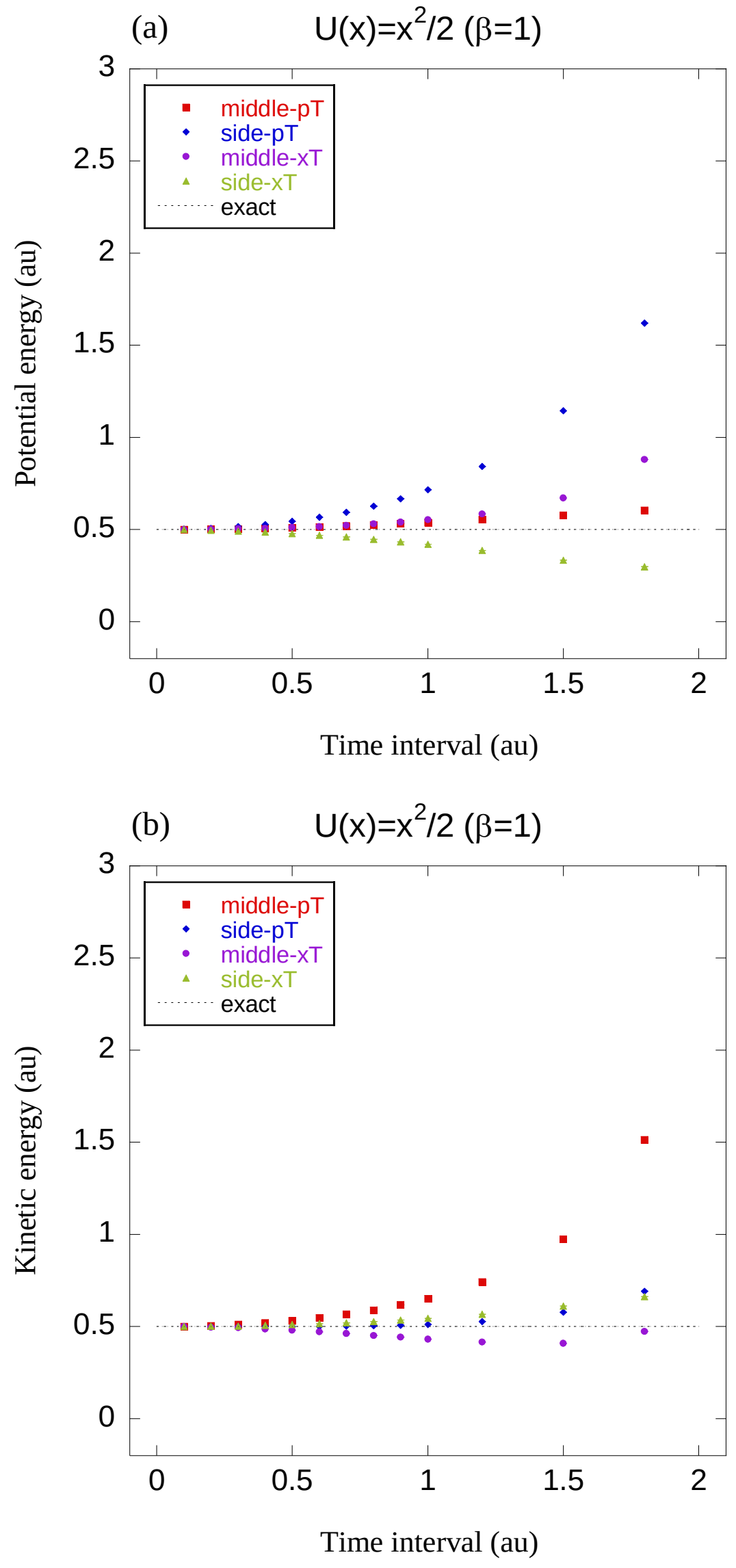

**Fig. S3**

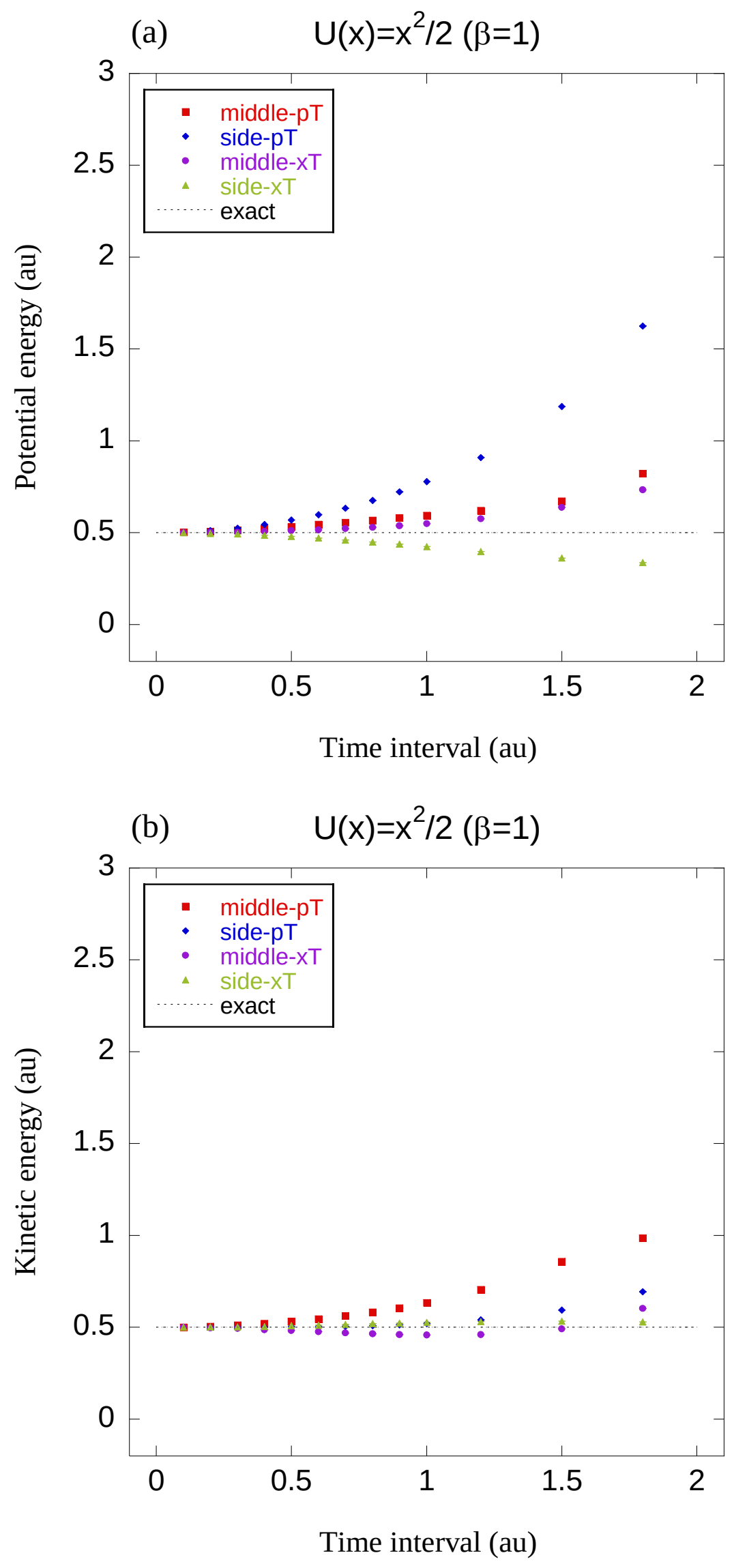

**Fig. S4**

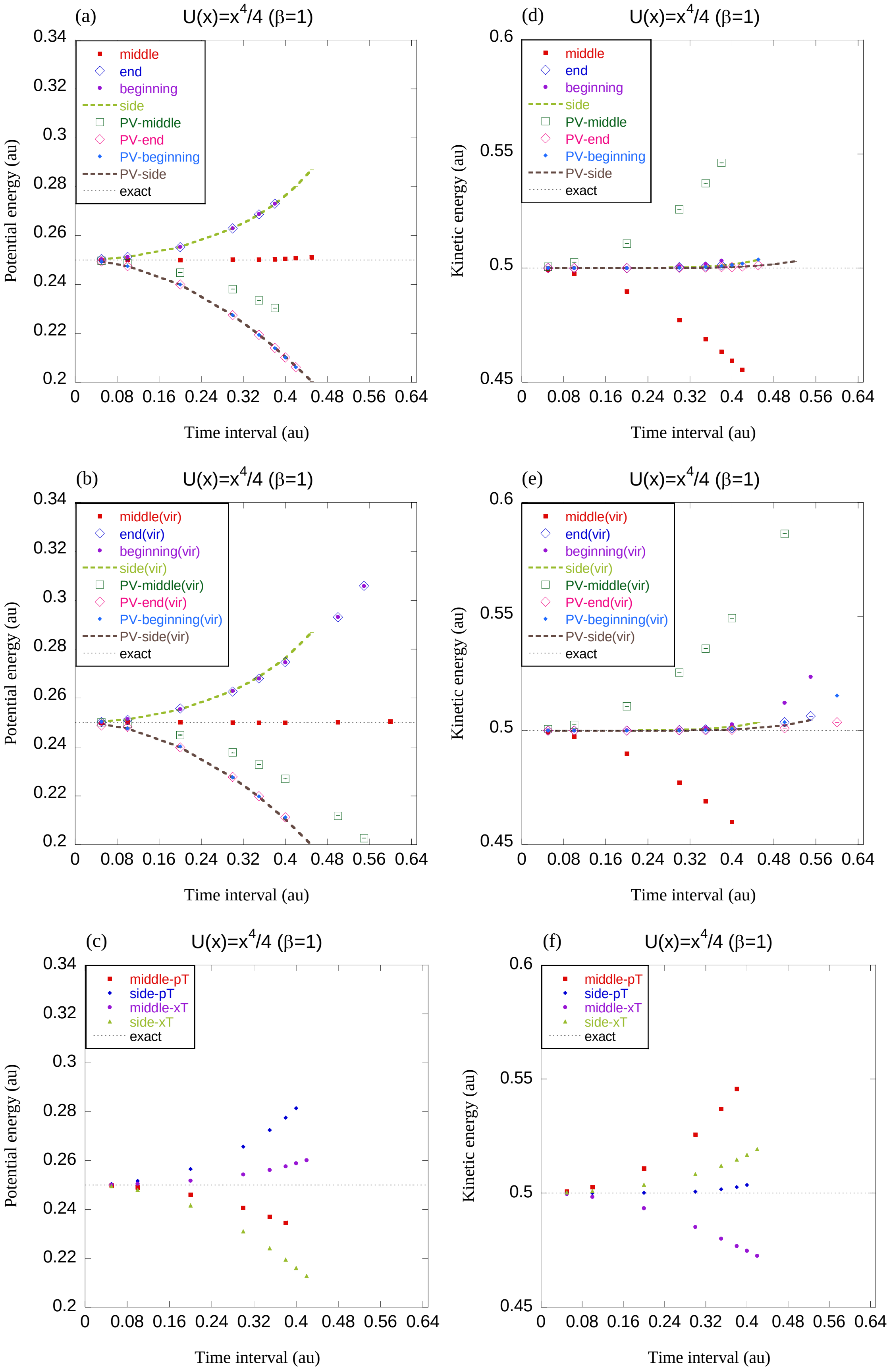

**Fig. S5**

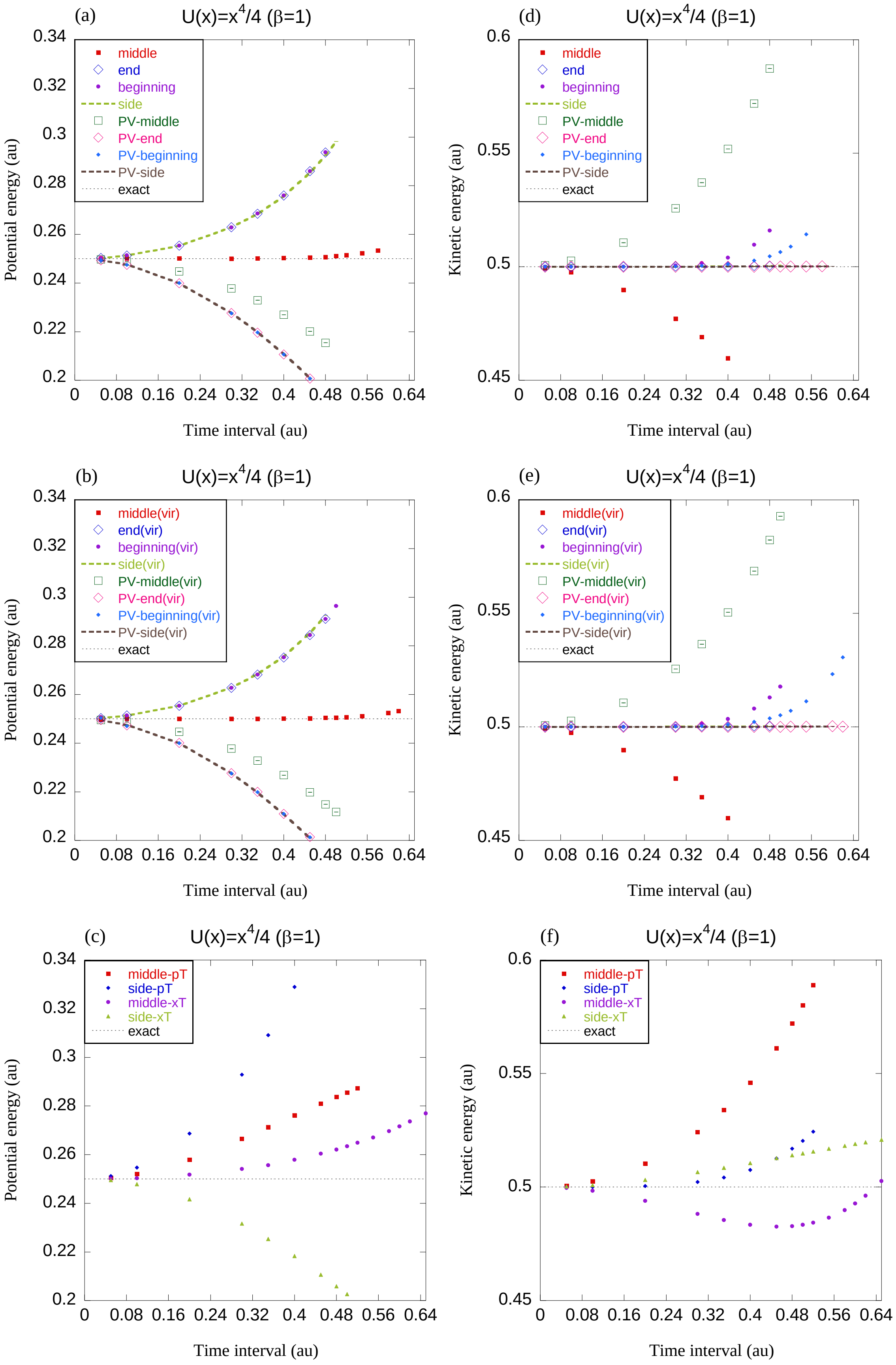

(a) U(x)=x^4/4 (β=1)
middle
end
beginning
side
PV-middle
PV-end
PV-beginning
PV-side
exact
Potential energy (au)
0.34
0.32
0.3
0.28
0.26
0.24
0.22
0.2
0 0.08 0.16 0.24 0.32 0.4 0.48 0.56 0.64
Time interval (au)
(d) U(x)=x^4/4 (β=1)
middle
end
beginning
side
PV-middle
PV-end
PV-beginning
PV-side
exact
Kinetic energy (au)
0.6
0.55
0.5
0.45
0 0.08 0.16 0.24 0.32 0.4 0.48 0.56 0.64
Time interval (au)
(b) U(x)=x^4/4 (β=1)
middle(vir)
end(vir)
beginning(vir)
side(vir)
PV-middle(vir)
PV-end(vir)
PV-beginning(vir)
PV-side(vir)
exact
Potential energy (au)
0.34
0.32
0.3
0.28
0.26
0.24
0.22
0.2
0 0.08 0.16 0.24 0.32 0.4 0.48 0.56 0.64
Time interval (au)
(e) U(x)=x^4/4 (β=1)
middle(vir)
end(vir)
beginning(vir)
side(vir)
PV-middle(vir)
PV-end(vir)
PV-beginning(vir)
PV-side(vir)
exact
Kinetic energy (au)
0.6
0.55
0.5
0.45
0 0.08 0.16 0.24 0.32 0.4 0.48 0.56 0.64
Time interval (au)
(c) U(x)=x^4/4 (β=1)
middle-pT
side-pT
middle-xT
side-xT
exact
Potential energy (au)
0.34
0.32
0.3
0.28
0.26
0.24
0.22
0.2
0 0.08 0.16 0.24 0.32 0.4 0.48 0.56 0.64
Time interval (au)
(f) U(x)=x^4/4 (β=1)
middle-pT
side-pT
middle-xT
side-xT
exact
Kinetic energy (au)
0.6
0.55
0.5
0.45
0 0.08 0.16 0.24 0.32 0.4 0.48 0.56 0.64
Time interval (au)

**Fig. S6**

**2B. Numerical results of the characteristic correlation time for the harmonic system**

We show the numerical results of the characteristic correlation time of the potential and the Hamiltonian for the "middle"/"side"/"middle-xT" schemes for the harmonic system $U(x)=\frac{1}{2}x^2$. While Fig. S7 demonstrates the results of the characteristic correlation time of the potential, Fig. S8 does so for that of the Hamiltonian.

**Fig. S7**. Characteristic correlation time of the potential energy for the harmonic system $U(x)=\frac{1}{2}x^2$. Three time intervals $\Delta t = 0.3$, $\sqrt{2}$, 1.9 are used. The unit of all the parameters is atomic unit (a.u.). Statistical error bars are included. (a). For the "middle" scheme. Hollow symbols: numerical results for the virtual dynamics case. Solid symbols: numerical results for the real dynamics case. Solid lines: analytic results for the virtual dynamics case. Dotted lines: analytic results for the real dynamics case. "vir-num-0.3" represents the numerical results from the virtual dynamics case for $\Delta t = 0.3$; "real-num-0.3" stands for those from the real dynamics case for $\Delta t = 0.3$; "vir-ana-0.3" corresponds to the analytical results from the virtual dynamics case for $\Delta t = 0.3$; "real-ana-0.3" indicates those from the real dynamics case for $\Delta t = 0.3$; *etc.* (b). Same as Panel (a), but for the "side" scheme. (c). Same as Panel (a), but for the "middle-xT" scheme. No virtual dynamics case in the scheme. "num-0.3" represents the numerical results for $\Delta t = 0.3$; "ana-0.3" stands for the analytical results for $\Delta t = 0.3$; *etc.*

**Fig. S8**. Same as **Fig. S7**, but for the characteristic correlation time of the Hamiltonian.

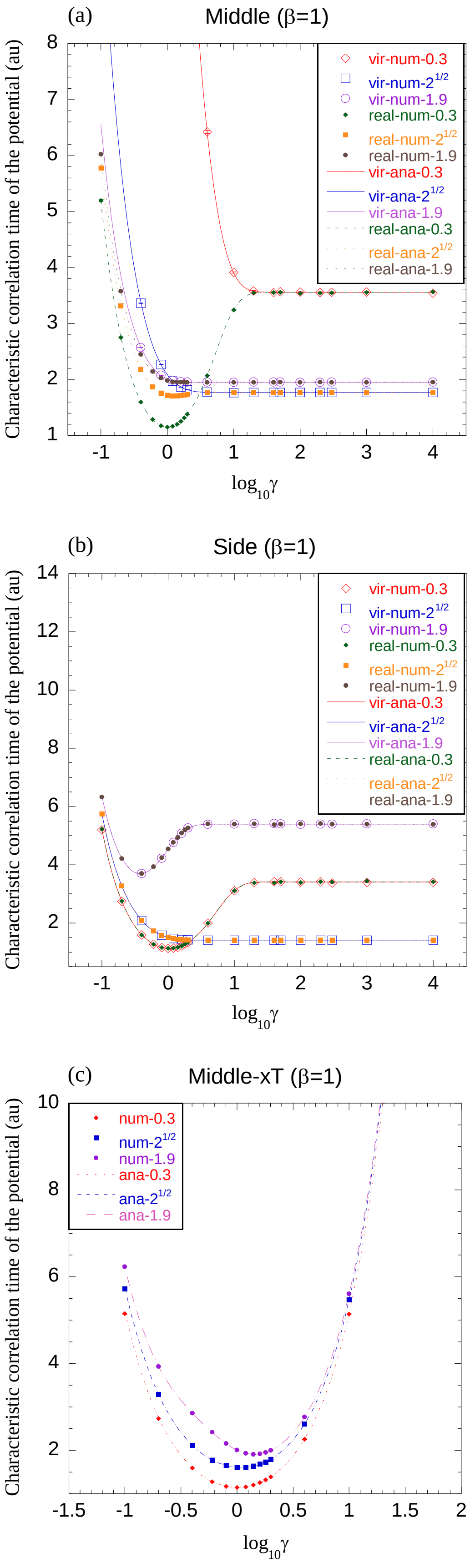

**Fig. S7**

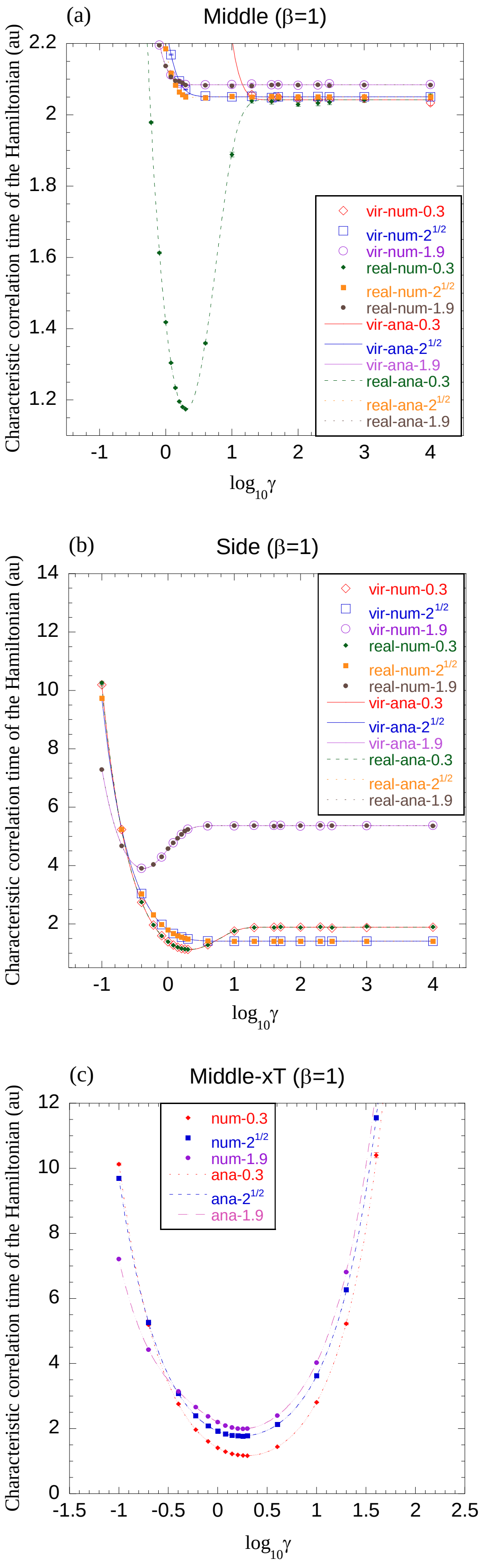

**Fig. S8**

## S3. More schemes that employ the first type of repartition

Here we propose eight more schemes that employ the first type of repartition [Eq. (10) in the paper]. The first four of these schemes can be reduced to the conventional velocity-Verlet algorithm when the Ornstein-Uhlenbeck (OU) process disappears.

**1) "pTxTp" scheme**

$$e^{\mathcal{L}\Delta t} \approx e^{\mathcal{L}^{\mathrm{pTxTp}}\Delta t} = e^{\mathcal{L}_{\mathrm{p}}\Delta t/2} e^{\mathcal{L}_{T}\Delta t/2} e^{\mathcal{L}_{\mathrm{x}}\Delta t} e^{\mathcal{L}_{T}\Delta t/2} e^{\mathcal{L}_{\mathrm{p}}\Delta t/2} \quad \text{(S44)}$$

The thermostat process for half an interval $\Delta t/2$ is arranged on both sides of the position propagator.

**2) "pTxp" scheme**

$$e^{\mathcal{L}\Delta t} \approx e^{\mathcal{L}^{\mathrm{pTxp}}\Delta t} = e^{\mathcal{L}_{\mathrm{p}}\Delta t/2} e^{\mathcal{L}_{T}\Delta t} e^{\mathcal{L}_{\mathrm{x}}\Delta t} e^{\mathcal{L}_{\mathrm{p}}\Delta t/2} \quad \text{(S45)}$$

The thermostat process is applied after the position propagator.

**3) "pxTp" scheme**

$$e^{\mathcal{L}\Delta t} \approx e^{\mathcal{L}^{\mathrm{pxTp}}\Delta t} = e^{\mathcal{L}_{\mathrm{p}}\Delta t/2} e^{\mathcal{L}_{\mathrm{x}}\Delta t} e^{\mathcal{L}_{T}\Delta t} e^{\mathcal{L}_{\mathrm{p}}\Delta t/2} \quad \text{(S46)}$$

The thermostat process is applied before the position propagator.

**4) "TpTxTpT" scheme**

$$e^{\mathcal{L}\Delta t} \approx e^{\mathcal{L}^{\mathrm{TpTxTpT}}\Delta t} = e^{\mathcal{L}_{T}\Delta t/4} e^{\mathcal{L}_{\mathrm{p}}\Delta t/2} e^{\mathcal{L}_{T}\Delta t/4} e^{\mathcal{L}_{\mathrm{x}}\Delta t} e^{\mathcal{L}_{T}\Delta t/4} e^{\mathcal{L}_{\mathrm{p}}\Delta t/2} e^{\mathcal{L}_{T}\Delta t/4} \quad \text{(S47)}$$

The thermostat process for a quarter of an interval $\Delta t/4$ is arranged at each of the two sides, and between each sub-step of the velocity-Verlet process.

Similarly, one can also obtain the four schemes that approach the position-Verlet algorithm as the OU process vanishes.

**5) "xTpTx" scheme**

$$e^{\mathcal{L}\Delta t} \approx e^{\mathcal{L}^{\mathrm{xTpTx}}\Delta t} = e^{\mathcal{L}_{\mathbf{x}}\Delta t/2} e^{\mathcal{L}_T \Delta t/2} e^{\mathcal{L}_{\mathbf{p}}\Delta t} e^{\mathcal{L}_T \Delta t/2} e^{\mathcal{L}_{\mathbf{x}}\Delta t/2} \tag{S48}$$

**6) "xTpx" scheme**

$$e^{\mathcal{L}\Delta t} \approx e^{\mathcal{L}^{\mathrm{xTpx}}\Delta t} = e^{\mathcal{L}_{\mathbf{x}}\Delta t/2} e^{\mathcal{L}_T \Delta t} e^{\mathcal{L}_{\mathbf{p}}\Delta t} e^{\mathcal{L}_{\mathbf{x}}\Delta t/2} \tag{S49}$$

**7) "xpTx" scheme**

$$e^{\mathcal{L}\Delta t} \approx e^{\mathcal{L}^{\mathrm{xpTx}}\Delta t} = e^{\mathcal{L}_{\mathbf{x}}\Delta t/2} e^{\mathcal{L}_{\mathbf{p}}\Delta t} e^{\mathcal{L}_T \Delta t} e^{\mathcal{L}_{\mathbf{x}}\Delta t/2} \tag{S50}$$

**8) "TxTpTxT" scheme**

$$e^{\mathcal{L}\Delta t} \approx e^{\mathcal{L}^{\mathrm{TxTpTxT}}\Delta t} = e^{\mathcal{L}_T \Delta t/4} e^{\mathcal{L}_{\mathbf{x}}\Delta t/2} e^{\mathcal{L}_T \Delta t/4} e^{\mathcal{L}_{\mathbf{p}}\Delta t} e^{\mathcal{L}_T \Delta t/4} e^{\mathcal{L}_{\mathbf{x}}\Delta t/2} e^{\mathcal{L}_T \Delta t/4} \tag{S51}$$

Using either the phase space propagator approach or the trajectory-based one described in the paper, it is straightforward to prove that the stationary density distribution for the one-dimensional harmonic system [Eq. (100) of the paper] of any one of the eight schemes shares the same form as Eq. (A34) in the paper, i.e.,

$$\rho(x,p) = \frac{1}{\tilde{Z}} \exp\left[ -\frac{1}{2} \left( \mathbf{R} - \bar{\mathbf{R}} \right)^T \mathbf{W}^{-1} \left( \mathbf{R} - \bar{\mathbf{R}} \right) \right] \tag{S52}$$

with $\bar{\mathbf{R}} = \left( x_{eq}, 0 \right)^T$, except that the corresponding fluctuation correlation matrix $\mathbf{W}$ is different ($\tilde{Z}$ is the normalization factor). The results are listed in Table S1. All the eight more schemes lead to the stationary distributions that depend on the friction coefficient $\gamma$ even in the harmonic limit. This suggests that results on thermodynamic properties are more sensitive to the parameter $\gamma$ (than the first eight schemes described in the paper).

**Table S1.** Fluctuation correlation matrices of the stationary state distributions [Eq. (S52)] for the eight more schemes (with the first type of reparation) for the one-dimensional harmonic system [Eq. (100) of the paper].

| Scheme | Fluctuation correlation matrix |
|---|---|
| pTxTp | $W_{xx}=\frac{1}{\beta A}\frac{e^{-\frac{1}{2}\gamma\Delta t}\left(1+e^{\gamma\Delta t}\right)^2}{\left[2\left(1+e^{\gamma\Delta t}\right)-e^{\frac{1}{2}\gamma\Delta t}\Delta t^2 AM^{-1}\right]}$ <br> $W_{xp}=W_{px}=\frac{1}{\beta}\frac{e^{-\frac{1}{2}\gamma\Delta t}\left(1-e^{2\gamma\Delta t}\right)\Delta t}{2\left[2\left(1+e^{\gamma\Delta t}\right)-e^{\frac{1}{2}\gamma\Delta t}\Delta t^2 AM^{-1}\right]}$ <br> $W_{pp}=\frac{M}{\beta}\frac{8\left(1+e^{\gamma\Delta t}\right)+e^{-\frac{1}{2}\gamma\Delta t}\left(1-6e^{\gamma\Delta t}+e^{2\gamma\Delta t}\right)\Delta t^2 AM^{-1}}{4\left[2\left(1+e^{\gamma\Delta t}\right)-e^{\frac{1}{2}\gamma\Delta t}\Delta t^2 AM^{-1}\right]}$ |
| pTxp | $W_{xx}=\frac{1}{\beta A}\frac{e^{-\gamma\Delta t}\left(1+e^{\gamma\Delta t}\right)^2}{\left[2\left(1+e^{\gamma\Delta t}\right)-e^{\gamma\Delta t}\Delta t^2 AM^{-1}\right]}$ <br> $W_{xp}=W_{px}=\frac{1}{\beta}\frac{e^{-\gamma\Delta t}\left(1-e^{2\gamma\Delta t}\right)\Delta t}{2\left[2\left(1+e^{\gamma\Delta t}\right)-e^{\gamma\Delta t}\Delta t^2 AM^{-1}\right]}$ <br> $W_{pp}=\frac{M}{\beta}\frac{\left(1+e^{-\gamma\Delta t}\right)\left[8e^{\gamma\Delta t}+\left(1-3e^{\gamma\Delta t}\right)\Delta t^2 AM^{-1}\right]}{4\left[2\left(1+e^{\gamma\Delta t}\right)-e^{\gamma\Delta t}\Delta t^2 AM^{-1}\right]}$ |
| pxTp | $W_{xx}=\frac{1}{\beta A}\frac{\left(1+e^{\gamma\Delta t}\right)^2}{\left[2\left(1+e^{\gamma\Delta t}\right)-\Delta t^2 AM^{-1}\right]}$ <br> $W_{xp}=W_{px}=\frac{1}{\beta}\frac{\left(1-e^{2\gamma\Delta t}\right)\Delta t}{2\left[2\left(1+e^{\gamma\Delta t}\right)-\Delta t^2 AM^{-1}\right]}$ <br> $W_{pp}=\frac{M}{\beta}\frac{\left(1+e^{\gamma\Delta t}\right)\left[8+\left(e^{\gamma\Delta t}-3\right)\Delta t^2 AM^{-1}\right]}{4\left[2\left(1+e^{\gamma\Delta t}\right)-\Delta t^2 AM^{-1}\right]}$ |
| TpTxTpT | $W_{xx}=\frac{1}{\beta A}\frac{4e^{-\frac{1}{4}\gamma\Delta t}\left(1+e^{\gamma\Delta t}\right)^2}{(1+e^{\frac{1}{2}\gamma\Delta t})\left[4\left(1+e^{\gamma\Delta t}\right)-e^{\frac{1}{4}\gamma\Delta t}(1+e^{\frac{1}{2}\gamma\Delta t})\Delta t^2AM^{-1}\right]}$ <br> $W_{xp}=W_{px}=\frac{1}{\beta}\frac{2(1-e^{\frac{1}{2}\gamma\Delta t})\left(1+e^{\gamma\Delta t}\right)\Delta t}{(1+e^{\frac{1}{2}\gamma\Delta t})\left[4\left(1+e^{\gamma\Delta t}\right)-e^{\frac{1}{4}\gamma\Delta t}(1+e^{\frac{1}{2}\gamma\Delta t})\Delta t^2AM^{-1}\right]}$ <br> $W_{pp}=\frac{M}{\beta}\frac{4\left[(1+e^{\frac{1}{2}\gamma\Delta t})\left(1+e^{\gamma\Delta t}\right)-e^{\frac{3}{4}\gamma\Delta t}AM^{-1}\right]}{(1+e^{\frac{1}{2}\gamma\Delta t})\left[4\left(1+e^{\gamma\Delta t}\right)-e^{\frac{1}{4}\gamma\Delta t}(1+e^{\frac{1}{2}\gamma\Delta t})\Delta t^2AM^{-1}\right]}$ |
| xTpTx | $W_{xx}=\frac{1}{\beta A}\cosh\left(\frac{1}{2}\gamma\Delta t\right)$ <br> $W_{xp}=W_{px}=0$ <br> $W_{pp}=\frac{M}{\beta}\frac{\cosh\left(\frac{1}{2}\gamma\Delta t\right)}{\left[\cosh\left(\frac{1}{2}\gamma\Delta t\right)-\frac{1}{4}\Delta t^2AM^{-1}\right]}$ |
| xTpx | $W_{xx}=\frac{1}{\beta A}\frac{1+e^{\gamma\Delta t}}{2}$ <br> $W_{xp}=W_{px}=0$ <br> $W_{pp}=\frac{M}{\beta}\frac{2\left(1+e^{\gamma\Delta t}\right)}{\left[2\left(1+e^{\gamma\Delta t}\right)-\Delta t^2AM^{-1}\right]}$ |
| xpTx | $W_{xx}=\frac{1}{\beta A}\frac{1+e^{-\gamma\Delta t}}{2}$ <br> $W_{xp}=W_{px}=0$ <br> $W_{pp}=\frac{M}{\beta}\frac{2\left(1+e^{-\gamma\Delta t}\right)}{\left[2\left(1+e^{-\gamma\Delta t}\right)-\Delta t^2AM^{-1}\right]}$ |
| TxTpTxT | $W_{xx}=\frac{1}{\beta A}\frac{\left[8e^{-\frac{1}{4}\gamma\Delta t}\left(1+e^{\gamma\Delta t}\right)\left(1+e^{\frac{1}{2}\gamma\Delta t}\right)^{2}-4\left(1+2e^{\frac{1}{2}\gamma\Delta t}+2e^{\gamma\Delta t}+e^{\frac{3}{2}\gamma\Delta t}\right)\Delta t^{2}AM^{-1}+e^{\frac{1}{4}\gamma\Delta t}\left(1+e^{\gamma\Delta t}\right)\Delta t^{4}\left(AM^{-1}\right)^{2}\right]}{4\left(1+e^{\frac{1}{2}\gamma\Delta t}\right)\left[4\left(1+e^{\gamma\Delta t}\right)-e^{\frac{1}{4}\gamma\Delta t}\left(1+e^{\frac{1}{2}\gamma\Delta t}\right)\Delta t^{2}AM^{-1}\right]}$ <br> $W_{xp}=W_{px}=\frac{1}{\beta}\frac{e^{\frac{1}{2}\gamma\Delta t}\left(e^{\frac{1}{2}\gamma\Delta t}-1\right)\Delta t^{3}AM^{-1}}{2\left(1+e^{\frac{1}{2}\gamma\Delta t}\right)\left[4\left(1+e^{\gamma\Delta t}\right)-e^{\frac{1}{4}\gamma\Delta t}\left(1+e^{\frac{1}{2}\gamma\Delta t}\right)\Delta t^{2}AM^{-1}\right]}$ <br> $W_{pp}=\frac{M}{\beta}\frac{\left(1+e^{\gamma\Delta t}\right)\left[4\left(1+e^{\frac{1}{2}\gamma\Delta t}\right)-e^{\frac{1}{4}\gamma\Delta t}\Delta t^{2}AM^{-1}\right]}{\left(1+e^{\frac{1}{2}\gamma\Delta t}\right)\left[4\left(1+e^{\gamma\Delta t}\right)-e^{\frac{1}{4}\gamma\Delta t}\left(1+e^{\frac{1}{2}\gamma\Delta t}\right)\Delta t^{2}AM^{-1}\right]}$ |

## S4. Fourth type of repartition

In the three types of repartition in the paper, the stochastic term and the friction one are arranged in the same part. The two terms may be separated in different parts. It then suggests the fourth type of repartition of the Langevin equation

$$\begin{bmatrix} d\mathbf{x}_t \\ d\mathbf{p}_t \end{bmatrix} = \underbrace{\begin{bmatrix} \mathbf{M}^{-1}\mathbf{p}_t \\ 0 \end{bmatrix} dt}_{\mathrm{x}} + \underbrace{\begin{bmatrix} 0 \\ -\nabla_{\mathbf{x}_t} U(\mathbf{x}_t) + \boldsymbol{\sigma}\mathbf{M}^{1/2} d\mathbf{W}_t \end{bmatrix} dt}_{\mathrm{pr}} + \underbrace{\begin{bmatrix} 0 \\ -\boldsymbol{\gamma}\mathbf{p}_t dt_t \end{bmatrix}}_{\mathrm{f}} \tag{S53}$$

for the Hamiltonian system defined by Eq. (1) of the paper. Note that in 2009 Davidchack *et al.* presented an algorithm[1] that also separates the stochastic term and the friction one, although their algorithm does not faithfully employ Eq. (S53).

When there is only the first term in the right hand side of Eq. (S53), then exact dynamics leads to the update relation

$$\begin{bmatrix} \mathbf{x}(t+\Delta t) \\ \mathbf{p}(t+\Delta t) \end{bmatrix} = \begin{bmatrix} \mathbf{x}(t) + \mathbf{M}^{-1}\mathbf{p}(t)\Delta t \\ \mathbf{p}(t) \end{bmatrix} \tag{S54}$$

Similarly, the other two solutions corresponding to the 2nd and 3rd terms respectively read

$$\begin{bmatrix} \mathbf{x}(t+\Delta t) \\ \mathbf{p}(t+\Delta t) \end{bmatrix} = \begin{bmatrix} \mathbf{x}(t) \\ \mathbf{p}(t) - \nabla U(\mathbf{x})\big|_{\mathbf{x}=\mathbf{x}(t)} \Delta t + \bar{\bar{\boldsymbol{\Omega}}}(t,\Delta t) \end{bmatrix} \tag{S55}$$

$$\begin{bmatrix} \mathbf{x}(t+\Delta t) \\ \mathbf{p}(t+\Delta t) \end{bmatrix} = \begin{bmatrix} \mathbf{x}(t) \\ e^{-\boldsymbol{\gamma}\Delta t}\mathbf{p}(t) \end{bmatrix} \tag{S56}$$

where

$$\bar{\bar{\boldsymbol{\Omega}}}(t,\Delta t) = \boldsymbol{\sigma}\mathbf{M}^{1/2} \int_t^{t+\Delta t} ds\ \boldsymbol{\eta}_s \quad . \tag{S57}$$

The white noise vector $\boldsymbol{\eta}_s$ in Eq. (S57) formally stands for $d\mathbf{W}_s/ds$, where $\mathbf{W}_s$ is a vector of $3N$-dimensional independent Wiener processes. In Eq. (S57) $\boldsymbol{\sigma} = \sqrt{2/\beta}\boldsymbol{\gamma}^{1/2}$ where $\boldsymbol{\gamma}$ is often a diagonal friction matrix with positive elements.

We only consider two symmetric schemes

$$e^{\mathcal{L}\Delta t} \approx e^{\mathcal{L}^{\text{f-pr-x-pr-f}}\Delta t} = e^{\mathcal{L}_{\text{f}}\Delta t/2} e^{\mathcal{L}_{\text{pr}}\Delta t/2} e^{\mathcal{L}_{\text{x}}\Delta t} e^{\mathcal{L}_{\text{pr}}\Delta t/2} e^{\mathcal{L}_{\text{f}}\Delta t/2} \quad \text{(S58)}$$

and

$$e^{\mathcal{L}\Delta t} \approx e^{\mathcal{L}^{\text{pr-x-f-x-pr}}\Delta t} = e^{\mathcal{L}_{\text{pr}}\Delta t/2} e^{\mathcal{L}_{\text{x}}\Delta t/2} e^{\mathcal{L}_{\text{f}}\Delta t} e^{\mathcal{L}_{\text{x}}\Delta t/2} e^{\mathcal{L}_{\text{pr}}\Delta t/2} \quad \text{(S59)}$$

Similar to the four schemes that employ the second and third type of repartition in the paper, it is straightforward to prove that the stationary density distribution for the one-dimensional harmonic system [Eq. (100) of the paper] of any one of the schemes shares the same form as Eq. (S52) of Section S2 [i.e., Eq. (A34) in the paper] except that the corresponding fluctuation correlation matrix is different. The results are listed in Table S2. These two schemes lead to the stationary distributions that depend on the friction coefficient $\gamma$ even in the harmonic limit. When the friction coefficient $\gamma$ approaches infinity, the stationary distribution is not well-defined. These suggest that numerical results on thermodynamic properties produced by the two schemes are sensitive to the parameter $\gamma$.

**Table S2.** Fluctuation correlation matrices in the stationary state distribution [Eq. (S52)] for the two schemes for the one-dimensional harmonic system [Eq. (100) of the paper].

| Scheme | Fluctuation correlation matrix |
|---|---|
| f-pr-x-pr-f | $W_{xx}=\frac{1}{\beta A}\frac{\left(1+e^{-2\gamma\Delta t}\right)\gamma\Delta t}{\left(1+e^{-\gamma\Delta t}\right)\left(1-e^{-\gamma\Delta t}\right)\left(1-\frac{1}{4}\Delta t^2 AM^{-1}\right)}$ <br> $W_{xp}=W_{px}=0$ <br> $W_{pp}=\frac{M}{\beta}\frac{\gamma\Delta t}{\sinh\gamma\Delta t}$ |
| pr-x-f-x-pr | $W_{xx}=\frac{1}{\beta A}\frac{\left(1+e^{-\gamma\Delta t}\right)\gamma\Delta t}{2\left(1-e^{-\gamma\Delta t}\right)\left(1-\frac{1}{4}\Delta t^2 AM^{-1}\right)}$ <br> $W_{xp}=W_{px}=\frac{1}{\beta}\frac{\gamma\Delta t^2}{4\left(-1+\frac{1}{4}\Delta t^2 AM^{-1}\right)}$ <br> $W_{pp}=\frac{M}{\beta}\frac{\left[8\left(1+e^{-2\gamma\Delta t}\right)-\left(1+e^{-\gamma\Delta t}\right)^2\Delta t^2 AM^{-1}\right]\gamma\Delta t}{8\left(1+e^{-\gamma\Delta t}\right)\left(1-e^{-\gamma\Delta t}\right)\left(1-\frac{1}{4}\Delta t^2 AM^{-1}\right)}$ |

**S5. Stationary state distribution for the one-dimensional harmonic system for the "Langevin A" algorithm of Davidchack *et. al***

The two theoretical approaches (the trajectory-based approach and the phase space propagator approach) introduced in the paper are useful to study the stationary state distribution for the harmonic system for numerical algorithms that employ the Lie-Trotter splitting. When a numerical algorithm does not involve the Lie-Trotter splitting, the phase space propagator approach is not convenient to deal with it, but the trajectory-based approach is still useful. An example is the "Langevin A" algorithm described in Ref. [1] of Davidchack *et. al*.

Below we use the trajectory-based approach to derive the stationary state distribution for the one-dimensional harmonic system [Eq. (100) of the paper].

Consider the harmonic system [Eq. (50) of the paper]. We may use the strategy in part 2 of section IV of the paper. Let $\mathbf{R}$ denote the phase-space point, namely

$$\mathbf{R}_n \equiv \begin{pmatrix} \mathbf{x}_n \\ \mathbf{p}_n \end{pmatrix} \quad . \tag{S60}$$

Similarly, define the averaged phased-space point $\bar{\mathbf{R}} \equiv \lim_{n\to\infty} \bar{\mathbf{R}}_n$ (here $\bar{\mathbf{R}}_n \equiv \langle \mathbf{R}_n \rangle$) and fluctuation correlation matrix $\mathbf{W} \equiv \lim_{n\to\infty} \left\langle \left(\mathbf{R}_n - \bar{\mathbf{R}}_n\right)\left(\mathbf{R}_n - \bar{\mathbf{R}}_n\right)^T \right\rangle$. The update of the phase-space point can be expressed as

$$\mathbf{R}_{n+1} = \tilde{\tilde{\mathbf{M}}}\mathbf{R}_n + \tilde{\mathbf{F}}_0 + \tilde{\tilde{\mathbf{\Omega}}}_n \tag{S61}$$

where

$$\tilde{\tilde{\mathbf{M}}} = \begin{pmatrix} \tilde{\tilde{\mathbf{M}}}_{\mathbf{xx}} & \tilde{\tilde{\mathbf{M}}}_{\mathbf{xp}} \\ \tilde{\tilde{\mathbf{M}}}_{\mathbf{px}} & \tilde{\tilde{\mathbf{M}}}_{\mathbf{pp}} \end{pmatrix} \quad , \tag{S62}$$

$$\tilde{\mathbf{F}}_0 = \begin{pmatrix} \frac{\Delta t^2}{2}\mathbf{M}^{-1}\mathbf{A}\mathbf{x}_{eq} \\ \Delta t e^{-\frac{1}{2}\gamma\Delta t}\left[\mathbf{1}-\frac{\Delta t^2}{4}\mathbf{A}\mathbf{M}^{-1}\right]\mathbf{A}\mathbf{x}_{eq} \end{pmatrix} , \tag{S63}$$

$$\tilde{\tilde{\boldsymbol{\Omega}}}_n = \begin{pmatrix} \Delta t\mathbf{M}^{-1}\sqrt{\frac{\mathbf{M}\gamma\Delta t}{2\beta}}\boldsymbol{\mu}_n \\ 2e^{-\frac{1}{2}\gamma\Delta t}\left[\mathbf{1}-\frac{\Delta t^2}{4}\mathbf{A}\mathbf{M}^{-1}\right]\sqrt{\frac{\mathbf{M}\gamma\Delta t}{2\beta}}\boldsymbol{\mu}_n \end{pmatrix} , \tag{S64}$$

with

$$\tilde{\tilde{\mathbf{M}}}_{\mathbf{xx}} = \mathbf{1}-\frac{\Delta t^2}{2}\mathbf{M}^{-1}\mathbf{A} \tag{S65}$$

$$\tilde{\tilde{\mathbf{M}}}_{\mathbf{xp}} = \frac{\Delta t}{2}\mathbf{M}^{-1}e^{-\frac{1}{2}\gamma\Delta t} \tag{S66}$$

$$\tilde{\tilde{\mathbf{M}}}_{\mathbf{px}} = -\Delta t\left[\mathbf{1}-\frac{\Delta t^2}{4}\mathbf{A}\mathbf{M}^{-1}\right]e^{-\frac{1}{2}\gamma\Delta t}\mathbf{A} \tag{S67}$$

$$\tilde{\tilde{\mathbf{M}}}_{\mathbf{pp}} = e^{-\gamma\Delta t}\left(\mathbf{1}-\frac{\Delta t^2}{2}\mathbf{A}\mathbf{M}^{-1}\right) \tag{S68}$$

and $\boldsymbol{\mu}_n$ is a standard-Gaussian-random-number vector as defined in Eq. (B4) in the paper. Given the initial condition $\mathbf{R}_0$, one may use the iteration relation Eq. (S61) to find the solution of $\mathbf{R}_n$,

$$\mathbf{R}_n = \tilde{\tilde{\mathbf{M}}}^n\mathbf{R}_0 + \sum_{j=0}^{n-1}\tilde{\tilde{\mathbf{M}}}^j\left(\tilde{\mathbf{F}}_0 + \tilde{\tilde{\boldsymbol{\Omega}}}_{n-1-j}\right) . \tag{S69}$$

Upon taking random average and letting $n \to \infty$, there yields the equilibrium phase-space point $\bar{\mathbf{R}}$ as

$$\bar{\mathbf{R}} = \left(\mathbf{1}-\tilde{\tilde{\mathbf{M}}}\right)^{-1}\tilde{\mathbf{F}}_0 = \begin{pmatrix} \mathbf{x}_{eq} \\ 0 \end{pmatrix} . \tag{S70}$$

We now turn to the one-dimensional harmonic system [Eq. (100) of the paper]. Note that now the block matrix $\tilde{\tilde{\mathbf{M}}}$ [Eq. (S62)] becomes an ordinary $2\times 2$ matrix with eigenvalues

$$\tilde{\varepsilon}_{1,2} = \frac{\tilde{T}}{2} \pm \sqrt{\frac{\tilde{T}^2}{4} - \tilde{D}} \tag{S71}$$

where

$$\begin{aligned} \tilde{T} &= \tilde{\tilde{M}}_{xx} + \tilde{\tilde{M}}_{pp} = \left(1 + e^{-\gamma\Delta t}\right)\left[1 - \frac{\Delta t^2}{2} A M^{-1}\right] , \\ \tilde{D} &= \tilde{\tilde{M}}_{xx}\tilde{\tilde{M}}_{pp} - \tilde{\tilde{M}}_{xp}\tilde{\tilde{M}}_{px} = e^{-\gamma\Delta t} \end{aligned} \tag{S72}$$

and eigenvectors

$$\tilde{\boldsymbol{v}}_{1,2} = \frac{1}{\sqrt{\tilde{\tilde{M}}_{xp}^2 + \left(\tilde{\varepsilon}_{1,2} - \tilde{\tilde{M}}_{xx}\right)^2}} \begin{pmatrix} \tilde{\tilde{M}}_{xp} \\ \tilde{\varepsilon}_{1,2} - \tilde{\tilde{M}}_{xx} \end{pmatrix} . \tag{S73}$$

Therefore, $\tilde{\tilde{\mathbf{M}}}$ can be diagonalized by $\tilde{\mathbf{V}} = (\tilde{\boldsymbol{v}}_1, \tilde{\boldsymbol{v}}_2)$, namely

$$\tilde{\mathbf{V}}^{-1}\tilde{\tilde{\mathbf{M}}}\tilde{\mathbf{V}} = \begin{pmatrix} \tilde{\varepsilon}_1 & 0 \\ 0 & \tilde{\varepsilon}_2 \end{pmatrix} \equiv \tilde{\boldsymbol{\Lambda}} , \tag{S74}$$

or $\tilde{\tilde{\mathbf{M}}} = \tilde{\mathbf{V}}\tilde{\boldsymbol{\Lambda}}\tilde{\mathbf{V}}^{-1}$. With this result the powers of $\tilde{\tilde{\mathbf{M}}}$ can be written explicitly,

$$\begin{aligned} \tilde{\tilde{\mathbf{M}}}^n &= \tilde{\mathbf{V}}\tilde{\boldsymbol{\Lambda}}^n\tilde{\mathbf{V}}^{-1} \\ &= \begin{pmatrix} \tilde{v}_{11}\tilde{u}_{11}\tilde{\varepsilon}_1^n + \tilde{v}_{12}\tilde{u}_{21}\tilde{\varepsilon}_2^n & \tilde{v}_{11}\tilde{u}_{12}\tilde{\varepsilon}_1^n + \tilde{v}_{12}\tilde{u}_{22}\tilde{\varepsilon}_2^n \\ \tilde{v}_{21}\tilde{u}_{11}\tilde{\varepsilon}_1^n + \tilde{v}_{22}\tilde{u}_{21}\tilde{\varepsilon}_2^n & \tilde{v}_{21}\tilde{u}_{12}\tilde{\varepsilon}_1^n + \tilde{v}_{22}\tilde{u}_{22}\tilde{\varepsilon}_2^n \end{pmatrix} , \end{aligned} \tag{S75}$$

where $\tilde{v}_{i,j}$ and $\tilde{u}_{i,j} (i, j = 1, 2)$ are the elements of $\tilde{\mathbf{V}}$ and $\tilde{\mathbf{V}}^{-1}$ respectively. Substituting Eq. (S75) into Eq. (S69), we may obtain the deviations of the position and momentum from their equilibrium values,

$$\Delta x_n = x_n - \overline{x}_n = \sum_{j=0}^{n-1}\left(\tilde{c}_{x1}\tilde{\varepsilon}_1^j + \tilde{c}_{x2}\tilde{\varepsilon}_2^j\right)\sqrt{\frac{M\gamma\Delta t}{2\beta}}\mu_{n-1-j} \tag{S76}$$

$$\Delta p_n = p_n - \overline{p}_n = \sum_{j=0}^{n-1}\left(\tilde{c}_{p1}\tilde{\varepsilon}_1^j + \tilde{c}_{p2}\tilde{\varepsilon}_2^j\right)\sqrt{\frac{M\gamma\Delta t}{2\beta}}\mu_{n-1-j} , \tag{S77}$$

where these coefficients are

$$\tilde{c}_{x1} = \frac{\Delta t}{M}\tilde{v}_{11}\tilde{u}_{11} + 2e^{-\frac{1}{2}\gamma\Delta t}\left(1 - \frac{\Delta t^2}{4}\frac{A}{M}\right)\tilde{v}_{11}\tilde{u}_{12}$$

$$\tilde{c}_{x2} = \frac{\Delta t}{M}\tilde{v}_{12}\tilde{u}_{21} + 2e^{-\frac{1}{2}\gamma\Delta t}\left(1 - \frac{\Delta t^2}{4}\frac{A}{M}\right)\tilde{v}_{12}\tilde{u}_{22}$$

$$\tilde{c}_{p1} = \frac{\Delta t}{M}\tilde{v}_{21}\tilde{u}_{11} + 2e^{-\frac{1}{2}\gamma\Delta t}\left(1 - \frac{\Delta t^2}{4}\frac{A}{M}\right)\tilde{v}_{21}\tilde{u}_{12}$$

$$\tilde{c}_{p2} = \frac{\Delta t}{M}\tilde{v}_{22}\tilde{u}_{21} + 2e^{-\frac{1}{2}\gamma\Delta t}\left(1 - \frac{\Delta t^2}{4}\frac{A}{M}\right)\tilde{v}_{22}\tilde{u}_{22} \quad .$$

With these expressions we are able to calculate the required fluctuation correlations

$$W_{x_n x_n} = \tilde{Q}\left[\tilde{c}_{x1}^2 \frac{1-\tilde{\varepsilon}_1^{2n}}{1-\tilde{\varepsilon}_1^2} + 2\tilde{c}_{x1}\tilde{c}_{x2}\frac{1-\tilde{\varepsilon}_1^n\tilde{\varepsilon}_2^n}{1-\tilde{\varepsilon}_1\tilde{\varepsilon}_2} + \tilde{c}_{x2}^2\frac{1-\tilde{\varepsilon}_2^{2n}}{1-\tilde{\varepsilon}_2^2}\right] \tag{S78}$$

$$\begin{aligned} W_{x_n p_n} &= W_{p_n x_n} \\ &= \tilde{Q}\left[\tilde{c}_{x1}\tilde{c}_{p1}\frac{1-\tilde{\varepsilon}_1^{2n}}{1-\tilde{\varepsilon}_1^2} + \left(\tilde{c}_{x1}\tilde{c}_{p2} + \tilde{c}_{x2}\tilde{c}_{p1}\right)\frac{1-\tilde{\varepsilon}_1^n\tilde{\varepsilon}_2^n}{1-\tilde{\varepsilon}_1\tilde{\varepsilon}_2} + \tilde{c}_{x2}\tilde{c}_{p2}\frac{1-\tilde{\varepsilon}_2^{2n}}{1-\tilde{\varepsilon}_2^2}\right] \end{aligned} \tag{S79}$$

$$W_{p_n p_n} = \tilde{Q}\left[\tilde{c}_{p1}^2 \frac{1-\tilde{\varepsilon}_1^{2n}}{1-\tilde{\varepsilon}_1^2} + 2\tilde{c}_{p1}\tilde{c}_{p2}\frac{1-\tilde{\varepsilon}_1^n\tilde{\varepsilon}_2^n}{1-\tilde{\varepsilon}_1\tilde{\varepsilon}_2} + \tilde{c}_{p2}^2\frac{1-\tilde{\varepsilon}_2^{2n}}{1-\tilde{\varepsilon}_2^2}\right] \quad , \tag{S80}$$

where $\tilde{Q} = \frac{M\gamma\Delta t}{2\beta}$. It should be stressed that we are only interested in the stable iteration of the phase-space points, which requires $\left|\tilde{\varepsilon}_{1,2}\right| < 1$. Otherwise, the fluctuation correlation matrix elements Eqs. (S78), (S79), and (S80) will be divergent. To complete the involved algebra, we may follow part 2B of section IV of the paper to parameterize the matrix $\tilde{\tilde{\mathbf{M}}}$. Then after a similar process, we may obtain the fluctuation matrix elements

$$W_{xx} \equiv \lim_{n\to\infty} W_{x_n x_n} = \frac{1}{\beta A}\frac{\left[\left(1+e^{-\gamma\Delta t}\right)^2 - e^{-\gamma\Delta t}\Delta t^2 A M^{-1}\right]\gamma\Delta t}{2\left(1+e^{-\gamma\Delta t}\right)\left(1-e^{-\gamma\Delta t}\right)\left(1-\frac{1}{4}\Delta t^2 A M^{-1}\right)} \tag{S81}$$

$$W_{xp} \equiv \lim_{n\to\infty} W_{x_n p_n} = W_{px} = \frac{1}{\beta}\frac{e^{-\frac{1}{2}\gamma\Delta t}\gamma\Delta t^2}{2\left(1+e^{-\gamma\Delta t}\right)} \tag{S82}$$

$$W_{pp} \equiv \lim_{n\to\infty} W_{p_n p_n} = \frac{M}{\beta}\frac{2e^{-\gamma\Delta t}\left(1-\frac{1}{4}\Delta t^2 A M^{-1}\right)\gamma\Delta t}{\left(1+e^{-\gamma\Delta t}\right)\left(1-e^{-\gamma\Delta t}\right)} \tag{S83}$$

As suggested by Eqs. (S81)-(S83), the stationary state distribution depends on the friction coefficient $\gamma$ for the harmonic system. As $\gamma \to \infty$ the stationary state distribution is not even well-defined.

**S6. Some comments on the order of the schemes in the paper**

All the algorithms of the three types of repartition in the paper employ the Lie-Trotter splitting. The splitting schemes are either symmetric (“middle”/“side”/“PV-middle”/“PV-side”/“middle-pT”/“side-pT”/“middle-xT”/“side-xT”) or asymmetric (“end”/“beginning”/“PV-end”/“PV-beginning”). We may follow the work of Leimkuhler *et al.*[2-4] to show that the symmetric versions lead to second-order splitting schemes with respect to $\Delta t$, while the asymmetric ones yield first-order splitting schemes. Regardless of whether first-order or second-order splitting schemes are involved, all the algorithms that we present in the three types of repartition produce stationary state distributions with second-order accuracy with respect to $\Delta t$. So we note all of them “second-order algorithms” in terms of accuracy in the paper.

**Acknowledgements**

This work was supported by the Ministry of Science and Technology of China (MOST) Grant No. 2016YFC0202803, by the 973 program of MOST No. 2013CB834606, by the National Natural Science Foundation of China (NSFC) Grants No. 21373018, No. 21573007, and No. 21421003, by the Recruitment Program of Global Experts, by Specialized Research Fund for the Doctoral Program of Higher Education No. 20130001110009, and by Special

Program for Applied Research on Super Computation of the NSFC-Guangdong Joint Fund (the second phase) under Grant No.U1501501. We acknowledge the Beijing and Tianjin supercomputer centers for providing computational resources. This research also used resources of the National Energy Research Scientific Computing Center, a DOE Office of Science User Facility supported by the Office of Science of the U.S. Department of Energy under Contract No. DE-AC02-05CH11231.